\documentclass[12pt]{article}

\usepackage{amsmath,amsfonts,latexsym,amssymb,amscd,amsbsy,dsfont}
\usepackage{pslatex}
\usepackage[latin1]{inputenc}
\usepackage[T1]{fontenc}
\usepackage{pspicture}
\usepackage{verbatim,amsthm,curves,graphics}
\usepackage{mathrsfs}
\usepackage[dvips]{graphicx}
\usepackage{pstricks}
\usepackage{color}

\usepackage{tikz}
\usetikzlibrary{shapes.geometric}
\usetikzlibrary{decorations.markings}
\usetikzlibrary{decorations.pathmorphing}
\def\f(#1){{\mathop{f}^{(#1)}}}
\def\m(#1){{\mathop{m}^{(#1)}}}
\def\C(#1){{\mathop{C}^{(#1)}}}
\def\p(#1){{\mathop{p}^{(#1)}}}

\def\ben{\begin{equation}}
\def\een{\end{equation}}
\def\bena{\begin{eqnarray}}
\def\eena{\end{eqnarray}}
\def\non{\nonumber}

\def\TT{{\mathscr T}}
\def\V{{\cal V}}

\def\I{{\mathfrak I}}
\def\R{{\mathfrak R}}
\def\V{{\mathcal V}}

\def\mr{{\mathbb R}}

\newcommand{\T}{{\mathfrak T}}

\newcommand{\mn}{{\mathbb N}}
\newcommand{\mz}{{\mathbb Z}}

\newcommand{\ren}{{\rm R}}
\newcommand{\mc}{{\mathbb C}}

\newcommand{\F}{{\mathcal F}}
\newcommand{\hF}{{}_2 F_1}

\renewcommand{\C}{{\mathcal C}}

\newcommand{\U}{{\mathcal U}}

\newcommand{\acon}{\operatorname*{anal.cont.}}
\newcommand{\E}{{\rm E}}
\newcommand{\e}{{\rm e}}
\newcommand{\pp}{\operatorname{pp}}
\newcommand{\pre}{{\rm P}}
\theoremstyle{definition}

\newtheorem{thm}{Theorem}

\begin{document}

\title{Correlators, Feynman diagrams, and quantum no-hair in deSitter spacetime}

\author{
Stefan Hollands$^{1}$\thanks{\tt HollandsS@Cardiff.ac.uk}\:,
\\ \\
{\it ${}^{1}$School of Mathematics,
     Cardiff University, UK}
}
\date{29 November 2010}
\maketitle

\begin{abstract}
We provide a parametric representation for position-space Feynman integrals of a massive, self-interacting
scalar field in deSitter spacetime, for an arbitrary graph. The expression is
given as a multiple contour integral over a kernel whose structure is determined by the
set of all trees (or forests) within the graph, and it belongs to a class of generalized hypergeometric
functions. We argue from this representation that connected deSitter $n$-point vacuum correlation functions
have exponential decay for large proper time-separation, and also decay for large spatial separation, to arbitrary orders in perturbation theory. Our results
may be viewed as an analog of the so-called cosmic-no-hair theorem in the context of a quantized test scalar field. This work has
significant overlap with a paper by Marolf and Morrison, which is being released simultaneously.
\end{abstract}

\pagebreak

\tableofcontents

\pagebreak


\sloppy

\section{Introduction}

DeSitter space is a model for an expanding universe that is believed to describe
both the earliest epoch of our universe, as well as the present epoch of accelerated expansion.
It is therefore important to understand well the dynamics of classical as well as quantized fields on
this spacetime. In quantum field theory, one wants to understand the behavior of correlation functions
of the field(s) of the theory, such as the 1-point function $\langle \phi_A(x) \rangle_\Psi$, 2-point function
$\langle \phi_A(x) \phi_B(y) \rangle_\Psi$, etc. where $\Psi$ is some quantum state, and $\phi_A$ are some
(possibly composite-) quantum fields.

When the points $x$ and $y$ are a proper distance or proper time apart which is not
larger than the Hubble radius, $H^{-1}$, then
one expects intuitively that the behavior of correlation functions in a generic state will not be substantially different from a ``corresponding'' state in Minkowski spacetime. This statement can be made somewhat more precise via the operator product expansion, which states that
\ben\label{ope}
\langle \phi_A(x) \phi_B(y) \rangle_\Psi \sim \sum_C C^C_{AB}(x,y) \ \langle \phi_C(y) \rangle_\Psi \ .
\een
Each coefficient $C_{AB}^C$ itself is known~\cite{hollands2}, at least in perturbation theory, to have an expansion in terms of the
geodesic distance between $x,y$ and curvature invariants at $y$. Thus, for distances or proper times much less than the Hubble
radius (and masses greater than $H$), the coefficients will essentially be equal to those in Minkowski spacetime.
As a consequence, for a state $\Psi$ on deSitter such that the form factors $\langle \phi_C(y) \rangle_\Psi$ at
point $y$ are approximately equal to those of a corresponding state on Minkowski spacetime, the deSitter 2-point correlation function
will not essentially differ from the Minkowski correlation function of that state near $y$. Thus, in this sense, one does
not expect large departures of deSitter correlators from corresponding Minkowski correlators at small distances.

For large distances, however, correlators could in principle show an unexpected behavior, and this intriguing
possibility has been discussed by many authors in the literature, and for various theories, see e.g.~\cite{woodard,higuchi,higuchi1,polyakov,bros1,tanaka,mottola,marolf1}, with partly contradicting claims. The simplest theory one can study is that of a free, minimally coupled Klein-Gordon field $\phi$ with mass parameter $m^2$. When the mass parameter is strictly positive, then there exists a unique deSitter invariant Hadamard-state~\cite{allen,mottola,bros}, sometimes called ``Bunch-Davies-'' or also ``Hartle-Hawking-'' or
``Euclidean-'' state. For large proper time-like separations $\tau$, the 2-point function of this state decays as $\e^{-M|\tau|}$ for some $M>0$ depending on $H$ and the mass, and a similar statement also applies to the connected higher $n$-point functions. Letting the Hilbert space vector representing the deSitter invariant state be $|0\rangle$, we may then also form a class of new states
\ben
|\Psi \rangle = \sum_n \int f_n(x_1, \dots, x_n) \ \phi(x_1) \dots \phi(x_n) \ |0\rangle
\een
by applying a smeared product of operators to it, where the $f_n$'s are smooth smearing functions vanishing outside some compact region. From the exponential decay of the
$n$-point functions of $|0\rangle$, it then follows that e.g.
\ben
\langle \phi(x) \rangle_\Psi = O(\e^{-M|\tau|})
\een
as the proper time separation $\tau$ of $x$ relative to any reference point goes to infinity. By the Reeh-Schlieder
theorem~\cite{stroh} in curved spacetime, the class of such states $|\Psi\rangle$ is dense in the Hilbert space
built upon $|0\rangle$. Therefore, this result states that the expectation value of $\phi$ decays exponentially in time in {\em generic} states. A similar statement also holds for higher $n$-point correlation functions.
An analogous result for the graviton field $h_{\mu\nu}$, with the appropriate value of $M$, could be viewed as a quantum analog of a ``cosmic no-hair theorem''.

Because the graviton field is interacting with itself, it is of interest to ask whether a similar exponential decay property holds for the (connected) $n$-point function in an interacting theory, e.g. when the self-interaction
$\lambda \phi^4$ is turned on in massive Klein-Gordon theory. This question has been analyzed at 1-loop level in~\cite{marolf1} by an explicit calculation, and it was found that the interaction does not change the exponential decay of the vacuum $2$-point function. In the present paper, we
consider this question for an arbitrary number of loops, for an arbitrary $n$-point function, and any dimension $D \ge 2$. We find that, again, the behavior does not change, so in this sense the quantum cosmic no hair theorem holds to all orders in
renormalized perturbation theory in the analog system of a self-interacting test scalar field on deSitter. The decay constant
$M$ is the same in the interacting theory as in the free theory, i.e. it does not receive perturbative corrections.
Our analysis is carried out at the level of individual Feynman graphs $G$, and the valence of the interaction
vertices ($=4$ in $\lambda \phi^4$-theory) is not important for our analysis. It therefore also applies
to any other polynomial interaction. Of course, for a general polynomial interaction, and general dimension $D$,
the theory is to be viewed then as an effective theory only, and the usual limitations concerning the interpretation apply.
The same result has been obtained simultaneously, using a different method, by~\cite{marolf2}\footnote{
We are grateful to the authors for making available to us their manuscript before publication.
}.

In order to demonstrate exponential decay in time for the perturbative corrections to the correlation functions, we first develop a parametric representation for arbitrary renormalized Feynman integrals which is also of some interest in its own right. Our representation is based on the use of a Mellin-representation of the free propagator, which is combined with a version of the Schwinger parameter trick, and with some results from graph theory. The renormalization procedure is essentially that
developed by~\cite{hollandswald1,hollandswald2,brunetti}, specialized to deSitter spacetime and taking also advantage of
many simplifications that occur due to the large symmetry group of this spacetime. Our final formula for the renormalized position space Feynman integrals $I_G$ is eq.~\eqref{ig1} as well as its Mellin-Barnes counterpart~\eqref{ig4}, which we repeat here:
\ben
 I_G(X_1, \dots, X_E) =  \ K_G \ \int_{\vec w}
\Gamma_G (  \vec w  ) \
\prod_{1 \le r \neq s \le E} ( 1-Z_{rs} )^{\sum_{F \in \T_E(r,s)}  w_F} \ .
\een
$\Gamma_G$ is a kernel made from products of gamma-functions, and the quantities $Z_{rs}$ are
deSitter space invariants related in a simple way to the signed squared geodesic distance between each pair
of deSitter points $X_r$ and $X_s$. There is one path of integration over a complex variable $w_F$ for each ``forest'' $F$ in the underlying
Feynman graph $G$, and the forests appearing in the exponent are precisely those connecting the
external leg $X_r$ with $X_s$. For example, for the following Feynman graph $G$ in $\lambda \phi^4$-theory
with four external legs labeled $X_1, \dots, X_4$,
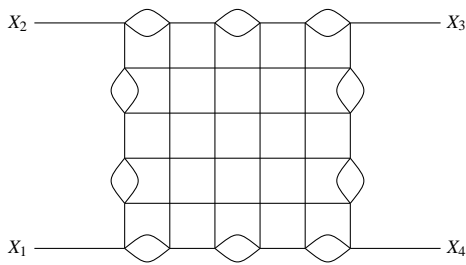
\begin{center}
\begin{tikzpicture}[scale=.6, transform shape]
\draw[black] (0,5) .. controls (.5,5.4) and (.5,5.4) .. (1,5);
\draw[black] (0,5) .. controls (.5,4.6) and (.5,4.6) .. (1,5);
\draw[black] (2,5) .. controls (2.5,5.4) and (2.5,5.4) .. (3,5);
\draw[black] (2,5) .. controls (2.5,4.6) and (2.5,4.6) .. (3,5);
\draw[black] (4,5) .. controls (4.5,5.4) and (4.5,5.4) .. (5,5);
\draw[black] (4,5) .. controls (4.5,4.6) and (4.5,4.6) .. (5,5);
\draw[black] (0,0) .. controls (.5,.4) and (.5,.4) .. (1,0);
\draw[black] (0,0) .. controls (.5,-.4) and (.5,-.4) .. (1,0);
\draw[black] (2,0) .. controls (2.5,.4) and (2.5,.4) .. (3,0);
\draw[black] (2,0) .. controls (2.5,-.4) and (2.5,-.4) .. (3,0);
\draw[black] (4,0) .. controls (4.5,.4) and (4.5,.4) .. (5,0);
\draw[black] (4,0) .. controls (4.5,-.4) and (4.5,-.4) .. (5,0);
\draw[black] (0,1) .. controls (-.4,1.5) and (-.4,1.5) .. (0,2);
\draw[black] (0,1) .. controls (.4,1.5) and (.4,1.5) .. (0,2);
\draw[black] (0,3) .. controls (-.4,3.5) and (-.4,3.5) .. (0,4);
\draw[black] (0,3) .. controls (.4,3.5) and (.4,3.5) .. (0,4);
\draw[black] (5,1) .. controls (4.6,1.5) and (4.6,1.5) .. (5,2);
\draw[black] (5,1) .. controls (5.4,1.5) and (5.4,1.5) .. (5,2);
\draw[black] (5,3) .. controls (4.6,3.5) and (4.6,3.5) .. (5,4);
\draw[black] (5,3) .. controls (5.4,3.5) and (5.4,3.5) .. (5,4);
\draw[black] (1,5) -- (2,5);
\draw[black] (3,5) -- (4,5);
\draw[black] (1,0) -- (2,0);
\draw[black] (3,0) -- (4,0);
\draw[black] (0,0) -- (0,1);
\draw[black] (0,2) -- (0,3);
\draw[black] (0,4) -- (0,5);
\draw[black] (5,0) -- (5,1);
\draw[black] (5,2) -- (5,3);
\draw[black] (5,4) -- (5,5);
\draw[black] (0,4) -- (5,4);
\draw[black] (0,3) -- (5,3);
\draw[black] (0,2) -- (5,2);
\draw[black] (0,1) -- (5,1);
\draw[black] (1,0) -- (1,5);
\draw[black] (2,0) -- (2,5);
\draw[black] (3,0) -- (3,5);
\draw[black] (4,0) -- (4,5);
\draw[black] (-2,5) node[black,left]{$X_2$} -- (0,5);
\draw[black] (-2,0) node[black,left]{$X_1$} -- (0,0);
\draw[black] (5,5) -- (7,5) node[black,right]{$X_3$};
\draw[black] (5,0) -- (7,0) node[black,right]{$X_4$};
\end{tikzpicture}
\end{center}
an associated forest \textcolor{red}{$F$} in $\T_{4}(2,4)$ connecting
$X_2$ with $X_4$ can look like this:
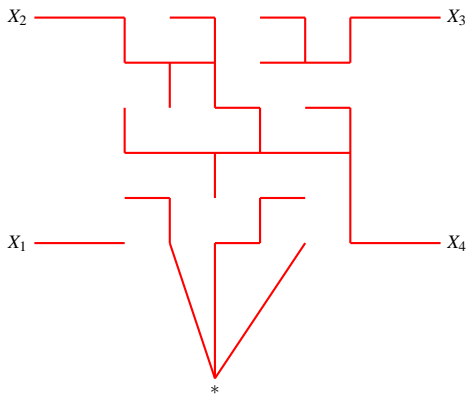
\begin{center}
\begin{tikzpicture}[scale=.6, transform shape]
\draw[red,thick] (0,2) -- (0,3);
\draw[red,thick] (0,4) -- (0,5);
\draw[red,thick] (1,0) -- (1,1);
\draw[red,thick] (1,3) -- (1,4);
\draw[red,thick] (2,1) -- (2,2);
\draw[red,thick] (2,3) -- (2,5);
\draw[red,thick] (3,0) -- (3,1);
\draw[red,thick] (3,2) -- (3,3);
\draw[red,thick] (4,4) -- (4,5);
\draw[red,thick] (5,0) -- (5,3);
\draw[red,thick] (5,4) -- (5,5);
\draw[red,thick] (-2,5) node[black,left]{$X_2$} -- (0,5);
\draw[red,thick] (1,5) -- (2,5);
\draw[red,thick] (3,5) -- (4,5);
\draw[red,thick] (0,4) -- (2,4);
\draw[red,thick] (3,4) -- (5,4);
\draw[red,thick] (2,3) -- (3,3);
\draw[red,thick] (4,3) -- (5,3);
\draw[red,thick] (0,2) -- (5,2);
\draw[red,thick] (0,1) -- (1,1);
\draw[red,thick] (3,1) -- (4,1);
\draw[red,thick] (-2,0) node[black,left]{$X_1$} -- (0,0);
\draw[red,thick] (2,0) -- (3,0);
\draw[red,thick] (5,0) -- (7,0) node[black,right]{$X_4$};
\draw[red,thick] (2,-3) node[black,below]{$*$} -- (1,0);
\draw[red,thick] (2,-3) -- (2,0);
\draw[red,thick] (2,-3) -- (4,0);
\draw[red,thick] (5,5) -- (7,5) node[black,right]{$X_3$};
\end{tikzpicture}
\end{center}

As we shall explain, the forests are closely related to graph polynomials of the underlying Feynman graph~\cite{tutte}. Such graph polynomials have also appeared previously in the study of Minkowski
space Feynman integrals. However, our setup is different from that in Minkowski spacetime (see e.g.~\cite{weinzierl2,rivasseau}) due to the fact that (a) we work in position space, that (b) there are additional
integrals of Mellin-Barnes type, and that (c) our graphs effectively involve an additional ``virtual vertex'', called
``$*$'' throughout this paper, as indicated in the above picture. The appearance of this virtual 
vertex is related to the fact that the loop integrations are over the deSitter hyperboloid, rather than flat space.
 To demonstrate the exponential decay of the connected correlation functions to arbitrary
orders in perturbation theory, we use the above Mellin-Barnes representation and the explicit knowledge
of the poles of the meromorphic kernel $\Gamma_G$. Our Mellin-Barnes represenation also shows that $I_G$ belongs to
a class of special functions that have been called generalized $H$-functions, see appendix~\ref{app:A}.

In this work, we explicitly only treat massive scalar fields in the so-called ``principal series.'' This assumption is made to simplify
the discussion; massive scalar fields in the ``complementary series'' could presumably be treated by our method as well,
and the only difference would be a correspondingly weaker exponential decay. However, our method would definitely fail for massless scalar fields, for which no deSitter invariant state exists even in the free theory~\cite{allen}. This case is a
physically very important one, because interesting IR-effects are most often found in massless theories, and also
because massless fields play a prominent role in various inflation scenarios. Some progress along these lines has been
made e.g. by~\cite{tanaka}, but it appears unknown whether correlators in a generic, non-deSitter-invariant state
decay exponentially in time or not. We also do not consider higher spin fields, such as the graviton field. For this field,
a deSitter invariant propagator exists, but one has to deal with issues related to gauge invariance, see e.g.
\cite{woodard,higuchi} for related discussions. We think that it would be very interesting to extend our analysis to such fields. Finally, it would be interesting to extend our results to non-perturbative constructions of the
interacting Klein-Gordon quantum field theory on deSitter spacetime, which are available in two spacetime dimensions~\cite{jaekel}.

\medskip
\noindent
{\bf Notations and Conventions:} A graph $G$ is a collection of vertices ${\rm V} G = \{1, \dots, V+E\}$, together with
a set of lines $\E G = \{l_1, \dots, l_r\}$, each of which connects a pair of distinct vertices. Our
graphs have no ``tadpoles''.
DeSitter points are typically denoted by capital letters such as $X$. $d\mu=|g|^{\frac{1}{2}} \ d^D x$ is the invariant integration
measure on $D$-dimensional deSitter or $S^D$, depending on the context. $\I(z), \R(z)$ denote the imaginary-
resp. real parts of a complex number.

\section{DeSitter space basics}\label{sec:basics}

DeSitter spacetime can be characterized in a variety of different ways.
Maybe the most natural description of this spacetime is as the embedded hyperboloid
\ben\label{1}
dS_D = \{ X \in \mr^{D+1} \mid X \cdot X \equiv -X_0^2+X_1^2+...+X_D^2 = H^{-2} \}
\een
in an ambient $(D+1)$-dimensional Minkowski spacetime, with the induced metric. As is evident from this description,
the isometry group of $dS_D$ is $O(D,1)$, and it is hence a space of maximal symmetry~\footnote{This means that
the number of Killing vectors, $\frac{1}{2} D(D+1)$, is the maximum number that any (pseudo-) Riemannian
manifold can have.}. DeSitter spacetime is
a solution to the Einstein equations  with a positive cosmological constant $\Lambda =
\frac{1}{2}(D-1)(D-2)H^2$, and with a vanishing Weyl tensor. It is entirely well-behaved
from the point of view of causality---technically speaking, it is
an example of a ``globally hyperbolic spacetime''. As a consequence, the initial value problem for an equation like the Klein-Gordon equation
\ben
(\nabla^2 - m^2) \ \phi = 0
\een
is well-posed, for any value of $m^2$, including negative.
Although deSitter space has the same number of Killing vector fields
as Minkowski space, it has no analog of time translations, i.e. does not possess a Killing vector that is timelike everywhere. While this does not indicate any kind of causal pathology, it means that one cannot form a positive definite conserved Hamiltonian (energy) from the conserved stress tensor $T_{\mu\nu}$ of the Klein-Gordon field, or similar other fields, although this is possible in certain sub-regions of deSitter spacetime, see below. In quantum field theory on deSitter spacetime, it also makes it impossible to define a reasonable global notion of particle with similar properties as in Minkowski spacetime.

The deSitter hyperboloid admits various natural slicings and coordinate systems, some of which we consider in this paper.
When we slice deSitter spacetime by a family of exponentially contracting, then expanding,
round $S^{D-1}$-spheres of constant $X_0$, the metric reads:
\ben
ds^2 = -d\tau^2 + H^{-2} \cosh^2 (H\tau) \ d\omega_{D-1}^2  \  .
\een
The coordinate $\tau$ is the proper time of the geodesic curves of constant angle on the sphere.
This coordinate system is useful e.g. to find the conformal diagram of deSitter
spacetime. By making
the coordinate transformation $t=2 \arctan(\e^{H\tau}) \in (-\pi, \pi)$, the line element is brought into the form
\ben
ds^2 = \Omega^{-2} \ (-dt^2 + d\omega^2_{D-1}) \equiv \Omega^{-2} d\tilde s^2 \ ,
\een
where $d\tilde s^2$ is the metric of the Einstein static universe, and where the conformal factor
is $\Omega(t) = H \cosh^{-1} (H\tau)$. It is evident from this description that we can add a boundary to the
deSitter manifold consisting of the two disjoint $(D-1)$-spheres labeled by $\mathscr{I}^\pm = \{t = \pm \pi\}$. While the
deSitter metric is singular on this boundary, the ``unphysical metric'' $d\tilde s^2$ is smooth.
For this reason, the boundaries are called ``conformal boundaries'' or ``scri''.

It is sometimes of interest to consider subregions of deSitter spacetime as spacetimes in their own right.
In cosmology, one is mostly interested in the ``cosmological chart'', which is the subregion of deSitter spacetime
sliced by flat sections and which is
covered by the coordinates $(t, \mathbf{x}) \in \mr^D$ defined by
\begin{eqnarray*}\label{cosmchart}
X_0 &=& H^{-1} \sinh Ht - \frac{1}{2}H \e^{tH} r^2\\
X_1 &=& \e^{tH} x_1\\
&\vdots& \\
X_{D-1} &=& \e^{tH} x_{D-2} \\
X_D &=& H^{-1} \cosh Ht + \frac{1}{2}H \e^{tH} r^2 \ .
\end{eqnarray*}
In this region, the metric takes the form
\ben\label{cosmc}
ds^2 = -dt^2 + \e^{2Ht} d\mathbf{x}^2
\een
where $d\mathbf{x}^2$ is the  Euclidean flat metric on $\mr^{D-1}$. The cosmological chart covers
the half $\{X_D - X_0 < 0\}$ of $dS_D$, and its boundary is sometimes called the (a) ``cosmological horizon''.
The cosmological horizon is also equal to the boundary $\partial J^+(i^-)$ of the causal future of a point $i^-$ of $\mathscr{I}^-$.
The conformal diagram for the cosmological chart is:

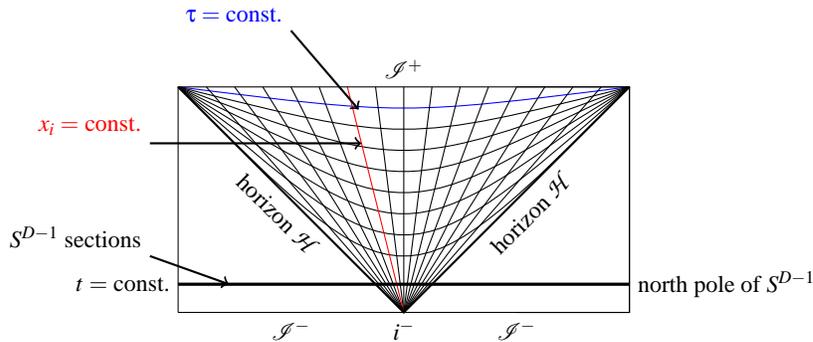
\begin{center}
\begin{tikzpicture}[scale=.75, transform shape]
\draw[blue] (-4,2) .. controls (0,1.5) and (0,1.5) .. (4,2);
\draw (-4,2) .. controls (0,1) and (0,1) .. (4,2);
\draw (-4,2) .. controls (0,0.5) and (0,0.5) .. (4,2);
\draw (-4,2) .. controls (0,0) and (0,0) .. (4,2);
\draw (-4,2) .. controls (0,-0.5) and (0,-0.5) .. (4,2);
\draw (-4,2) .. controls (0,-1) and (0,-1) .. (4,2);
\draw (-4,2) .. controls (0,-1.5) and (0,-1.5) .. (4,2);
\draw (-4,2) .. controls (0,-2) and (0,-2) .. (4,2);
\draw (0,-2) -- (0,2);
\draw (0,-2) -- (-0.5,2);
\draw[red] (0,-2) -- (-1,2);
\draw (0,-2) -- (-1.5,2);
\draw (0,-2) -- (-2,2);
\draw (0,-2) -- (-2.5,2);
\draw (0,-2) -- (-3,2);
\draw (0,-2) -- (-3.5,2);
\draw (0,-2) -- (0.5,2);
\draw (0,-2) -- (1,2);
\draw (0,-2) -- (1.5,2);
\draw (0,-2) -- (2,2);
\draw (0,-2) -- (2.5,2);
\draw (0,-2) -- (3,2);
\draw (0,-2) -- (3.5,2);
\draw[very thick] (-4,-1.5) -- (4,-1.5);
\draw[->, thick] (-4.5,-1) node[above left] {$S^{D-1}$ sections} -- (-3.1,-1.5);
\draw[->, thick] (-4.5,1) node[red, above left] {$x_i=$ const.} -- (-0.75,1);
\draw[->, thick] (-2,3) node[blue,above left] {$\tau=$ const.} -- (-.8,1.6);
\draw[thick] (-4,2) -- node[black,below, sloped]{horizon $\mathcal H$} (0,-2) --
node[black,below, sloped]{horizon $\mathcal H$} (4,2);
\draw (-4,-2) -- (-4,2) -- node[black,above]{${\mathscr I}^+$} (4,2) --
(4,-2) -- node[black,below]{${\mathscr I}^-$} (0,-2) node[black,below]{$i^-$} -- node[black,below]{${\mathscr I}^-$} (-4,-2);
\draw (2,-2) -- (4,-2) --  (4,2) --  (2,2);
\draw  (4,-1.5) node[black,right]{north pole of $S^{D-1}$} ;
\draw  (-4,-1.5) node[black,below,left]{$t=$ const.};
\end{tikzpicture}
\end{center}

Another subregion of interest is the ``static chart''. It can be defined as the
intersection of $dS_D$ with a wedge $\{X_0 \pm X_1 > 0\}$ in the ambient $\mr^{D+1}$.

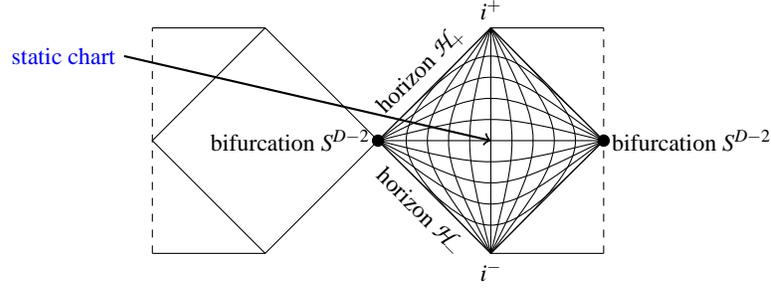
\begin{center}
\begin{tikzpicture}[scale=.75, transform shape]
\draw (0,0) node[black,left]{bifurcation $S^{D-2}$};
\draw (0,0) -- (-2,2) -- (-4,0) -- (-2,-2) --   (0,0);
\draw (-2,-2) -- (-4,-2);
\draw[dashed] (-4,-2)  --  (-4,2);
\draw (-4,2) -- (-2,2) ;
\draw (0,0) -- node[black,above, sloped]{horizon ${\mathcal H}_+$}(2,2) -- (4,0) -- (2,-2) -- node[black,below, sloped]{horizon ${\mathcal H}_-$}(0,0);
\draw (2,-2)node[black,below]{$i^-$} -- (4,-2);
\draw[dashed] (4,-2)  -- node[black,right]{bifurcation $S^{D-2}$} (4,2);
\draw (4,0) node[draw,shape=circle,scale=0.5,fill=black]{};
\draw (0,0) node[draw,shape=circle,scale=0.5,fill=black]{};
\draw (4,2)  -- (2,2)node[black,above]{$i^+$};
\draw (0,0) .. controls (2,2) and (2,2) .. (4,0);
\draw (0,0) .. controls (2,1.5) and (2,1.5) .. (4,0);
\draw (0,0) .. controls (2,1) and (2,1) .. (4,0);
\draw (0,0) .. controls (2,0.5) and (2,0.5) .. (4,0);
\draw (0,0) .. controls (2,0) and (2,0) .. (4,0);
\draw (0,0) .. controls (2,-0.5) and (2,-0.5) .. (4,0);
\draw (0,0) .. controls (2,-1) and (2,-1) .. (4,0);
\draw (0,0) .. controls (2,-1.5) and (2,-1.5) .. (4,0);
\draw (0,0) .. controls (2,-2) and (2,-2) .. (4,0);
\draw[->, thick] (-4.5,1.5) node[blue,left] {static chart} -- (2,0);
\draw (2,-2) .. controls (1.5,0) and (1.5,0) .. (2,2);
\draw (2,-2) .. controls (1,0) and (1,0) .. (2,2);
\draw (2,-2) .. controls (.5,0) and (.5,0) .. (2,2);
\draw (2,-2) .. controls (0,0) and (0,0) .. (2,2);
\draw (2,-2) .. controls (2.5,0) and (2.5,0) .. (2,2);
\draw (2,-2) .. controls (3,0) and (3,0) .. (2,2);
\draw (2,-2) .. controls (3.5,0) and (3.5,0) .. (2,2);
\draw (2,-2) .. controls (4,0) and (4,0) .. (2,2);
\draw (2,-2) -- (2,2);
\end{tikzpicture}
\end{center}
The static chart is again a globally hyperbolic spacetime in its own right, and can also be
defined as the intersection $J^+(i^-) \cap J^-(i^+)$ of two points $i^\pm \in \mathscr{I}^\pm$
which are at the ``same angle''. It can be covered by the coordinate
system $(\eta, r, \hat x)$ defined for $\eta \in \mr, |r| < H^{-1}, \hat x \in S^{D-2}$ by
\begin{eqnarray*}
X_0 &=& (H^{-2} - r^2)^{1/2} \sinh H\eta\\
X_1 &=& (H^{-2} - r^2)^{1/2} \cosh H\eta\\
X_2 &=& r \hat x_1\\
&\vdots & \\
X_D &=& r \hat x_{D-2} \ .
\end{eqnarray*}
In those coordinates, the line element takes the form
\ben\label{static}
ds^2 = - f \ d\eta^2 + f^{-1}\ dr^2 + r^2 \ d\omega^2_{D-2} \ ,
\quad f = 1 - H^2 r^2 \ .
\een
It can be seen from this form of the line element that,
within this chart--but of course not
in the full deSitter space--the
metric is static, with timelike Killing field $k =
\frac{\partial}{\partial \eta}$. The corresponding flow
$\eta \mapsto \eta + T$ defines a 1-parameter group of isometries
in the static chart, which correspond to a boost in the $X_0$-$X_1$ plane in the ambient
$\mr^{D+1}$. The boundary $\mathcal{H} = \mathcal{H}_+ \cup \mathcal{H}_-$ is formed from two intersecting
cosmological horizons, and is an example of a ``bifurcate Killing horizon'', with Killing field $k$ and
with surface gravity $\kappa = H$. This setup
also occurs in the context of the black holes, or wedges in Minkowski spacetime, and is closely related with
thermal effects in the corresponding quantum field theory on such backgrounds~\cite{kaywald,gibbons,buchholz}.

Given two points $X,X'\in dS_D$ one can define
\ben\label{Zdef}
Z \equiv H^2 \ X \cdot X' \ .
\een
The quantity $Z$ is sometimes called ``point-pair invariant'' because it
is evidently invariant under the action of $O(D,1)$. It is the analogue of the invariant distance squared
in Minkowski spacetime. The causal relationships between points can be put in correspondence with values of
$Z$, see the table~\ref{table1} and the following conformal diagram:
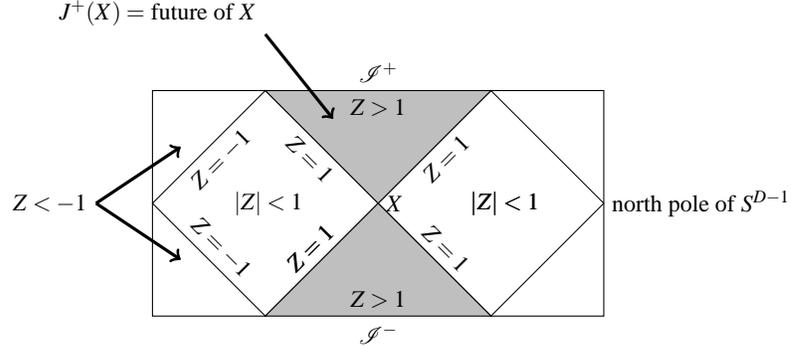
\begin{table}
\begin{center}
\begin{tabular}{|c|c|}
\hline
$Z$ & Relationship between $X$ and $X'$\\
\hline
\hline
$<1$ & spacelike $X' \notin J^\pm(X) \cup \partial J^\pm(X)$ \\
$=1$ & lightlike $X' \in \partial J^\pm(X)$ \\
$>1$ & timelike $X' \in J^\pm(X)$\\
\hline
\end{tabular}
\end{center}
\caption{Shown here is the relation between $Z$ and the
causal relationship between $X,X'$. }
\label{table1}
\end{table}

\begin{center}
\begin{tikzpicture}[scale=.75, transform shape]
\draw (0,0) -- (-2,2) -- node[black,below, sloped]{$Z=-1$} (-4,0) -- node[black,above, sloped]{$Z=-1$} (-2,-2) -- node[black,above, sloped]{$Z=1$} (0,0) node[right]{$X$};
\draw (-2.7,0) node[black,right]{$|Z|<1$};
\draw (3,0) node[black,left]{$|Z|<1$};
\draw (-2,-2) -- (-4,-2) --  (-4,2) -- (-2,2);
\draw (0,0) -- (2,2) -- (4,0) -- (2,-2) -- node[black,above, sloped]{$Z=1$}(0,0);
\filldraw[fill=gray!50] (0,0) -- node[black,below, sloped]{$Z=1$}(-2,2) --
node[below,black]{$Z>1$} node[above,black]{${\mathscr I}^+$} (2,2) -- node[black,below, sloped]{$Z=1$} (0,0);
\filldraw[fill=gray!50] (0,0) -- node[black,above,sloped]{$Z=1$} (-2,-2) -- node[black,above]{$Z>1$} node[below,black]{${\mathscr I}^-$}(2,-2)  --  (0,0);
\draw[->, very thick] (-2,3) node[above left] {$J^+(X) =$ future of $X$} -- (-.8,1.5);
\draw[->, very thick] (-5,0) node[left] {$Z<-1$} -- (-3.5,1);
\draw[->, very thick] (-5,0) -- (-3.5,-1);
\draw (3,0) node[black,left]{$|Z|<1$};
\draw (2,-2) -- (4,-2) -- node[black,right]{north pole of $S^{D-1}$} (4,2) -- (2,2);
\end{tikzpicture}
\end{center}

The relation to the signed squared geodesic distance $\sigma$ is given by
\ben\label{ztau}
Z =
\begin{cases}
\cos (H\sqrt{|\sigma|}) & \text{if points are spacelike},\\
\cosh (H\sqrt{|\sigma|}) & \text{if points are timelike}.
\end{cases}
\een
Table~\ref{table2} records the expressions for the point-pair invariant $Z$ in the various charts of
deSitter spacetime described above.

\begin{table}
\begin{center}
\begin{tabular}{|c||c|c|}
\hline
Chart & Metric & $Z$ \\
\hline\hline
Spherical & $-d\tau^2 + H^{-2} \cosh^2 (H\tau) \ d\omega_{D-1}^2$ & $\cosh H(\tau-\tau') - (\hat x -\hat x')^2 \cosh H\tau' \cosh H\tau $ \\
\hline
Cosmological & $-dt^2 + \e^{2Ht} d\mathbf{x}^2$ & $1 + \dfrac{(\e^{-Ht} - \e^{-t'H})^2 - H^2 (\mathbf{x} - \mathbf{x}')^2}{2 \e^{-tH} \e^{-t'H}} $\\
\hline
Static & $-f d\eta^2 + f^{-1}dr^2 + r^2 \ d\omega^2_{D-2}$ & $\cosh [H(\eta-\eta')] \ (f(r)f(r'))^{\frac{1}{2}} + H^2 rr' \ \hat x \cdot \hat x'$ \\
\hline
Hyperboloid & $-dX_0^2 + dX_1^2 + \dots + dX_D^2$ & $ H^2 X \cdot X'$ \\
\hline
\end{tabular}
\end{center}
\caption{Shown here are the expressions for the point-pair invariant in various charts. }
\label{table2}
\end{table}
 A spacetime metric
whose corresponding unphysical metric $\tilde g_{\mu\nu} = \Omega^2 g_{\mu\nu}$ is smooth near $\mathscr{I}^\pm$ is
called ``locally asymptotically deSitter''\footnote{A stronger notion of asymptotically deSitter
would be to require that the restrictions of $\tilde g_{\mu\nu}$ to $\mathscr{I}^\pm$ is that
of a round sphere.}. Such metrics have the property that they approach deSitter spacetime
within a domain of fixed physical volume as that domain is moved towards $\mathscr{I}^\pm$.
The {\em cosmic no hair theorem} asserts that if we pick generic initial data for a metric on a spatial slice of
the deSitter manifold and solve the Einstein equations with cosmological constant $\Lambda$,
then such a metric will be locally asymptotically deSitter (away from any black holes that may form).
In this generality, the cosmic no hair theorem is still a conjecture, but a proof has been given
when black holes are absent by
\cite{friedrich} in $D=4$, and by \cite{anderson} in higher dimensions. At the linearized level,
the cosmic no-hair theorem says that, if $h_{\mu\nu}$ is a perturbation of the deSitter metric,
then we have (in a suitable gauge, see e.g.~\cite{ishibashi})
\ben
h_{\mu\nu} = \Omega^2 \tilde h_{\mu\nu} \ ,
\een
where $\tilde h_{\mu\nu}$ is smooth (i.e. in particular finite) near $\mathscr{I}^\pm$. Thus, in this sense, the
cosmic no hair theorem is a statement about the growth/decay of perturbations near infinity. For other fields
such as the Klein-Gordon field $\phi$ on a deSitter background, a similar decay property holds. This is
immediately evident for a conformally coupled field ($m^2=\frac{1}{4}D(D-2)H^2$); such a field has the behavior
\ben\label{phidecay}
\phi = \Omega^{(D-2)/2} \tilde \phi \, , \quad \text{hence} \quad \phi = O(\e^{-M|\tau|})
\een
where $\tilde \phi$ is smooth near $\mathscr{I}^\pm$, and $M=H(D-2)/2$ for the conformally coupled field.
For a general massive minimally coupled KG-field, $M$ would be $H(D-1)/2 - \sqrt{H^2(D-1)^2/4-m^2}$.

\section{QFT in general curved spacetimes and deSitter space}\label{sec1}

\subsection{General spacetimes}
The topic of this paper are quantum field theory correlators
on deSitter spacetime in renormalized perturbation theory. However, to put things into a general
perspective, we will begin by recalling some general key concepts regarding QFT in curved spacetimes.
What we will say in this section is
of a general nature and should apply to any (reasonable) QFT. For definiteness, the reader may
think of the field theory described by the classical Lagrange density
\ben\label{L}
L = d\mu \ [(\nabla \phi)^2 + m^2\phi^2 + \lambda \phi^4]
\een
where  $d\mu = \sqrt{-g} \ d^D x$ is the invariant
integration element that is defined by the metric of a $D$-dimensional spacetime $M$.
This is the theory that we will focus on
in perturbation theory in the next sections, and our general remarks in this section are also
going to apply to this theory, at least in the perturbative sense.

A general curved spacetime does not have any isometries, so a quantum field theory on such a background
will not have any of the features that are normally associated with these. In particular, one does not have
conserved quantities like a hamiltonian, (angular) momenta etc., because their definition involves a
time-like Killing field, translational (or rotational) Killing field, etc.
For the same reason, one cannot define a vacuum- or ground state of the field, because its definition involves the
hamiltonian. Without a vacuum (state of no particles), it is clearly also impossible to define in an unambiguous fashion
particles, and in a spacetime without asymptotic regions approaching Minkowski spacetime, it is also clearly impossible
to define an $S$-matrix, amplitudes, etc.

While this means that one is forced to give up many familiar concepts when passing from Minkowski quantum field theory
to a quantum field theory on a general curved spacetime, this does not mean that quantum field theory on a curved
spacetime as such is ill defined. In fact, in any spacetime with reasonable causal behavior (e.g. ``globally  hyperbolic''),
we expect to be able to define a quantum field theory by (see~\cite{hollands2,hollandswald3} for details)
\begin{enumerate}
\item
The set of {\em correlation functions} $\langle \phi(x_1) \cdots \phi(x_n) \rangle_\Psi$ of the basic
quantum field $\phi$ (or more generally, any composite field) in a given state $\Psi$.
\item
The {\em operator product expansion} [cf. eq.~\eqref{ope}] (OPE) between an arbitrary set of composite fields, whose
coefficients are generally covariant.
Because a quantum field theory will have many admissible states in
general, the OPE can be used to determine that a given set of correlators belongs to the ``same theory'',
but ``different states'', in that theory.
\end{enumerate}
An individual state should be thought of as being {\em defined} by the collection of its correlation functions.
A quantum field theory will have infinitely many states, and their correlation functions will
differ from each other substantially. But they will have the same OPE  coefficients, as well as
a number of other common properties which reflect general properties of any relativistic quantum field theory.
The most important ones are that of ``unitarity'' or ``positivity'', and
that of ``causality''. Positivity states that if $A$ is any expression of the form
\ben\label{Adef}
A = \sum_n \int_{M^n} \phi(x_1) \dots \phi(x_n) \ f_n(x_1, \dots, x_n) \ d\mu(x_1) \dots d\mu(x_n)
\een
with any finite number of smooth compactly supported testfunctions $f_n$, then we should have $\langle A^* A \rangle_\Psi \ge 0$,
and we should also have the normalization $\langle 1 \rangle_\Psi = 1$.
The condition of causality is that, if $x_1$ and $x_2$ are spacelike to each other, then any correlator
containing $[\phi(x_1), \phi(x_2)]$ should vanish. The condition of causality is required in order
that the theory has ``propagation of information inside the light-cone''\footnote{Note that this statement is entirely
unconnected to the statement that the theory has ``no tachyons''. The latter statement is one about the stability
of the theory, whereas the former is one about the propagation properties of signals.}.
The condition of positivity is required for the probability interpretation of the state. This interpretation is
as follows. Let $A$ be a hermitian observable of the above form~\eqref{Adef}.
The probability distribution $P_\Psi(a) da$ of this
observable $A$ in the state $\Psi$ must satisfy, for all $n$,
\ben\label{Pdef}
\langle A^n \rangle_\Psi = \int_{\mr} a^n \ P_\Psi(a) \ da \, .
\een
Under normal circumstances (i.e. when the moments $\mu_n = \langle A^n \rangle_\Psi$ grow no faster
than $|\mu_n| \le {\rm cst.}^n \ n!$), there indeed exists a unique probability distribution satisfying this
formula for all $n \ge 0$, by the classical ``Hamburger moment problem''. The growth condition can be shown to be true to finite
but arbitrary loop order in $\lambda \varphi^4$-theory, and also non-perturbatively in flat spacetime
in $D=2$ dimensions \cite{rivasseau2}.
The positivity requirement for the state implies that $\mu_n \ge 0$ for all $n$,
and this in turn means that $P_\Psi(a) \ge 0$ as required for a probability distribution. Furthermore,
the normalization requirement of the state implies that the probability of measuring any value of $a$ is equal to $1$, again as required for a probability distribution.

There are also, less obvious, regularity conditions that should be imposed on the correlation functions.
These conditions state that, in a sense, the correlation functions are not just generic distributions, but
ones that are suitable ``boundary values'' of analytic functions\footnote{If $(M,g)$ is (real) analytic.}.
For a free Klein-Gordon field on deSitter,
this condition is given below in sec.~\ref{sec2}, see items~1),~2), and for the interacting $\lambda \phi^4$-theory
in sec.~\ref{sec:anal}. For Minkowski spacetime, such conditions follow
from the ``spectrum condition'', i.e. the requirement that the Hamiltonian--or more properly, the energy momentum operator--has non-negative spectrum. For a free field on Minkowski space, or a spacetime admitting a conserved non-negative
Hamiltonian, another way of saying this is that the mode-functions defining the state are ``positive frequency''.
While no conserved Hamiltonian exists in a generic curved spacetime, such regularity conditions can nevertheless be formulated
in a very elegant and geometrical way using the mathematical machinery of microlocal analysis~\cite{bfk}. For the 2-point
function, this condition is stated below in sec.~\ref{sec2}, item 3). The microlocal condition may be viewed as
a ``tangent space'' replacement of the usual spectrum condition and plays an important role in the renormalization
theory on generic curved spacetimes~\cite{hollandswald1,hollandswald2,brunetti}.

\subsection{DeSitter spacetime}

If the spacetime admits any isometries, then particular states $\Psi$ may be singled out by requiring that
the correlators be invariant under the symmetries. Evidently, such conditions are most stringent in
spacetimes with a maximally large isometry group, such as Minkowski or (anti-) deSitter spacetime.
The invariance requirement is that (in the distributional sense)
$
\langle \phi(g X_1) \cdots \phi(g X_n) \rangle_\Psi  = \langle \phi(X_1) \cdots \phi( X_n) \rangle_\Psi\, ,
$
for any element in the isometry group of the spacetime (or possibly the connected component of the identity if there
are higher spin fields present), any $X_i$ and any $n$. In deSitter space, on which we focus from now on\footnote{
In a theory like $\lambda \phi^4$ which is additionally separately $P,T$-invariant, the invariance group is
enlarged to $O(D,1)$.},
$g \in SO(D,1)_0$. Because the deSitter group is non-compact, it is not
clear that any such states will exist, and if so, if they will be unique. The free field theory example
$(\lambda = 0)$ is instructive in this respect, and will be discussed in more detail below in sec.~\eqref{sec2}.

Because deSitter spacetime is an exponentially expanding spacetime, one might suspect that correlators of generic states in deSitter will decay exponentially for large time-like separation.
More precisely, let us define the connected $n$-point functions as
\ben
\langle \phi(X_1) \dots \phi(X_n) \rangle_\Psi^C :=
\sum_{\{S_1,...,S_r\}} (-1)^r \langle \prod_{j \in S_1} \phi(X_j) \rangle_\Psi \cdots
\langle \prod_{j \in S_r} \phi(X_j) \rangle_\Psi \, ,
\een
where the sum is over all partitions of the $n$-points into ordered disjoint subsets. 
As we show in sec.~\ref{sec:anal} in the context of perturbation theory, the correlation 
functions are analytic for configurations $(X_1, \dots, X_n)$ such that no point is 
on the lightcone of another point. Consider a sequence of such configurations where one point $X_r$--or more generally 
several points--go(es) to infinity in a time-like direction, i.e. in such a way that 
$\e^{H\tau_{rs}} \sim Z_{rs} \to \infty$ for all $s \neq r$,  (hence $|\tau_{rs}| \to \infty$). 
Then we say that the state satisfies exponential clustering at large time-like distances
if there exists a constant $M>0$ such that
\ben
\langle \phi(X_1) \dots \phi(X_n) \rangle_\Psi^C
=O( \e^{-M|\tau_{rs}|}) \, ,
\een
for any $s \neq r$.
For a free Klein-Gordon field with $m^2>0$, exponential clustering indeed holds for the vacuum
state and ``most'' other states, and the constant
$M$ is related to the mass and $H$, see sec.~\ref{sec2}.
However, we note that it certainly not obvious that such a condition
will still hold for general, i.e. interacting, QFT's. We will demonstrate this in sec.~\ref{sec:nohair} in the
context of perturbation theory for a massive Klein-Gordon field with self-interaction.

As stated, the clustering property does not apply to configurations of points that are null-related,
because for such configurations, the $n$-point functions are distributions. To obtain a sensible statement, one should smear the points out. One can easily formulate smeared versions of the exponential clustering property. One such condition may e.g. be given as follows. Let $A,B,C$ be observables as in eq.~\eqref{Adef}, with testfunctions $f_n$ etc. supported in the static chart of deSitter spacetime, see eq.~\eqref{static}. For $T \in \mr$,
let us denote by $\alpha_T(A)$ the time-shifted observable~\eqref{Adef}, which is defined by replacing $f_n$
with $f_n^T$, where we are setting $f_n^T(\{\eta_i, r_i, \hat x_i\}) = f_n(\{\eta_i-T, r_i, \hat x_i\})$. Then
a natural alternative formulation of the time-like clustering property is that
\ben
\label{expcl}
\langle A \ \alpha_T(B) \ C \rangle^C_\Psi =O( \e^{-M|T|} ) \quad \text{as $|T| \to \infty$,}
\een
for any $A,B,C$ as in eq.~\eqref{Adef}.

This form of the time-like clustering property implies a quantum field theory analog of the cosmic no hair
theorem, in the sense that all states approach the vacuum state at late times, at an exponential rate.
Indeed, let $\langle \, . \, \rangle_0$ be a deSitter invariant (i.e. vacuum-) state, which we assume satisfies the
expontial clustering property as in eq.~\eqref{expcl}. For some arbitrary but fixed $A$ as in
eq.~\eqref{Adef}, let $\langle \, . \, \rangle_\Psi$ be the state whose $n$-point functions are given by\footnote{
In the GNS-representation of the state $\langle \ . \ \rangle_0$, this state would be given by
$|\Psi\rangle = A |0\rangle$ up to normalization.
}
\ben
\langle \phi(X_1) \cdots \phi(X_n)\rangle_\Psi := \frac{\langle A^* \ \phi(X_1) \cdots \phi(X_n)
\ A\rangle_0}{\langle A^* A \rangle_0} \, .
\een
Let $X_i$ be in the static chart [cf. eq.~\eqref{static}] with time coordinate $\eta_i$ in the static chart,
and let $X_i^T$ be the point which is time-translated by $T$ into the future, i.e. it has time-coordinate
$\eta_i + T$.
Then it is easily seen from eq.~\eqref{expcl} that, we have (in the sense of distributions in the static chart)
\ben
\langle \phi(X_1^T) \cdots \phi(X_n^T)\rangle_\Psi \to \langle \phi(X_1) \cdots \phi(X_n) \rangle_0
\quad \text{as $T \to \infty$}
\een
at an exponential rate. In particular, for the 1-point function, we have $\langle \phi(X^T) \rangle_\Psi
= O(\e^{-M|T|})$, which is the analog of eq.~\eqref{phidecay} in the classical context.

The clustering property can also be interpreted as saying that the probability distributions for two observables with large time-like separation become independent. Indeed, let $A_-$ again be as in eq.~\eqref{Adef}, with $f_n$ supported within some bounded region in the static chart. Let $\Psi$ be a state with exponential clustering as in eq.~\eqref{expcl}, and let
$A_+ = \alpha_T(A_-)$ be the time-translated observable.
Then from the clustering property, the $n$-th moment
$\langle (A_+ + A_-)^n \rangle_\Psi$ factorizes into the moments of $A_\pm$ up to an exponentially
small correction.
This implies that, for large times $T$,
the probability distributions $P^\pm_\Psi$ for $A_\pm$ [cf. eq.~\eqref{Pdef}] become independent in the standard
sense that the probability distribution of $A_+ + A_-$ is given by the convolution of $P^+_\Psi$ and $P^-_\Psi$.

\medskip

DeSitter spacetime is a real Lorentzian section of the complex deSitter spacetime $dS^\mc_D$, defined by
allowing complex vectors $X$ in eq.~\eqref{1},
\ben
dS^\mc_D = \{ X \in \mc^{D+1} \mid X \cdot X = H^{-2} \} \ .
\een
The complex deSitter manifold also contains the
$D$-sphere $S^D$ as a real Riemannian section. It is thus natural to ask what might be the relation between
the QFT's associated with these two sections. In Minkowski spacetime, such a connection is well known and
is described by the Osterwalder-Schrader (OS)-theorem~\cite{osterwalder},
which clarifies how to switch back and fourth between these QFT's. It essentially
states that, under reasonable circumstances, the correlators are related by analytic continuation.
There is also a corresponding theorem for the deSitter case~\cite{birke}, and the main requirement for
the correlators on the sphere is as in Minkowski spacetime a ``reflection positivity'' property. This reflection
positivity property is the Euclidean counterpart of the Hilbert-space positivity condition mentioned above.
The analytic continuation from Euclidean deSitter space will also be important in this paper, but rather than
appealing to the OS-theorem\footnote{The OS-theorem works as it stands only for
quantum field theories that are fully defined in the non-perturbative sense, whereas in this paper,
we work in the perturbative setting.}, we will perform the analytic continuation ``by hand'' based on an explicit
expression for the Euclidean correlators, see sec.~\ref{sec:anal}.

\subsection{The free field on deSitter}\label{sec2}

The free Klein-Gordon field provides a simple but instructive example to some of the general remarks made
in the previous two subsections.  It is defined classically by setting the interaction coupling
$\lambda = 0$ in eq.~\eqref{L}. The quantum field theory may be defined by its operator product expansion,
but because it is such a simple theory, one can also directly give the relations
in terms of commutators:
\begin{enumerate}
\item
The correlation functions of the free field in deSitter should satisfy
\ben
\langle \phi(X_1) \dots [\phi(X_i), \phi(X_{i+1})] \dots \phi(X_n) \rangle_\Psi
= i \Delta(X_i,X_{i+1}) \ \langle \phi(X_1)  \dots \phi(X_n) \rangle_\Psi
\een
for any state $\Psi$, where $\Delta$ is the commutator function on deSitter spacetime,
equal to the advanced minus retarded fundamental solution for the Klein-Gordon
operator $(\nabla^2 - m^2)$, see e.g.~\cite{friedlaender}.
\item
We should have
\ben
(\nabla^2 - m^2)_i \ \langle \phi(X_1) \dots \phi(X_i) \dots \phi(X_n) \rangle_\Psi = 0 \ .
\een
\item
The state should satisfy positivity, $\langle A^* A \rangle_\Psi \ge 0$ for any $A$ as
in \eqref{Adef}.
\end{enumerate}
These conditions must be supplemented by a regularity condition in order to rule out
states with pathological behavior such as the well-known ``$\alpha$-vacua''\footnote{
These deSitter invariant states~\cite{allen,mottola} have many pathological
features such as infinite fluctuations for the stress tensor~\cite{holfred}. For a discussion
of related problems, see e.g.~\cite{goldstein}. The main deficiency of these states from the point of
view of the present paper is that they are not boundary values of analytic functions, see 2.) below. \label{alpha}} or ``instantaneous ground states'',
for which e.g. the OPE does not hold. A common requirement which is essentially equivalent to
the validity of an OPE in the case of a free field theory is that the state be ``Hadamard''.
By this, one means that (i)
the connected $n$-point functions are smooth ($C^\infty$) for $n\neq 2$, and that (ii) the
$2$-point function satisfies the first one of the following conditions:
\begin{enumerate}
\item The 2-point function behaves like $\langle \phi(X_1) \phi(X_2) \rangle_\Psi \sim \sum u_n \sigma^{-n}_\epsilon +
v \log \sigma_\epsilon + {\rm smooth}$, where $u_n, v$ are the Hadamard-deWitt coefficients, see e.g.~\cite{moretti}, and where $\sigma_\epsilon = \sigma + it\epsilon$ with $t$ the time-difference
between the coordinates relative to an arbitrary foliation of deSitter by spacelike slices indicates the
distributional boundary value ($\epsilon \to 0+$).
\item The 2-point function is the boundary value, in the distributional sense, of
an analytic function defined on the neighborhood $T_2=\{(X_1,X_2) \in (dS_D^\mc)^2 \mid
\I(X_1-X_2) \in V^+ \}$, where
$V^\pm$ is the future/past interior lightcone in $\mr^{D+1}$, and where
$dS_D^\mc = \{ X \in \mc^{D+1} \mid X \cdot X = H^{-2}\}$ is the complexified
deSitter manifold.
\item The analytic wave front set of the 2-point function is contained in $
\{(X_1, k_1; X_2, k_2) \mid k_{1/2} \ {\rm future/past \ directed} \}$~\cite{bfk,hollands1}. This version
of the Hadamard property is a special case of the ``microlocal spectrum condition''~\cite{bfk}.
\end{enumerate}
The second and third conditions are essentially equivalent formulations of the first one~\cite{rad,sanders}.

It is also customary to restrict attention to so-called ``Gaussian states'', which are
defined by demanding that the connected $n$-point functions actually vanish for $n \neq 2$. The $n$-point
functions of such states are by definition determined by a version of ``Wick's theorem'' in terms of the 2-point function alone.
It turns out that, for a massive Klein-Gordon field ($m^2 > 0$),
there is a {\em unique}\footnote{If we drop the condition that the state be Hadamard, then
there are in addition the $\alpha$-vacua of~\cite{allen,mottola}, see also footnote~\ref{alpha}.} Gaussian Hadamard state which is invariant under the full deSitter group $O(D,1)$. This state, often
called ``Bunch-Davies state'', has 2-point function~\cite{allen,bros}
\ben\label{vac}
\langle \phi(X_1) \phi(X_2) \rangle_{0} = \frac{H^{D-2}}{(4\pi)^{D/2}} \frac{\Gamma(-c)\Gamma(c+D-1)}{\Gamma(D/2)} \ \hF \ \left(-c, D-1+c; D/2; \frac{1+Z}{2} \right) \, ,
\een
where $Z$ is the point-pair invariant, and
where the dimensionless constant $c$ is defined by
\ben
c = -\frac{D-1}{2} + \sqrt{\frac{(D-1)^2}{4} - \frac{m^2}{H^2}}
\een
As with any correlation function, this is distributional in nature, so it needs to be defined with
some care as the boundary value (in the distributional sense) of an analytic function. The boundary prescription
understood here is that explained in items 1.),~2.) above\footnote{\label{foot1} Namely we should replace
$X_1 \to X_1 + i \epsilon e \ , X_2 \to X_2$, with $e \in V^+, \epsilon > 0$, then smear out the distribution,
and then let $\epsilon \to 0$.}.
The Bunch-Davies state is sometimes referred to as a ``vacuum state'' because it has a comparable amount of symmetry
than the ordinary free field vacuum state in flat Minkowski spacetime. However, one ought to be cautious when taking this analogy to literally. For example, in the static chart, where there is a natural notion of time-translation symmetry and a similar natural notion of Hamiltonian, the restriction of the Bunch-Davies state is actually a thermal
state (a ``KMS-state''), rather than a ground state, of this Hamiltonian~\cite{kaywald}.

It turns out that the Bunch-Davies state can also be characterized as the unique analytic continuation (in the sense of
boundary values of distributions) of the Euclidean Green's function defined on the real Riemannian section
$S^D \subset dS_D^\mc$ to the real Lorentzian section $dS_D \subset dS_D^\mc$. This analytic continuation property
is a manifestation of the OS-reconstruction theorem mentioned in the previous section in this simple quantum field theory. Later, we will see that an analogous statement remains true for the $n$-point correlation functions of an
interacting theory described by \eqref{L}, to all orders in $\lambda$.
An alternative ``momentum space type-'' expression for the Lorentzian 2-point function of the free KG-field which manifestly displays its positivity and analyticity properties has been provided by~\cite{bros}. Another alternative representation displaying explicitly the Hadamard
property is given below in eq.~\eqref{hadarep}, yet a further representation is given below in eq.~\eqref{mellinrep}.

As in Minkowski space, there is an intimate relation to the representation theory of symmetry group of the spacetime, namely $O(D,1)$ and the mass parameter, encoded here by $c$. We will not
 go into this topic here, but we note that the relation between the classification of the representations and $c$ is
 (see e.g. \cite{vilenken})
 \begin{enumerate}
 \item
 $c =
 -\frac{D-1}{2} + i\rho  \quad (\rho \in \mr)$, the principal series.
 \item $c \in [-(D-1)/2, 0)$, the complementary series.
 \item
 $c=
 1,2,\dots$, the discrete series.
 \end{enumerate}
The properties of the theory at large distances are different in each case;
for positive $m^2$, which is the case considered in this paper, only the first two cases can occur.
For  values $m^2 \le 0$, no deSitter invariant vacuum state exists~\cite{allen}.

The Bunch-Davies state satisfies the time-like clustering property (see above), which in the present free field theory
reduces to the statement that for points $X_1,X_2$ which are timelike separated by a large proper time $\tau$,
\ben
\langle \phi(X_1) \phi(X_2) \rangle_{0} = O(\e^{-M|\tau|})
\een
where $M>0$. For the principal series $M= H(D-1)/2$; this follows from a standard transformation formula for the hypergeometric function, or from the contour-integral representation given below in eq.~\eqref{mellinrep}. For the complimentary series,
the decay rate is only $M = H(D-1)/2 - [H^2(D-1)^2/4 - m^2]^{1/2} > 0$.

\section{Interacting fields on dS-spacetime}\label{sec3}

\subsection{Basic setup}
In this section, we come to the main purpose of this paper, which is to give a prescription for defining the
correlation functions $\langle \phi(X_1) \dots \phi(X_n) \rangle_\Psi$ for the interacting QFT corresponding
to the classical Lagrangian~\eqref{L}. We must face the following issues:
\begin{enumerate}
\item We have to deal with any UV divergences in such a way that the renormalized correlators satisfy the
desired general properties explained in sec.~\ref{sec1}.
\item We must specify which state $\Psi$ we would like to use.
\item We must specify how to deal with any potential IR-divergence.
\item (Optional) We would like to check whether the time-like clustering holds.
\end{enumerate}
1) The renormalization problem has been dealt with for general curved spacetimes in~\cite{hollandswald1,hollandswald2},
and we will apply this general procedure to the special case of deSitter spacetime in subsec.~\ref{sec:ren}. 2) For $m>0$, it seems reasonable to use an $O(D,1)$-invariant state, which in perturbation theory will be obtained from the
Bunch-Davies state of the underlying free theory. 3) For this state, we can analytically continue the propagators
from deSitter space $dS_D$ to the sphere $S^D$. Working on the sphere will obviously eliminate any IR-divergences,
and we will argue in sec.~\ref{sec:anal} that the result can be continued back without introducing any new IR-divergences. 4) will be demonstrated in sec.~\ref{sec:nohair}.

Following standard perturbation theory, our starting point will
hence be the, as yet formal, expression
\bena\label{correlators}
&&\bigg\langle \phi(X_1) \cdots \phi(X_E) \bigg\rangle_{0,\lambda} := \non \\
&&N^{-1} \ \acon_{X_i \to dS_D}
\sum_{V \ge 0} \ \frac{1}{V!} \ \left\langle
\phi(X_1) \cdots \phi(X_E) \ \left( \lambda
\int_{S^D} \phi^4(Y) \ d\mu(Y) \right)^V
\right\rangle_0
\eena
The ``expectation value'' on the right side is intended to mean the Euclidean (analytically continued)
Bunch-Davies state on the sphere in the free theory, and the expectation value on the left side that in the interacting theory on deSitter. $N$ is an overall normalization factor making
$\langle 1 \rangle_{0,\lambda} = 1$, and the fourth power $\phi^4$ in the free theory is defined via normal
ordering\footnote{As explained in~\cite{hollandswald1}, normal ordering is actually not a renormalization prescription that
is in general ``local and covariant'', nor ``analytic in the coupling constants/metric''. The correct definition
of $\phi^4$ would be via the operator product expansion in the free field theory. This will differ from
the normal ordering prescription by lower order terms, but can be compensated by making the re-definition $\phi^4 \to \phi^4 +
6 H^2 A \ \phi^2 + 3 H^4 A^2$, where, with $\psi = \Gamma'/\Gamma$ and $c$ as in the main text, $A=(\log H^2/\mu^2 + \psi(-c) + \psi(c+3) )/16\pi^2$ for some arbitrary
$\mu$ of dimension [mass] (in $D=4$). The advantage of normal ordering is that we can straightforwardly apply Wick's theorem.}.
Then we can (formally) apply Wick's theorem on the right side and get the
standard expression\footnote{The normalization factor cancels as usual the contributions from bubble diagrams.}
\ben\label{ac}
\langle \phi(X_1) \cdots \phi(X_E) \rangle_{0,\lambda}^C =
\acon_{X_i \to dS_D} \ \sum_{V \ge 0} \ \lambda^V \ \sum_{G \ {\rm with} \ V+E \ {\rm vertices}} \ {\rm Sym}(G) \ I_G(X_1, \dots, X_E)
\een
for the connected correlators,
where $I_G$ is the contribution from a connected Feynman graph $G$ with $V$ interior vertices and
$E$ external legs, weighted by the appropriate
symmetry factor (equal to the order of the automorphism group of the graph, $|{\rm Aut}(G)|^{-1}$). Formally, $I_G$ is computed by writing down a factor of
the Euclidean Green's function $\langle \phi(X_i) \phi(X_j) \rangle_0$ for each line $(ij) \in G$, and then
integrating the $V$ ``interaction vertices'' over $S^D$. These integrations lead to UV-divergences, and
hence have to be defined in a more sophisticated way. In a general curved spacetime, this has
been explained in~\cite{hollandswald1,hollandswald2}, and we now start explaining how the general procedure
may be implemented concretely in the Euclidean deSitter spacetime, $S^D$. For the $\lambda \phi^4$-theory,
the graphs $G$ have external vertices $(X_1, \dots, X_E)$ of valence 1, and interaction vertices $(X_{E+1}, \dots, X_{E+V})$ 
of valence 4. But for most of
what follows in the paper, the latter fact will not be so important for our discussion, so we may think of $G$ as any (connected) graph with $E$ external lines (without ``tadpoles'').

\medskip
\noindent

The first step is to bring the free Euclidean 2-point function~\eqref{vac} on the sphere $S^D$ of radius $H^{-1}$, into a form that will be more suitable for our purposes. In expression~\eqref{vac}, the two-point function is expressed in terms of $(1+Z)/2 = H^2 (X_1 + X_2)^2/4$, so the singularity of the Euclidean Green's function at $X_1 = X_2$ is reflected by the singularity of the hypergeometric function at the value $1$ of the argument.
It is desirable to have a representation of this hypergeometric function which displays explicitly this singularity. A representation which is particularly useful for our purposes is given  by\footnote{We obtained this formula for even $D$.
The formula for all $D$ (including odd) was obtained first by~\cite{marolf2}.}
\ben\label{mellinrep}
\langle \phi(X_1)\phi(X_2) \rangle_0 =  K_{c,D} \ \int_{C} \frac{dz}{2\pi i} \left( \frac{H^2(X_1-X_2)^2}{4} \right)^z
\ \frac{\Gamma(D-1+c+z) \Gamma(-c+z) \Gamma(-z)}{\sin[\pi (z+(D-2)/2)] \ \Gamma(D/2 + z)} \ .
\een
Here,
\ben
K_{c,D}= \frac{\sin [\pi (c+(D-2)/2)] \ H^{D-2}}{(4\pi)^{D/2}}
\een
the contour $C$ is parallel to the imaginary axis near infinity, and it leaves the poles
\textcolor{green}{$-(D-2)/2 + \mn_0$} and \textcolor{green}{$\mn_0$} to the right, whereas it leaves the poles
\textcolor{red}{$c-\mn_0$} and \textcolor{red}{$-(D-1+c)-\mn_0$} to the left.
The contour $C$ is visualized in the following figure, where we assume for simplicity that we have
a principal series scalar field with $c = -(D-1)/2 + i\rho$, and even $D$:

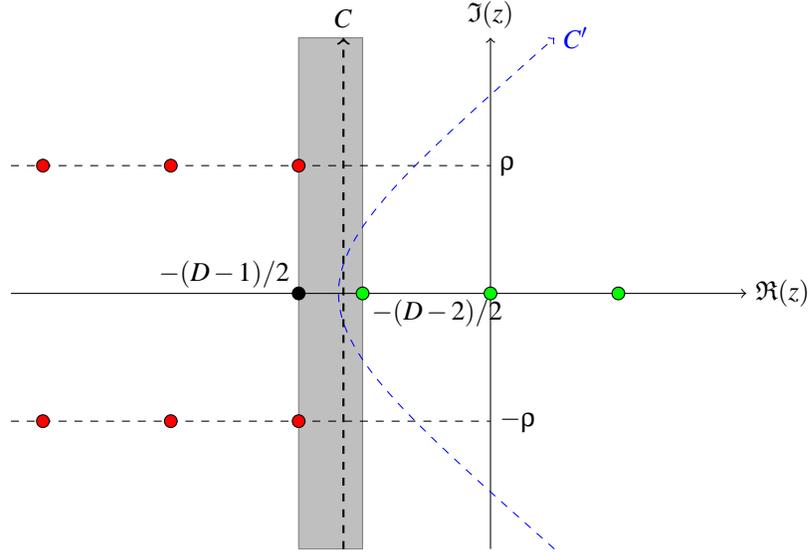
\begin{center}
\begin{tikzpicture}[scale=.85, transform shape]
\filldraw[gray,fill=gray!50] (-3,-4) -- (-3,4) -- (-2,4) -- (-2,-4) -- (-3,-4);
\draw[->] (-7.5,0) -- (4,0) node[black,right]{$\R(z)$};
\draw (-3,0.3) node[black,left]{$-(D-1)/2$};
\draw (-2,-0.3) node[black,right]{$-(D-2)/2$};
\draw[->] (0,-4)  -- (0,4) node[black,above]{$\I(z)$};
\draw[->,thick,black,dashed] (-2.3,-4) -- (-2.3,4) node[black,above]{$C$};
\draw (-3,0) node[draw,shape=circle,scale=0.5,fill=black]{};
\draw (-2,0) node[draw,shape=circle,scale=0.5,fill=green]{};
\draw (0,0) node[draw,shape=circle,scale=0.5,fill=green]{};
\draw (2,0) node[draw,shape=circle,scale=0.5,fill=green]{};
\draw[dashed] (-7.5,2) -- (0,2) node[black,right]{$\rho$};
\draw[dashed] (-7.5,-2) -- (0,-2) node[black,right]{$-\rho$};
\draw (-3,-2) node[draw,shape=circle,scale=0.5,fill=red]{};
\draw (-5,-2) node[draw,shape=circle,scale=0.5,fill=red]{};
\draw (-7,-2) node[draw,shape=circle,scale=0.5,fill=red]{};
\draw (-3,2) node[draw,shape=circle,scale=0.5,fill=red]{};
\draw (-5,2) node[draw,shape=circle,scale=0.5,fill=red]{};
\draw (-7,2) node[draw,shape=circle,scale=0.5,fill=red]{};
\draw[->,dashed,blue] (1,-4) .. controls (-3.5,0) and (-3.5,0) .. (1,4) node[blue,right]{$C'$};
\end{tikzpicture}
\end{center}

The above formula can be derived by noting that, for $H^2(X_1 -X_2)^2/4<1$, the contour $C$ can be deformed to
a contour \textcolor{blue}{$C'$} parallel to the real axis that encircles the
(double when $D$ is even) poles \textcolor{green}{$-(D-2)/2 + \mn_0$} of
the integrand. This integral can then be evaluated using the residue theorem, and it is seen to reproduce the
expression~\eqref{hadarep} for the Green's function. We note that the line integral is easily absolutely convergent,
as the integrand decays as $\e^{-2\pi |\mathfrak{I}(z)|}$ (here and later we use the fact that $|\Gamma(z)| \sim
\e^{-\pi|\mathfrak{I}(z)|/2} |\mathfrak{I}(z)|^{\mathfrak{R}(z) - 1/2}$).

To obtain the Feynman integral $I_G$, we now need to multiply one factor of the Euclidean Green's function~\eqref{mellinrep}
for each line $l$ of the graph $G$, and we need to integrate over
the internal vertices $X_{E+1}, \dots, X_{E+V}$, while leaving the external points $X_1, \dots, X_E$
alone. We end up with an expression of the form
\bena\label{ig}
I_G(X_1, \dots, X_E) &=& K_{c,D}^{L} \ \bigg( \prod_{i=E+1}^{V+E} \int_{S^D} d\mu(X_i) \bigg) \
\bigg(  \prod_{l \in \E G} \int_{C} \frac{dz_l}{2\pi i} \bigg) \
t_{\vec z}(X_1, \dots, X_{E+V}) \non \\
&&\times \  \prod_{l \in \E G} 2^{-2z_l} \ \frac{\Gamma(-c+z_{l}) \Gamma(c+D-1+z_{l}) \Gamma(-z_l)}{\sin[\pi (z_l+(D-2)/2)] \Gamma(D/2 + z_{l})}
\eena
where $L$ ($=2V+E/2$ in $\phi^4$-theory) is equal to the number of lines $l$ in the graph $G$ , and where $t_{\vec z}$ is the product
\ben\label{tdef}
t_{\vec z}(X_1, \dots, X_{V+E}) = \prod_{i,j=1}^{V+E} \left[ H^2(X_i-X_j)^2 \right]^{z_{ij}} \ .
\een
In this expression, $i,j$ is a pair of vertices, and for each such pair we have set
\ben\label{zij}
z_{ij} = \sum_{l \in G: l = (ij)} z_l \ ,
\een
i.e. $z_{ij}$ is the sum of all the parameters associated with lines that connect a given pair of
vertices $i,j = 1, \dots, V+E$. If there are no lines in the graph $G$ connecting $i,j$, then
the corresponding factor in eq.~\eqref{tdef} is understood to be absent.

\subsection{Renormalization}\label{sec:ren}

The $X_i$-integrations in the Feynman integral $I_G$ [cf. eq.~\eqref{ig}] are not absolutely
convergent for the values of $z_l$ that we need in the subsequent contour integrals
for $D>2$. The potential problems come from the divergence of the integrand $t_{\vec z}$ near
configurations of points such that $X_i = X_j$ for some $i,j$. The first task in this section
will be to define these integrals over the $X_i$ properly, by a method of extending
distributions. This method is an application of the more general constructions in~\cite{brunetti,hollandswald2,hollands2}
(based in turn on earlier work of~\cite{eg}) to the special case of deSitter spacetime, taking into account
many of its special features.

To simplify the discussion, let us suppose that there are no external points, $E=0$, so that $I_G$ is simply
a number. The basic type of quantity that we need to look at is (dropping $H$ etc.)
\ben\label{tdef1}
t_G(f) :=
\bigg( \prod_{i=1}^{V} \int_{S^D} d\mu(X_i) \bigg) \ \prod_{i,j=1}^V \ [(X_i-X_j)^2]^{z_{ij}} \
f(X_1, \dots, X_{V}) \ .
\een
Here, we have put a superscript  ``$G$'' to indicate the dependence
on the graph $G$ whose lines $l \in \E G$ carry the labels $z_l \in \mc$.
$z_{ij} \in \mc$ in \eqref{tdef} is given in terms of the parameters $z_l$  by  eq.~\eqref{zij}.  If there is no line in $G$ between a pair $i,j$ of vertices, then the corresponding factor is understood to be absent. $f$ is some smooth test function on $(S^D)^V$.

Of course, we will be interested in the case when $f \equiv 1$, but of course, this is a priori not
well defined. But one notes that the integrand is actually perfectly smooth as long
as $X_i \neq X_j$ for all $i,j$, so we can integrate it against any smooth weighting function
$f(X_1, ..., X_{V})$ whose support (i.e. the closure of all points where it is non-zero) is contained within
the subset
\ben\label{diag}
\{(X_1, \dots, X_V) \in (S^D)^{V} \mid X_i \neq X_j \ \text{for all $i\neq j$} \} \,
\een
of $(S^D)^V$. Since we would like to take $f \equiv 1$, the problem at hand can be viewed as that
of extending the distribution $t_G$, initially only defined on the above domain, to all of $(S^D)^V$.
The extension can then, by definition, be integrated against $f \equiv 1$. In this view,
UV-renormalization corresponds to extension of distributions, as originally proposed in~\cite{eg}.
Because we are not changing the integrand
of $t_G$ at the points~\eqref{diag} where it is already defined, it is intuitively clear---and
can be rigorously proved~\cite{eg,brunetti,hollandswald1}---that
this strategy for UV-renormalization corresponds to ``local counterterms''.

To illustrate how the extension process works concretely,
let us consider the following example in $D=4$, and
for $V=2$ points:
\ben
t(f) \equiv \int_{S^4 \times S^4} d\mu(X_1) d\mu(X_2) \ [(X_1-X_2)^2]^{z} \ f(X_1, X_2) \, .
\een
As a distribution, this is defined a priori only
for $f$ which are nonzero only strictly within the set $\{X_1 \neq X_2\}$,
and the task is to extend it to a distribution called $t^\ren$ (``R'' for ``renormalized'') on the entire
product manifold $(S^4)^2$, so that we can in particular set $f \equiv 1$.
In order to deal with the problematic configuration $X_1=X_2$, let us introduce Riemannian normal coordinates
around $X_2$, i.e. we identify the tangent space $T_{X_2} S^4$ with $\mr^4$ via a choice of orthonormal tetrad,
and use the exponential map to identify a neighborhood of the origin in $T_{X_2} S^4$ with a neighborhood 
of $X_2$ in $S^4$. The Riemann normal coordinates are denoted $x=(x^1, \dots, x^4)$ and are defined in an open neighborhood of $0$ in $\mr^4$. As usual, the squared geodesic distance is given by $\sigma = x^2$. Then we have, using eq.~\eqref{Zdef}
\bena
[(X_1-X_2)^2]^z &=& 2^{z} H^{-2z} [1-\cos (H \sqrt{x^2}) ]^{z} =
(x^2)^z \left(1 - 2\sum_{j=1}^\infty \frac{(-H^2 x^2)^j}{(2j+2)!} \right)^{z} \non\\
&=& \sum_{j=0}^\infty a_j \ H^{2j} (x^2)^{z+j} \ ,
\eena
with easily computable coefficients $a_j$ depending analytically on $z$. Similarly, we can expand
the integration measure as
\ben
d\mu(X_1) = \sum_{j = 0}^\infty b_j \ H^{2j} (x^2)^j \ d^4 x \ ,
\een
with easily computable coefficients $b_j$,
and together this gives an expansion of the form
\ben
t(X_1, X_2) \ d\mu(X_1) = \sum_{j=0}^\infty H^{2j} \ u_{z,2j}(x) \ d^4 x \ .
\een
The homogeneous distributions $u_{z,2j}$ of degree $2z+2j$ are of the form
\ben
u_{z,2j}(f) = c_j \int_{\mr^4} d^4 x \ (x^2)^{z+j} \ f(x) \ ,
\een
for some easily computable numerical coefficients $c_j$ depending analytically on $z$. $f$ is a testfunction
that a priori has to vanish in an open neighborhood of $x = 0$. But
for sufficiently large values of the exponent, $\R(2z+2j) > -4$, this distribution is automatically well-defined
on all of $\mr^4$, in the sense that it has a unique extension to all of $\mr^4$.
So we need to worry only about the case when $\mathfrak{R}(2z+2j) \le -4$, and in that case we simply proceed by analytically continuing the result from $\mathfrak{R}(2z+2j)>-4$. By lemma~6 and the following discussion of~\cite{hollands2}, we obtain in this way a family of extensions of $u_{z,2j}$ that is analytic in $z \in \mc$,
except at the simple poles $2z \in -4-\mn_0$. Around those, $u_{z,2j}$ has the Laurent expansion
\ben
u_{z,2j}(x) = \sum_{n=4}^N \frac{M^{-2z-2j-n}}{2z+2j+n} \ d_{n,j} \ (\partial^2)^{(n-4)/2} \delta^4(x) \
+ \text{regular,}
\een
for some computable numerical constants $d_{j,n}$, and an arbitrary constant $M > 0$ with the dimension of
mass. The regular part is determined by the condition that it is analytic in the half space $\R(2z+2j) > -N$.

Because we have now extended the $u_{z,2j}$ for $z$ such that $2z+2j \notin -\mn_0$, we also get a
corresponding family of extensions of $t$ to all of $(S^4)^2$, which has simple poles for certain
integer values of $z$. In fact, substituting the Laurent expansion for $u_{z,2j}$ into  $t$
gives, now in $D$ dimensions:
\bena\label{pole}
t(X_1,X_2) \ &=& \sum_{n=D}^N \sum_{j=0}^{(n-D)/2} \frac{M^{-2z-n}H^{2j}}{2z+n} \
{\mathcal D}_{n,j,D}(\nabla^2, H^2) \ \delta(X_1,X_2) +
\text{regular part} \non \\
&=:& \pp_N [t(X_1, X_2)] + \text{regular part} \ .
\eena
Here, ${\mathcal D}_{n,j,D}$ denotes a homogeneous polynomial in $H^2, \nabla^2$ of
degree $(n-D)/2-j$ with computable coefficients related to the $d_{n,j}$.
The ``regular part''by definition has the property that it
has no poles for $\R(2z) > -N$. The ``pole part'' $\pp_N$ is the sum of delta functions on the $D$-sphere and its derivatives.
In practice, we need to worry only about finitely many pole terms, i.e. finite $N$, so
convergence is not an issue.
The appearance of the $\delta$-function on the sphere is clear because, if $f$ vanishes near coincident points, then there is no divergence under the integral $t(f)$, and hence no divergence at $2z=-D-\mn_0$ etc. either.
Hence, a possible prescription
for defining the extension of $t(f)$ when $\R(2z) > -N$ is to simply compute the integrals for sufficiently large
$\R(z)$, then analytically continue, and then subtract the pole part~\eqref{pole}, after
which one can simply set $z$ to any desired value $\R(2z) > -N$. In formulae,
\ben
t^\ren(X_1, X_2) := t(X_1, X_2) - \pp_N [t(X_1, X_2)] \ .
\een

Let us now continue this procedure to extend the distribution $t_G$ of the form~\eqref{tdef} for general $V \ge 2$.
Again we are interested in the case $f \equiv 1$. For this choice, the integral is singular,
and we have to view it first as a distributional kernel that is, a priori, only defined on
test-functions $f$ which are nonzero only strictly within the subset
$\{ X_i \neq X_j \ \text{for all $i \neq j$} \}$ of $(S^D)^V$,
where it is perfectly smooth. Again, we would like to extend this to the entire product manifold $(S^D)^V$, so that
we can integrate against $f \equiv 1$ in particular, and so that the result is smooth (in fact analytic) in $z_{ij}$
 in a suitable domain of $\mc$. We could again try to simply subtract the pole part as in the case of two vertices, but this would give the wrong answer, i.e. it would not be an extension but a genuine modification of the distribution even at points where it is already defined. Furthermore, since the distribution $t$ is now dependent on many complex variables
 $z_{l}, l \in \E G$ [related to the $z_{ij}$'s in eq.~\eqref{tdef1} by eq.~\eqref{zij}], it is not so clear how the pole part etc. should be defined.

 The problem is that, in general, there are poles coming from any subgraph $\gamma \subset G$ (and subgraphs thereof etc.), and
 these have to be disentangled. One way of proceeding is by induction. Let us associate a {\em subgraph} $\gamma(I)$ of $G$ with any
 subset $I \subset \{1, \dots, V\}$, which is formed from the vertices in $I$, together with the set of {\em all} the
 lines in $G$ joining them. If $I_1, \dots, I_r$ is a partition
 of $\{1, \dots, V\}$ into pairwise disjoint subsets, then we let $\gamma(I_1), \dots, \gamma(I_r)$ be the corresponding subgraphs of $G$. Furthermore, we denote by $G/\cup_j \gamma(I_j)$ the new graph with $r$ vertices obtained
  by shrinking the vertices in each $\gamma(I_j)$ to a single vertex, and by $G \setminus \cup_j \gamma(I_j)$ the complement.
  For each $I \subset \{1, \dots, V\}$, let us furthermore choose a number $N_I \in \mn_0$ (to be further specified later except
  for $N_I = 0$ if $I$ has only one element), and
 \ben
 z_I = \sum_{l \in \E \gamma(I)} z_l \, .
 \een
 For $\gamma(I)$ the subgraph associated with $I = \{i,j\}$, the quantity $z_I$ coincides with $z_{ij}$ defined above in eq.~\eqref{zij}.
 We define a finite subset $\Delta_G \subset \mc$ by induction in $V$, with the following properties:
 \begin{enumerate}
\item If $G$ has only 2 vertices $i,j$, then $\Delta_G = \{2z_{ij}\}$.
\item Denoting ${\mathbb P}(V)$ the set of all partitions $I_1, \dots, I_r$ of $\{1, \dots, V\}$ excluding
$\{1\}, \dots, \{V\}$ and $\{1, \dots, V\}$, we have
\ben
\Delta_G = \{2z_I\} \cup \bigcup_{P \in {\mathbb P}(V), P = \cup_j I_j} \{ \Delta_{G/\cup_j \gamma(I_j)} - \sum_j N_{I_j} \} \ .
\een
 \end{enumerate}
For example, for the graph with $V=3$ vertices and lines $(12),(23),(13)$, we have $\Delta_G =
\{2(z_{12}+z_{23}+z_{13}), 2z_{12} - N, 2z_{23}-N, 2z_{13}-N\}$. where $N \equiv N_I$ is assumed to be the same
for all $I = \{1,2\},\{1,3\},\{2,3\}$.

Next, we define iteratively distributions $t^\pre_G$, which should be thought of as ``pre-renormalized'' versions of
$t_G$, where all poles from any subgraph have been subtracted. The distribution $t^\pre_G(X_1, \dots, X_V)$
is  characterized by the following properties:
\begin{enumerate}
\item For $G$ consisting only of $V=2$ vertices $\{1,2\}$, we have $t^\pre_G(X_1, X_2) = t_G(X_1, X_2)$.
\item If $G$ has $V$ vertices, $t^\pre_G$ is a well-defined distribution on $(S^D)^V \setminus \{X_1=\dots= X_V\}$.
\item $t^\pre_G$ has a scaling expansion of the form
\ben
t^\pre_G(X_1, \dots, X_V) \prod_{j=1}^{V-1} d\mu(X_j) \sim \sum_{\rho \in \Delta_G} \sum_{k=0}^\infty M^{-2z+\rho} H^{2k} u_{\rho,2k}(x_1, \dots, x_{V-1}) \ \prod_{j=1}^{V-1} d^D x_j \ ,
\een
where $x_1, \dots, x_{V-1} \in (T_{X_V} S^D)^{V-1}$ are the Riemannian normal coordinates, identified with vectors
in a neighborhood of the origin in $\mr^D$, of $X_1, \dots, X_{V-1}$ for fixed $X_V$. Each $u_{\rho,2j}$ is a homogeneous
distribution of degree $\rho + 2j$.
\item If $f$ has its support in $(S^D)^V \setminus \{X_1=\dots =X_V\}$, then $t^\pre_G(f)$ is analytic in the
variables $z_l, l \in  \E G$ subject to the condition that, for any proper subset $I \subset \{1, \dots, V\}$ we have
\ben\label{condI}
\Delta_{\gamma(I)} \subset \{ w \in \mc  \mid \R(w) > -N_I \} \ .
\een
\item $t^\pre_G$ agrees with $t_G$ for configurations $(X_1, \dots, X_V)$
such that $X_i \neq X_j$ for all $i \neq j$.
\end{enumerate}
From items 2),3), one can show by the same arguments as in sec.~3.3 of \cite{hollands2}
that $t^\pre_G$ has a unique extension to all of $(S^D)^V$. This extension is analytic in the $z_l \in \mc, l \in \E G$
subject to condition~\eqref{condI} in 4) for all $I$, except possibly for those configurations $z_l \in \mc$ such that $\Delta_G \cap (-\mn_0) \neq \emptyset$, where it can have  poles at $\rho \equiv \rho(\vec z) \in \Delta_G \cap (-\mn_0)$. The residues at
these poles are proportional to derivatives of the delta-function $\delta(X_1, \dots, X_V)$. These correspond to the ``superficial divergence(s)'' of the graph $G$, and we have in fact
\bena\label{pole1}
t^\pre_G(X_1,\dots,X_V) &=& \sum_{\rho \in \Delta_G} \ \sum_{n=(V-1)D}^N \sum_{j=0}^{n-(V-1)D}
\frac{M^{-2z-n} H^{2j} }{(\rho(\vec z)+n)^{m_{j,n,\rho}}} {\mathcal D}_{n,j,D,\rho} \delta(X_1,\dots,X_V) + \text{regular part}\non \\
&=:& \pp_N [t^\pre_G(X_1, \dots, X_V)] + \text{regular part}  .
\eena
Here ${\mathcal D}_{n,j,D,\rho}$ is a $O(D+1)$-invariant partial differential operator of degree $n-2j-D(V-1)$
on $(S^D)^V$ which is homogeneous in $\nabla, H$, which may depend analytically on the $z_l \in \mc$. The regular part has, by definition, no poles for any configuration of $z_l \in \mc$ such that $\R(\Delta_{\gamma(I)}) > -N_I$ for any subset $I$, including now $I = \{1,\dots, V\}$ itself (i.e. $\gamma(I)=G$), where $N_{\{1, \dots, V\}}:=N$. The extension $t_G$ is then defined as this regular part, i.e. by
performing the ``MS-subtraction''
\ben
t_G^\ren(X_1,\dots,X_V) := t^\pre_G(X_1,\dots,X_V) - \pp_{N}[t^\pre_G(X_1,\dots,X_V)] \ .
\een
What is left is to give a recursive definition for the pre-extension. The recursion starts with $V=2$, for which
the pre-extension is known, and then increases $V$. It is in fact uniquely determined in terms of
1)--5) above and is given as\footnote{When we iterate the extension procedure that we have sketched to general $V$, one ends up
with a compact formula which is called ``Zimmermann forest formula'', because
a similar formula was derived for momentum space Feynman integrals in Minkowski space in a famous paper by Zimmermann~\cite{zimmermann}.
In position space, a formula of this kind has been obtained by~\cite{keller2,garcia,pinter}.}:
\bena
&& t^\pre_G(X_1, \dots, X_V) := t_G(X_1, \dots, X_V) - \\
&&\vspace{2cm} - \sum_{P \in {\mathbb P}(V), P = \cup_j I_j}
t^\pre_{G \setminus \cup_j \gamma(I_j)}(X_{1}, \dots, X_{V}) \prod_j \pp_{N_{I_j}} [t^\pre_{\gamma(I_j)}(X_k)_{k \in I_j}] \, . \non
\eena
This formula requires some comment. First, since the partition $I = \{1, \dots, V\}$ is by definition excluded
in the sum over partitions, the unknown term $\pp_{N_{I}} [t^\pre_{G}(X_k)_{k \in I}]$ does not appear on the right side.
Furthermore, the partition $P: \{1\}, \dots, \{V\}$ is excluded as well, so at least one $I_j$ in each partition
$P: I_1, \dots, I_r$ has more than one element. Therefore, the unknown term $t_G^\pre(X_1, \dots, X_V)$ also 
does not appear in the right side. Hence this formula defines $t^\pre_G$ in terms of quantities that 
are already known inductively. 

Because $\pp_{N_{I_j}} [t^\pre_{\gamma(I_j)}(X_k)_{k \in I_j}]$ is proportional to derivatives of the $\delta$-function
with $|I_j|$ entries, this effectively means that the vertices from $I_j$ in $t^\pre_{G \setminus \cup_j \gamma(I_j)}(X_{1}, \dots, X_{V})$ are ``merged''. In fact, if there were no derivatives at all, we could replace this
by $t^\pre_{G /\cup_j \gamma(I_j)}(X_{j_1}, \dots, X_{j_r})$, where $G / \cup_j \gamma(I_j)$ is
the graph with $r<V$ vertices obtained by shrinking all graphs $\gamma(I_j)$ to a single point. If there are derivatives,
the story is similar, except that the derivatives will first act on the points $X_k, k \in I_j$ for each $j=1,\dots,r$.
This will produce again a distribution of the form $t^\pre_{G /\cup_j \gamma(I_j)}(X_{j_1}, \dots, X_{j_r})$
which has slightly shifted $z_l$ due to the action of the invariant operators ${\mathcal D}_j$.
One may be shown that the iterative definition
satisfies 1)--5) by induction---the proof of this is similar to that given in~\cite{hollandswald2}.

As we have described above, the
analyticity domain of $t_G^\ren$ as a function of the parameters $z_l, l \in \E G$, is given by those 
$z_l$ for which 
\ben
\R(\Delta_{\gamma(I)}) > -N_I \quad \text{for all $I \subset \{1, \dots, V\}$} \ .
\een
Thus, by taking all the $N_I$ sufficiently large, i.e. by subtracting enough of the poles in the MS-subtraction
step at each order, we can achieve that $t^\ren_G$ has an analyticity domain in the $z_l \in \mc$ that is
as large as we desire.

\medskip
\noindent
In summary, we have outlined the proof of the following:

\begin{thm}\label{thmren1}
The distribution $t_G(X_1, \dots, X_V)$, initially only defined as a distribution non-coinciding configurations of points
in $S^D$, i.e. $X_i \neq X_j$ for all $i \neq j$ has an extension to a distribution $t^\ren_G$ defined for all
configurations $(X_1, \dots, X_V) \in (S^D)^V$. For any arbitrary but fixed $N \in \mr$, we can define this extension so that it is analytic for $z_{l} \in \{ w \in \mc \mid \R(w) > -N\}$ for all $l \in \E G$.
\end{thm}

\noindent
{\bf Remark:}
1) Although this result has been stated for graphs $G$ without external legs ($E=0$), it is easy to generalize
the above argument to the case where external legs $X_1, \dots, X_E$ are present, as long as these
do not coincide, $X_i \neq X_j$ for all $0< i \neq j \le E$.

\noindent
2) In our formula for $I_G$, we need to integrate the complex parameters $z_l$ associated with the lines $l$ of
the Feynman graph $G$ over certain contours, and the location of these contours determines our choice of
the numbers $N_I$. The contours are located for $\R(z_l)$ between
$-(D-1)/2$ and $-(D-2)/2$ for a principal series scalar field.
Our extension $t_G$ must be analytic for such $z_l$.
and we must make a choice of the $N_I$ such that this domain contains the contours of interest.

\medskip
\noindent
Given that we have now an extension $t^\ren_G$ defined in terms of $t_G$ via subsequent MS-subtractions,
we may ask whether we can go back and write $t_G$ in terms of $t_G^\ren$. For this, we just have to
undo the subtractions at each step. This is a problem of basically combinatorical type, the solution to
which is known in the literature on renormalization theory as ``Zimmermann forest formula'', for
a derivation of this formula in position space, see~\cite{keller2}. To state this formula in reasonably
compact form, let $P: I_1 \cup \dots \cup I_r = \{1, \dots, V\}$ be a partition of the set of integration vertices.
Then define the inverse of the ``MS-subtraction operator'' by
\ben
-T_P \ t^\ren_G (X_1, \dots, X_V) := t_{G\setminus\cup_{I \in P} \gamma(I)}^\ren (X_1, \dots, X_V) \ \prod_{I \in P}
\begin{cases}
\pp_{N_I} \ [-t_{\gamma(I)}^\pre(X_j)_{j \in I}] & \text{if $|I| \ge 2$}\\
1 & \text{if $|I|=1$.}
\end{cases} 
\een
A ``forest'' $\Phi$ over $\{1, \dots, V\}$ is, in this context, a nested system of
partitions, i.e. the first element $P$ of the forest is just a partition as above, the second layer is
a partition $P_i$ further subdividing each set $I_i$ in the partition $P$, etc. The formula expressing $t_G$ 
in terms of $t^\ren_G$ is:
\ben\label{forest}
t_G(X_1, \dots, X_V) =  \ \sum_{{\rm forests} \ \Phi} \ \bigg( \prod_{P \in \Phi} (-T_P)  \bigg) \ t_G^\ren(X_1, \dots, X_V)
\ .
\een
In the product of the subtraction operators, the factors are ordered in such a way that
the coarsest partition stands to the left. The forests are over partitions of the $V$ interaction
vertices of the graph $G$ into subsets, which together with their lines can be thought of as
sub-graphs of $G$.

The right side of this is well-defined as a distribution
on all of $(S^D)^V$ apart from the poles $z_l$'s for which some $\rho = \rho(\vec z) \in \Delta_{\gamma(I)}$ becomes a negative
natural number for some $I \subset \{1, \dots, V\}$. Using the construction of the sets $\Delta_{\gamma}$, it is seen that this situation can occur at most when the $z_l$'s are ``{\em in resonance}''
\ben\label{conditionz}
\sum_{l \in \E G} n_l z_l \in \mz \quad \text{for some $n_i \in \mz$.}
\een
So we conclude:

\begin{thm} \label{renthm2}
The distribution $t_G(X_1, \dots, X_V)$, initially only defined as a distribution non-coinciding configurations of points in $S^D$, i.e. $X_i \neq X_j$ for all $i \neq j$ has an extension to a distribution to all
configurations, for $z_l$'s such that the absence of singularities condition~\eqref{conditionz} is fulfilled.
This extension is an analytic continuation in the $z_l$'s of the distribution $t_G$ in eq.~\eqref{tdef1} defined
automatically for sufficiently large $\R(z_l)$'s.
\end{thm}


Thus, to properly define $I_G$ in formula~\eqref{ig}, we have two possibilities: We could
use the extensions $t^\ren_G$ provided by thm.~\ref{thmren1}. Alternatively, if the $z_l$-integrations are chosen parallel to the imaginary axis, so that the absence of resonance condition
\ben
\sum_{l \in \E G} n_l \R(z_l) \notin \mz \quad \text{for any $n_i \in \mz$,}
\een
holds [in addition to the earlier restriction on the contour $C$ described in the picture after eq.~\eqref{mellinrep}], then we may also use the extension $t_G$ supplied by thm.~\ref{renthm2}. The difference between these two prescriptions for defining $I_G$ differs from the contribution of the pole terms.
The pole terms
contain $\delta$-functions and their derivatives in the $X_i$'s, see e.g. eq.~\eqref{pole1}, and hence would give rise to additional finite residues in the $z_l$-integrations proportional to $\delta$-functions and their derivatives. Hence, by the ``main theorem of perturbative renormalization''~\cite{hollands2,freddue}, the results for $I_G$ would differ from each 
 other by terms that can be absorbed as local counterterms in the Lagrangian~\eqref{L}.

The upshot of the entire discussion is hence that {\em we may simply use the formula~\eqref{ig} for the renormalized $I_G$,
provided that the $d\mu(X_i)$
integrations are performed first, and the $dz_l$ integrations afterwards,} where the analytic continuation
of $t_G$ in the $z_l$'s is understood, and where it is understood that the contours of the $z_l$-integrations satisfy the absence of resonance condition. That condition
is satisfied if $1$ and $\R(z_l), l \in \E G$ are linearly independent over ${\mathbb Q}$.
This automatically incorporates a renormalization
prescription compatible with adding/subtracting {\em finite} local counterterms\footnote{In $D=4$ dimensions, these local
counterterms would be of the standard form $A (\nabla \phi)^2 + B \ H^2 \phi^2 + C \ \phi^4$. Each $A,B,C$ is
a formal power series in $\lambda$, whose coefficients can depend on $c$ and $\log H^2/M^2$. It is important to
note that our prescription described here is not one that would in general have an ``analytic dependence on
the metric'', in the sense of~\cite{hollands1}. However, by adding appropriate finite local counterterms, we
can cancel out this non-analytic dependence (in the present context, it would arise through powers of $\log H^2/M^2$.)
This follows because we know from~\cite{hollands2} that a prescription with an analytic dependence of the metric
exists, and that it will differ from the given one by local counterterms.
}
.

\subsection{Parametric representation}\label{sec:parametric}

We next derive a parametric representation of $I_G$, cf. eq.~\eqref{ig}. As explained in the previous
subsection, our formula $I_G$ has to be understood in the sense that we first perform the integral
\ben\label{masterintegral}
\left( \prod_{i=E+1}^{V+E} \int_{S^D} d\mu(X_i) \right) \ t_{\vec z}(X_1, \dots, X_{V+E})
\een
for appropriate values of $\vec z$ such that it is well-defined, then analytically continue, and
then integrate over $\vec z$ along certain contours parallel to the imaginary axis subject to
the absence of resonances condition~\eqref{conditionz}. In this section, we explain how one can do this in practice using the familiar trick of Schwinger parameters. The derivation is only formal, and the resulting integral over the parameters is not well-defined a priori. But it can be given sense via analytic continuation in $\vec z$ and a technique called ``sector decomposition''~\cite{binoth1,binoth2,hepp,weinzierl1}. Another, this time rigorously derived, parametric representation based on the technique of Mellin-Barnes integrals will be provided below in subsec.~\ref{sec:mb}.

The first step is to replace the integrations
over a $D$-sphere rather to ones Euclidean space. This can be achieved using the identity,
\ben
d\mu(X) = \frac{H}{2} \ \delta(H^2 X^2-1) \ d^{D+1} X \ ,
\een
together with
\ben
\delta(H^2 X^2-1) = \delta(\log H^2 X^2) =
\int_{} \frac{dz}{2\pi i} (H^2 X^2)^{z}
\een
where the integration path is running parallel to the imaginary $z$-axis, so that the second expression is
up to a prefactor just the ordinary Fourier transform of the delta-function. We would like to
substitute this identity for
each integration measure $d\mu(X_i), i=E+1, \dots, V+E$ in the master integral, with new complex integration variables
which we call $z_{*i}$. We get, up to a prefactor $(H/2)^V$,
\bena\label{masterintegral1}
&&  \left( \prod_{i=E+1}^{V+E} \int_{S^D} d\mu(X_i) \right) \ t_{\vec z}(X_1, \dots, X_{V+E}) \\
&=& \left( \prod_{j=E+1}^{V+E} \int_{z_{*j}}  \right) \left( \prod_{i=E+1}^{V+E} \int_{\mr^{D+1}} d^{D+1} X_i \ (H^2 X_i^2)^{z_{*i}} \right) \ t_{\vec z}(X_1, \dots, X_{V+E}) \  \ . \non
\eena
We next introduce Schwinger parameters
for each factor under the integral~\eqref{masterintegral1} [cf. eq.~\eqref{tdef}] and use the standard formula
\ben
[ H^2 \ (X_i-X_j)^2 ]^{z} = \frac{1}{\Gamma(-z)} \int_0^\infty d\alpha \ \alpha^{-1-z} \ \exp [-\alpha H^2 \ (X_i-X_j)^2] \, .
\een
Inserting this into $t_{\vec z}$ in all places clearly reduces the integrand to a Gaussian.
We have one Schwinger parameter for each complex variable $z_{ij}$ or $z_{*i}$, which
we denote accordingly by $\alpha_{ij}$ or $\alpha_{i*}$. The collection of all Schwinger
parameters is denoted $\vec \alpha$.
The exponential in the Gaussian is minus
\bena\label{qform}
&& \sum_{i,j=1}^{V+E} \alpha_{ij} H^2 (X_i-X_j)^2 + \sum_{j=E+1}^{V+E} \alpha_{j*} H^2 X_j^2 \\
&=:& \sum_{i,j=E+1}^{V+E} Q_{ij} \ H^2 X_i \cdot X_j + 2 \sum_{j=E+1}^{V+E} B_j \cdot HX_j +
C \ . \non
\eena
The matrix, vector and scalar
quantities $Q,B,C$ are defined by the last equation. They are functions of the parameters $\vec \alpha$ and $B$
is additionally a function of the points $\vec X = (X_{1}, \dots, X_{E})$ that are not integrated in $I_G$, i.e. the external legs. Explicitly, we have
\ben
Q_{ij} =
\begin{cases}
\sum_{k} \alpha_{jk} & \text{if $i=j$, sum over $k$ connected to $j$,}\\
-\alpha_{ij} & \text{if $i\neq j$ and $i$ connected to $j$,}\\
0 & \text{otherwise,}
\end{cases}
\ \ \ i,j \in \{E+1, \dots, V+ E\} \ ,
\een
as well as
\ben
B_i = \sum_{j \in \{1,...,E\}: (ij) \in G} \alpha_{ij} \ HX_j \ , \quad
C = \sum_{i=1}^E \sum_{j \in \{1,...,E\}: (ij) \in G} \alpha_{ij} \ .
\een
These integration parameters can be decomposed conveniently as $R \cdot \vec \alpha$, where $R$ is a non-negative
radial coordinate, and where $\vec \alpha$ is now allowed to range through the standard simplex $\Delta$, defined by
the conditions $\alpha_{ij} \ge 0$ and  $\sum_{i,j} \alpha_{ij}  = 1$
where the sum is over those $i,j$ such that either $i=*, j = E+1,\dots,E+V$, or 
over $i,j$ such that there is a line in $G$ connecting $i,j$.

Carrying out the $d^{D+1} X_i$ and the $dR$ integrations is now straightforward, with the final result up to a constant not
depending on $\vec z$:
\bena\label{mz}
&& \left( \prod_{i=E+1}^{V+E} \int_{S^D} d\mu(X_i) \right) \ t_{\vec z}(X_1, \dots, X_{V+E}) =  \frac{\Gamma(-z - (D+1)V/2)}{{\displaystyle \prod_{i,j}} \Gamma(-z_{ij})} \\ &&\hspace{1cm} \times \ \bigg( \prod_j \int_{z_{*j}} \bigg) \bigg( \prod_{i,j} \int_0^1 \frac{d\alpha_{ij}}{\alpha_{ij}} \bigg) \delta\bigg( 1 - \sum_{i,j} \alpha_{ij} \bigg)
  \ \frac{\mathcal{F}^{(D+1)V/2+z}}{\mathcal{U}^{(D+1)(V+1)/2+z}}
\prod_{i,j}  \ \alpha_{ij}^{-z_{ij}} \
  \ . \non
\eena
Here and in the following,
the quantity $z$ is defined as
\ben
z = \sum_{i,j } z_{ij}
\een
and here, as well as in all the sums/products in eq.~\eqref{ig1}, a sum $\sum_{i,j}$ (or a product)
runs over $i,j$ such that either $i=*, j = E+1,\dots,E+V$, or
over $i,j$ such that there is a line in $G$ connecting $i,j$.
 In the notation from the previous
subsection, $z = z_I, I=\{1, \dots, V\}$.
The polynomials $\mathcal{U,F}$ depend on the graph $G$ and
are defined by\footnote{Polynomials similar to $\mathcal{U},\mathcal{F}$ also appear in the context of loop integrals on Minkowski space, and
are sometimes called ``Symanzik polynomials'' there, see e.g. \cite{iz}. However, we note that our polynomials are not identical to
these, because we work on deSitter space (actually $S^D$ in this section), and because we are in position space.}
\ben
\mathcal{U} = {\rm det} \ Q \ , \quad \mathcal{F} = {\rm det} \ Q \ (C- \sum (Q^{-1})_{ij}\ B_i \cdot B_j) \ .
\een
We will give another prescription how to obtain the polynomials $\U,\F$ in the next section, where we
also discuss many of the special properties that these polynomials have. Here we only note that
the polynomials $\mathcal{U}, \mathcal{F}$ are homogeneous in the variables $\alpha_{ij}$, and that each monomial
of $\U$ has coefficient $+1$, whereas each monomial in $\F$ has coefficient $+1$ or coefficient $(1-Z_{kl})$,
where $Z_{kl}$ are the point-pair invariants formed from the external points $X_1, \dots, X_E$. Hence, in the Euclidean domain where $Z_{kl} \in [-1,1)$, the coefficients of both $\U,\F$ are positive, and singularities in the integration over the parameters $\vec \alpha$ in eq.~\eqref{ig1} can therefore only arise from the boundary of the integration region $\Delta$ when
at least of the $\alpha_{ij} = 0$. Because of these singularities, the integral eq.~\eqref{mz} is actually
ill defined, and this is a consequence of the fact that our manipulations given above were in part only formal.

We ignore this important point however for the moment, and we insert
our expression~\eqref{mz} for the master integral into $I_G$, leading to
\bena\label{ig1}
&& I_G(\{Z_{ij}\}) =
K_G \ \int_{\vec z} \ \bigg( \prod_{i,j} \int_0^1 \frac{d\alpha_{ij}}{\alpha_{ij}} \bigg) \delta\bigg( 1 - \sum_{i,j} \alpha_{ij} \bigg) \non\\
&&\hspace{1cm} \times \  \ \frac{\Gamma(-z - (D+1)V/2)}{{\displaystyle \prod_{i,j}} \Gamma(-z_{ij})} \ \prod_{l \in \E G} 2^{-2z_l} \
\frac{ \Gamma((D-1)/2+i\rho+z_l)\Gamma((D-1)/2-i\rho+z_l)\Gamma(-z_l)}{\sin[\pi (z_l+(D-2)/2)]   \Gamma(D/2+z_l)} \  \non \\
&&\hspace{1cm} \times  \ \frac{\mathcal{F}^{(D+1)V/2+z}}{\mathcal{U}^{(D+1)(V+1)/2+z}}
\prod_{i,j}  \ \alpha_{ij}^{-z_{ij}} \
  \ .
\eena
This is the desired parametric representation for $I_G$. $\vec z$ stands for $z_l, l \in \E G$, or $z_{*i}, i \in \{E+1, \dots, E+V\}$, and $\int_{\vec z}$ means a multiple contour integral. For $l \in G$ the contours are parallel to the imaginary axis with
\ben\label{condi3}
-(D-1)/2 < \mathfrak{R}(z_l) < -(D-1)/2 + \epsilon
\een
for an arbitrarily small $\epsilon > 0$ . 
The remaining contours  can be chosen at this stage to be any contour parallel to the imaginary axis.
 $K_G$ is the constant
\ben\label{kgdef}
K_G = H^{L(D-2)-VD} \ \left[ \frac{\cosh \pi \rho}{(4\pi)^{D/2}} \right]^{L} \
2^{-V} \pi^{V(D+1)/2} \ ,
\een
with $L=|\E G|$ the number of lines in the graph $G$. For the graphs in $\lambda \phi^4$-theory, we
have $L=2V+E/2$, so the dimensionful term is $H^{V(D-4)+E\frac{D-2}{2}}$, which in $D=4$
reduces to $H^E$.

It is possible to make sense of the singular integrals over the Schwinger parameters via a method
which has been called ``iterated sector-decomposition'' in the literature. It is normally carried out
in the context of momentum space loop-integrals in flat space QFT. However, expression~\eqref{mz} is
of a form to which this formalism applies. Later in subsec.~\ref{sec:mb}, we will present another
parametric representation which avoids these problems, so will not discuss these ingenious constructions here
and refer the reader to the literature~\cite{binoth1,binoth2,hepp,weinzierl1}.

We close this section by pointing out another more geometrical (but still formal-) way of writing $I_G$. The $\vec \alpha$-integration can be viewed as an integration  on the simplex $\Delta$, where the integration measure is simply
\ben
[d \vec \alpha] = \delta \left[ 1-\sum_{ij} \alpha_{ij} \right] \prod_{ij} d\alpha_{ij}  \, .
\een
Alternatively, let us give the simplex $\Delta$ an orientation by
picking an order $e_1 < \dots < e_r$ of the edges in $\E G^*$,
and let $\Omega_{\vec z}$ be the
differential form in $\vec \alpha$ defined by
\ben
\Omega_{\vec z} =  \frac{\mathcal{F}^{(D+1)V/2+z}}{\mathcal{U}^{(D+1)(V+1)/2+z}} \bigg(
\prod_{i=1}^r  \ \alpha_{e_i}^{-1-z_{e_i}} \bigg)
\sum_{k=1}^r (-1)^k \alpha_{e_k} \ d\alpha_{e_1} \wedge \dots \wedge \widehat{d\alpha_{e_k}} \dots \wedge
d\alpha_{e_r}
 \ .
\een
 Then a straightforward
calculation based on the homogeneity of $\U,\F$ shows that
\ben
d\Omega_{\vec z} = 0 \ .
\een
Let $\tilde \Delta$ be any smooth manifold such that $\partial \tilde \Delta = \partial \Delta$, and
such that $\tilde \Delta$ is homotopic to $\Delta$. Then the
parametric form of the Feynman integral can be written as
\bena\label{ig12}
&& I_G(\{Z_{ij}\}) =
K_G \ \int_{\vec z} \ \int_{\tilde \Delta} \Omega_{\vec z}  \ \frac{\Gamma(-z - (D+1)V/2)}{\prod_{i,j} \Gamma(-z_{ij})} \non\\
&&\hspace{1cm} \times \  \ \prod_{l \in \E G} 2^{-2z_l} \
\frac{ \Gamma((D-1)/2+i\rho+z_l)\Gamma((D-1)/2-i\rho+z_l) \Gamma(-z_l)}{\sin[\pi (z_l+(D-2)/2)]  \Gamma(D/2+z_l)} \  .
\eena

\section{Analytic continuation and Mellin-Barnes represenation}

\subsection{The graph polynomials $\mathcal{U,F}$}\label{app:c}

In this subsection, we describe begin describing
the properties of the graph polynomials\footnote{In this section, we put a subscript ``$G$'' in order to indicate the
dependence on the graph.} $\U_G,\F_G$ introduced in the previous section,
which appear in our parametric formula \eqref{ig1}. Some of the properties of these polynomials will
be used in the next section.

Let $G$ be a Feynman graph, with internal vertices $i= E+1, \dots, V+E$, and $E$ external lines.
In the following, we consider these external lines as part of the graph, and
we let the external vertices on which they end be labeled by $i=1, \dots, E$, and each
such vertex is associated with a $X_i \in S^D$. A pair $i,j$ of internal
vertices can be connected by none or several lines. If they are connected by at least one line, then we associate
a Schwinger parameter $\alpha_{ij} \in \mr$ to it. Since all that matters in the following is the association
of the graph $G$ with the Schwinger parameters, it is convenient to replace $G$ with a graph $G^*$ such that
\begin{itemize}
\item $G^*$ has the same vertices as $G$, and an additional vertex called ``$*$''.
\item Two vertices $i,j \in \{1, \dots, V+E\}$ in $G^*$ are connected by a single line if these vertices are connected by at least one line in $G$, and they are not connected otherwise.
\item Each of the vertices $i=E+1,\dots,E+V$ is connected to $*$ with a line.
\end{itemize}
With each of the last type of line, we associate a
Schwinger parameter $\alpha_{i*}$. An example of a graph $G$ and the corresponding graph $G^*$ is given in the following pictures:

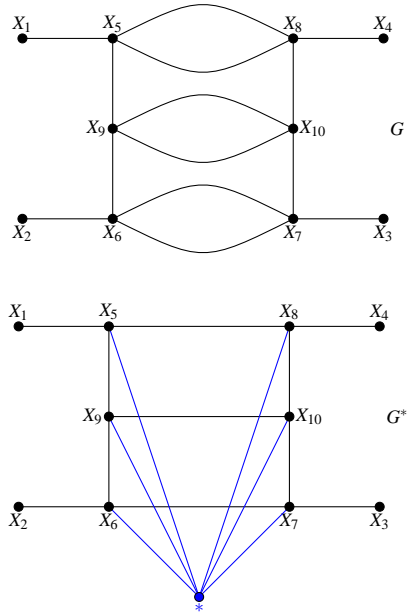
\begin{center}
\begin{tikzpicture}[scale=.6, transform shape]
\draw  (-4,4) node[black,above]{$X_1$} --  (-2,4) node[black,above]{$X_5$};
\draw  (-4,0) node[black,below]{$X_2$}--  (-2,0) node[black,below]{$X_6$};
\draw  (2,4) node[black,above]{$X_8$} --  (4,4) node[black,above]{$X_4$};
\draw  (2,0) node[black,below]{$X_7$} --  (4,0) node[black,below]{$X_3$};
\draw (-2,4) -- (-2,0);
\draw (2,4) -- (2,0);
\draw (-2,0) .. controls (0,1) and (0,1) .. (2,0);
\draw (-2,0) .. controls (0,-1) and (0,-1) .. (2,0);
\draw (-2,2) .. controls (0,1) and (0,1) .. (2,2);
\draw (-2,2) .. controls (0,3) and (0,3) .. (2,2);
\draw (-2,4) .. controls (0,3) and (0,3) .. (2,4);
\draw (-2,4) .. controls (0,5) and (0,5) .. (2,4);
\draw (-2,2) node[black,left]{$X_9$};
\draw (2,2) node[black,right]{$X_{10}$};
\draw (4,2) node[black,right]{$G$};
\draw (-4,0) node[draw,shape=circle,scale=0.5,fill=black]{};
\draw (-2,0) node[draw,shape=circle,scale=0.5,fill=black]{};
\draw (-2,2) node[draw,shape=circle,scale=0.5,fill=black]{};
\draw (-2,4) node[draw,shape=circle,scale=0.5,fill=black]{};
\draw (-4,4) node[draw,shape=circle,scale=0.5,fill=black]{};
\draw (2,2) node[draw,shape=circle,scale=0.5,fill=black]{};
\draw (2,4) node[draw,shape=circle,scale=0.5,fill=black]{};
\draw (4,4) node[draw,shape=circle,scale=0.5,fill=black]{};
\draw (4,0) node[draw,shape=circle,scale=0.5,fill=black]{};
\draw (2,0) node[draw,shape=circle,scale=0.5,fill=black]{};
\end{tikzpicture}
\end{center}

\begin{center}
\begin{tikzpicture}[scale=.6, transform shape]
\draw  (-4,4) node[black,above]{$X_1$} --  (-2,4) node[black,above]{$X_5$};
\draw  (-4,0) node[black,below]{$X_2$}--  (-2,0) node[black,below]{$X_6$};
\draw  (2,4) node[black,above]{$X_8$} --  (4,4) node[black,above]{$X_4$};
\draw  (2,0) node[black,below]{$X_7$} --  (4,0) node[black,below]{$X_3$};
\draw (-2,4) -- (-2,0);
\draw (2,4) -- (2,0);
\draw (-2,0) -- (2,0);
\draw (-2,2) -- (2,2);
\draw (-2,4) -- (2,4);
\draw (-2,2) node[black,left]{$X_9$};
\draw (2,2) node[black,right]{$X_{10}$};
\draw (4,2) node[black,right]{$G^*$};
\draw[blue] (-2,4) -- (0,-2);
\draw[blue] (-2,2) -- (0,-2);
\draw[blue] (-2,0) -- (0,-2);
\draw[blue] (2,4) -- (0,-2);
\draw[blue] (2,2) -- (0,-2);
\draw[blue] (2,0) -- (0,-2);
\draw (0,-2) node[blue,below]{$*$};
 \draw (-4,0) node[draw,shape=circle,scale=0.5,fill=black]{};
 \draw (-2,0) node[draw,shape=circle,scale=0.5,fill=black]{};
 \draw (-2,2) node[draw,shape=circle,scale=0.5,fill=black]{};
 \draw (-2,4) node[draw,shape=circle,scale=0.5,fill=black]{};
 \draw (-4,4) node[draw,shape=circle,scale=0.5,fill=black]{};
 \draw (2,2) node[draw,shape=circle,scale=0.5,fill=black]{};
 \draw (2,4) node[draw,shape=circle,scale=0.5,fill=black]{};
 \draw (4,4) node[draw,shape=circle,scale=0.5,fill=black]{};
 \draw (4,0) node[draw,shape=circle,scale=0.5,fill=black]{};
 \draw (0,-2) node[draw,shape=circle,scale=0.5,fill=blue]{};
 \draw (2,0) node[draw,shape=circle,scale=0.5,fill=black]{};
\end{tikzpicture}
\end{center}

Following standard constructions in graph theory~\cite{tutte}, let us define the {\em Laplacian} of the graph $G^*$ to be
the $V+E+1$-dimensional square matrix (for $i,j = *,1,\dots, V+E$)
\ben
L_{ij} = \begin{cases}
\sum_{k} \alpha_{jk} & \text{if $i=j$, sum over $k$ connected to $j$,}\\
-\alpha_{ij} & \text{if $i\neq j$ and $i$ connected to $j$,}\\
0 & \text{otherwise.}
\end{cases}
\een
Furthermore, if $I,J \subset \{*,1,\dots, V+E\}$ given as $I = \{i_1 < \dots < i_k\}, J = \{j_1 < \dots < j_k\}$, then
we define the matrix $L[I,J]$ to be $L$, with columns $i_1, \dots, i_k$ removed, and with rows $j_1, \dots, j_k$ removed.
It is straightforward to see that, if $I = \{*,1, \dots,E\}$, then $L[I, I] = Q$, where $Q$ is as in eq.~\eqref{qform}, so
because $\U_G = \det Q$,
\ben
\U_G = \det  L[I, I] \ .
\een
Obviously, since each matrix entry of $L[I,I]$ is linear in the Schwinger parameters, it follows that
$\U_G$ is a homogeneous polynomial of degree $V$. The polynomial $\F_G$ can similarly be written in terms of
determinants of the Laplacian. Let us define, for
$i \notin \{*,1,...,E\}$
\ben\label{Ii}
I_i = \{*, 1, \dots, E,i\} \ .
\een
The matrix $\det Q \ Q^{-1}_{ij}$ is of course formed from $(-1)^{i+j}$ times the $ij$-minors of $Q$,
which are given by $(-1)^{i+j} \det L[I_i,I_j]$. So the polynomial $\F_G$ is                                                                                                        the graph polynomial $\F_G$ is given by
\bena
\F_G &=& \det Q \left[ - \sum_{i,j=E+1}^{V+E}  (Q^{-1})_{ij} \ B_i \cdot B_j + C \right] \non\\
&=& -\sum_{i,j=E+1}^{V+E} (-1)^{i+j} \det L[I_i, I_j] \!\!
\sum_{
{\tiny
\begin{array}{c}
k,l \in \{1,...,E\}:\\
(ik),(jl) \in \E G
\end{array}
}
}
\alpha_{ik} \alpha_{jl} \ Z_{kl} \non\\
&& + \ \det L[I,I] \sum_{i=1}^E \sum_{
{\tiny
\begin{array}{c}
j \in \{1,...,E\}:\\
(ij) \in \E G
\end{array}
}
} \alpha_{ij} \ .
\eena

The determinants of $L[I,J]$ are closely related to the trees within the graph $G^*$ by the so-called (generalized)
``tree-matrix-theorem''~\cite{tutte,weinzierl1,rivasseau}. To state this theorem, let us first give some notation.
A tree $T$ in $G^*$ is a connected subgraph of $G^*$ without loops. It is called a ``spanning tree'' if it has the same
vertices as $G^*$. A forest $F$ is a collection of disjoint trees, i.e. a subgraph of $G^*$ containing no loops. We
denote by $\TT$ the set of all forests, and by $\TT_k$ the set of all forests
\ben
(T_1, \dots, T_k) \in \TT_k
\een
consisting of precisely $k$ trees. Such a forest is called a ``$k$-forest''. A forest \textcolor{red}{$F$} for the graph $G^*$ above is
drawn in the following picture:
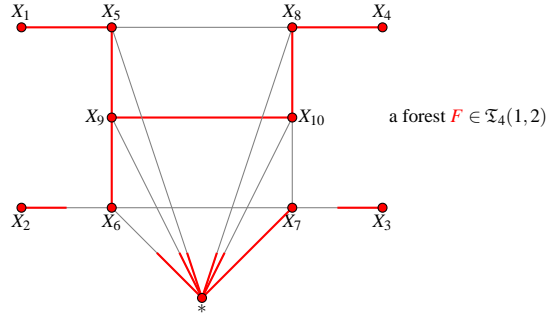
\begin{center}
\begin{tikzpicture}[scale=.6, transform shape]
\draw[red,thick]  (-4,4) node[black,above]{$X_1$} --  (-2,4) node[black,above]{$X_5$};
\draw[gray]  (-4,0) node[black,below]{$X_2$}--  (-2,0) node[black,below]{$X_6$};
\draw[red,thick]  (2,4) node[black,above]{$X_8$} --  (4,4) node[black,above]{$X_4$};
\draw[gray]  (2,0) node[black,below]{$X_7$} --  (4,0) node[black,below]{$X_3$};
\draw[red,thick] (-2,4) -- (-2,0);
\draw[red,thick] (2,4) -- (2,2);
\draw[gray] (2,2) -- (2,0);
\draw[gray] (-2,0) -- (2,0);
\draw[red,thick] (-2,2) -- (2,2);
\draw[gray] (-2,4) -- (2,4);
\draw (-2,2) node[black,left]{$X_9$};
\draw (2,2) node[black,right]{$X_{10}$};
\draw (4,2) node[black,right]{a forest \textcolor{red}{$F$} $\in \T_4(1,2)$};
\draw[gray] (-2,4) -- (0,-2);
\draw[gray] (-2,2) -- (0,-2);
\draw[gray] (-2,0) -- (0,-2);
\draw[gray] (2,4) -- (0,-2);
\draw[gray] (2,2) -- (0,-2);
\draw[red,thick] (2,0) -- (0,-2);
\draw (0,-2) node[black,below]{$*$};
\draw[red,thick] (0,-2) -- (-1,-1);
\draw[red,thick] (0,-2) -- (-0.5,-1);
\draw[red,thick] (0,-2) -- (-0.33,-1);
\draw[red,thick] (0,-2) -- (0.33,-1);
\draw[red,thick] (0,-2) -- (0.5,-1);
\draw[red,thick] (-4,0) -- (-3,0);
\draw[red,thick] (4,0) -- (3,0);
\draw (-4,0) node[draw,shape=circle,scale=0.5,fill=red]{};
\draw (-2,0) node[draw,shape=circle,scale=0.5,fill=red]{};
\draw (-2,2) node[draw,shape=circle,scale=0.5,fill=red]{};
\draw (-2,4) node[draw,shape=circle,scale=0.5,fill=red]{};
\draw (-4,4) node[draw,shape=circle,scale=0.5,fill=red]{};
\draw (2,2) node[draw,shape=circle,scale=0.5,fill=red]{};
\draw (2,4) node[draw,shape=circle,scale=0.5,fill=red]{};
\draw (4,4) node[draw,shape=circle,scale=0.5,fill=red]{};
\draw (4,0) node[draw,shape=circle,scale=0.5,fill=red]{};
\draw (0,-2) node[draw,shape=circle,scale=0.5,fill=red]{};
\draw (2,0) node[draw,shape=circle,scale=0.5,fill=red]{};
\end{tikzpicture}
\end{center}
With a forest such as \textcolor{red}{$F$}, we associate a monomial $m_{\textcolor{red}{F}}$ formed from the product of all Schwinger parameters
associated with each line. In the above example,
\ben
m_{\textcolor{red}{F}}(\vec \alpha) = \prod_{(kl) \in \textcolor{red}{F}} \textcolor{red}{\alpha_{kl}} = \textcolor{red}{\alpha_{15} \ \alpha_{59} \ \alpha_{9 \, 10} \ \alpha_{96} \ \alpha_{*7} \ \alpha_{84} \ \alpha_{8 \, 10}} \ .
\een
The graph polynomials $\U_G,\F_G$ are linear combinations of such monomials, where $F$ runs through certain forests of $G^*$, as we will explain.

If $I,J$ are as above, then  denote by
$\TT^{I,J}_k$ the set of all $k$ spanning forests in $G^*$ with the property that each tree $T$ within a forest $F \in \TT^{I,J}_k$ contains
precisely one vertex $i_\alpha$ from $I$ and one vertex $j_\beta$ from $J$. The forest clearly defines a bijective mapping from $I \to J$,
i.e. a permutation, and we call ${\rm sgn} \ \pi_F$ the sign of this permutation. Finally, we call $|I| = i_1 + ... + i_k$, and
similarly for $J$. With this notation in place, the tree-matrix theorem
states that
\ben
\det  L[I, J] = (-1)^{|I| + |J|} \sum_{F \in \TT^{I,J}_k} {\rm sign}(\pi_F) \prod_{(ij) \in F} \alpha_{ij} \ .
\een
Applying this identity to the set $I = J = \{*, 1, ..., E\}$ gives, with $\T_{E+1} \equiv \TT^{I,I}_{E+1}$,
\ben\label{utree}
\U_G = \sum_{F \in \T_{E+1}} m_F(\vec \alpha) = \sum_{F \in \T_{E+1}}  \prod_{(ij) \in F} \alpha_{ij} \ .
\een
One can also obtain a formula of this nature for $\F_G$. For this, we decompose $\F_G$ as
\bena
\F_G &=& \det Q \left[ - \sum_{i,j=E+1}^{V+E} (Q^{-1})_{ij} \ B_i \cdot B_j + C \right] \non\\
&\equiv& \F_0 + \V_G \ ,
\eena
where the polynomial $\V_G$ has been defined as $\V_G = \F_G(\{1\})$.
We now apply the matrix tree theorem to each polynomial. Letting $\T_{E}(i,j)$ be the $E$-forests such that precisely
one tree is connecting the vertices $i,j$, and such that the remaining $E-1$ trees each contain precisely
one of the points $\{*,1,...,E\} \setminus \{i,j\}$, we can write
\ben
\F_0
=\sum_{1 \le i \neq j \le E} (1-Z_{ij}) \sum_{F \in \T_E(i,j)}  \prod_{(kl) \in F} \alpha_{kl} \ .
\een
Furthermore, we obtain, taking into account various cancelations of negative terms,
\ben
\V_G = \sum_{i=1}^E \sum_{F \in \T_{E}(i,*)}   \prod_{(kl) \in F} \alpha_{kl} \, .
\een
Putting these equations together, we get
\ben\label{ftree}
\F_G = \sum_{1 \le i \neq j \le E} (1-Z_{ij}) \sum_{F \in \T_E(i,j)}  \prod_{(kl) \in F} \alpha_{kl}
+\sum_{1 \le i \le E} \ \sum_{F \in \T_E(i,*)}   \prod_{(kl) \in F} \alpha_{kl} \ .
\een
To illustrate the formula, we give the polynomials $\U_G,\F_G$ when $G$ is the setting sun graph.
The graph in question is:
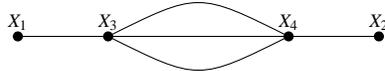
\begin{center}
\begin{tikzpicture}[scale=.6, transform shape]
\draw  (-4,0) node[black,above]{$X_1$} --  (-2,0) node[black,above]{$X_3$};
\draw  (2,0) node[black,above]{$X_4$} --  (4,0) node[black,above]{$X_2$};
\draw (-2,0) .. controls (0,1) and (0,1) .. (2,0);
\draw (-2,0) .. controls (0,-1) and (0,-1) .. (2,0);
\draw (-2,0) -- (2,0);
 \draw (-4,0) node[draw,shape=circle,scale=0.5,fill=black]{};
 \draw (-2,0) node[draw,shape=circle,scale=0.5,fill=black]{};
 \draw (2,0) node[draw,shape=circle,scale=0.5,fill=black]{};
 \draw (4,0) node[draw,shape=circle,scale=0.5,fill=black]{};
\end{tikzpicture}
\end{center}
The corresponding graph $G^*$ is:
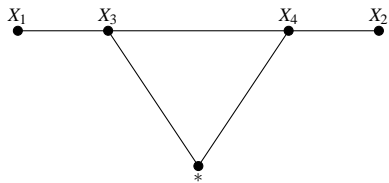
\begin{center}
\begin{tikzpicture}[scale=.6, transform shape]
\draw  (-4,0) node[black,above]{$X_1$} --  (-2,0) node[black,above]{$X_3$};
\draw  (2,0) node[black,above]{$X_4$} --  (4,0) node[black,above]{$X_2$};
\draw (-2,0) -- (2,0);
 \draw (-4,0) node[draw,shape=circle,scale=0.5,fill=black]{};
 \draw (-2,0) node[draw,shape=circle,scale=0.5,fill=black]{};
 \draw (2,0) node[draw,shape=circle,scale=0.5,fill=black]{};
 \draw (4,0) node[draw,shape=circle,scale=0.5,fill=black]{};
\draw (-2,0) -- (0,-3);
\draw (2,0) -- (0,-3) node[black,below]{$*$};
 \draw (0,-3) node[draw,shape=circle,scale=0.5,fill=black]{};
\end{tikzpicture}
\end{center}
The external points are $X_1, X_2$, and the only point-pair invariant is
hence $Z_{12}$.
The only spanning forest in $\textcolor{red}{F} \in \T_2(1,2)$, drawn in red is:
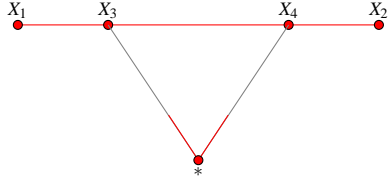
\begin{center}
\begin{tikzpicture}[scale=.6, transform shape]
\draw[red]  (-4,0) node[black,above]{$X_1$} --  (-2,0) node[black,above]{$X_3$};
\draw[red]  (2,0) node[black,above]{$X_4$} --  (4,0) node[black,above]{$X_2$};
\draw[red] (-2,0) -- (2,0);
 \draw (-4,0) node[draw,shape=circle,scale=0.5,fill=red]{};
 \draw (-2,0) node[draw,shape=circle,scale=0.5,fill=red]{};
 \draw (2,0) node[draw,shape=circle,scale=0.5,fill=red]{};
 \draw (4,0) node[draw,shape=circle,scale=0.5,fill=red]{};
\draw[gray] (-2,0) -- (0,-3);
\draw[gray] (2,0) -- (0,-3) node[black,below]{$*$};
\draw[red] (0,-3) -- (.66,-2);
\draw[red] (0,-3) -- (-.66,-2);
 \draw (0,-3) node[draw,shape=circle,scale=0.5,fill=red]{};
\end{tikzpicture}
\end{center}
The other 8 spanning forests in $\textcolor{red}{F} \in \T_2:= \cup_{r,s \in \{*,1,2\}} \T_2(r,s)$ are:
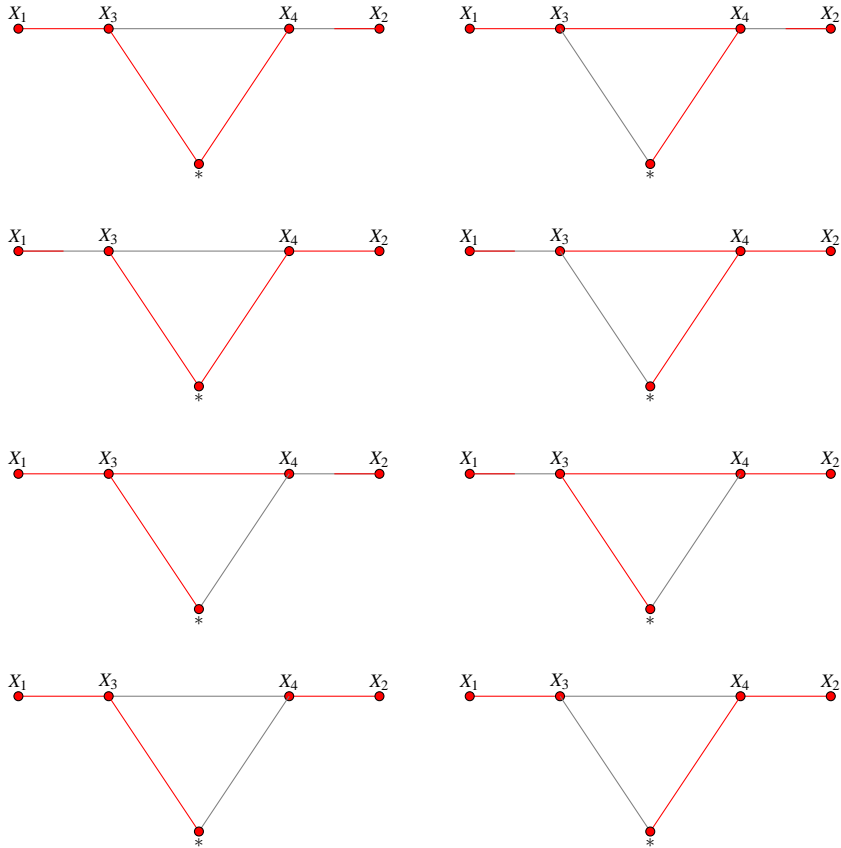
\begin{center}
\begin{tikzpicture}[scale=.6, transform shape]
\draw[red]  (-4,0) node[black,above]{$X_1$} --  (-2,0) node[black,above]{$X_3$};
\draw[gray]  (2,0) node[black,above]{$X_4$} --  (4,0) node[black,above]{$X_2$};
\draw[gray] (-2,0) -- (2,0);
 \draw (-4,0) node[draw,shape=circle,scale=0.5,fill=red]{};
 \draw (-2,0) node[draw,shape=circle,scale=0.5,fill=red]{};
 \draw (2,0) node[draw,shape=circle,scale=0.5,fill=red]{};
 \draw (4,0) node[draw,shape=circle,scale=0.5,fill=red]{};
\draw[red] (-2,0) -- (0,-3);
\draw[red] (2,0) -- (0,-3) node[black,below]{$*$};
\draw[red] (4,0) -- (3,0);
 \draw (0,-3) node[draw,shape=circle,scale=0.5,fill=red]{};
\draw[red]  (6,0) node[black,above]{$X_1$} --  (8,0) node[black,above]{$X_3$};
\draw[gray]  (12,0) node[black,above]{$X_4$} --  (14,0) node[black,above]{$X_2$};
\draw[red] (8,0) -- (12,0);
 \draw (6,0) node[draw,shape=circle,scale=0.5,fill=red]{};
 \draw (8,0) node[draw,shape=circle,scale=0.5,fill=red]{};
 \draw (12,0) node[draw,shape=circle,scale=0.5,fill=red]{};
 \draw (14,0) node[draw,shape=circle,scale=0.5,fill=red]{};
\draw[gray] (8,0) -- (10,-3);
\draw[red] (12,0) -- (10,-3) node[black,below]{$*$};
\draw[red] (14,0) -- (13,0);
 \draw (10,-3) node[draw,shape=circle,scale=0.5,fill=red]{};
\end{tikzpicture}
\end{center}
\begin{center}
\begin{tikzpicture}[scale=.6, transform shape]
\draw[gray]  (-4,0) node[black,above]{$X_1$} --  (-2,0) node[black,above]{$X_3$};
\draw[red]  (2,0) node[black,above]{$X_4$} --  (4,0) node[black,above]{$X_2$};
\draw[gray] (-2,0) -- (2,0);
 \draw (-4,0) node[draw,shape=circle,scale=0.5,fill=red]{};
 \draw (-2,0) node[draw,shape=circle,scale=0.5,fill=red]{};
 \draw (2,0) node[draw,shape=circle,scale=0.5,fill=red]{};
 \draw (4,0) node[draw,shape=circle,scale=0.5,fill=red]{};
\draw[red] (-2,0) -- (0,-3);
\draw[red] (2,0) -- (0,-3) node[black,below]{$*$};
\draw[red] (-4,0) -- (-3,0);
 \draw (0,-3) node[draw,shape=circle,scale=0.5,fill=red]{};
\draw[gray]  (6,0) node[black,above]{$X_1$} --  (8,0) node[black,above]{$X_3$};
\draw[red]  (12,0) node[black,above]{$X_4$} --  (14,0) node[black,above]{$X_2$};
\draw[red] (8,0) -- (12,0);
 \draw (6,0) node[draw,shape=circle,scale=0.5,fill=red]{};
 \draw (8,0) node[draw,shape=circle,scale=0.5,fill=red]{};
 \draw (12,0) node[draw,shape=circle,scale=0.5,fill=red]{};
 \draw (14,0) node[draw,shape=circle,scale=0.5,fill=red]{};
\draw[gray] (8,0) -- (10,-3);
\draw[red] (12,0) -- (10,-3) node[black,below]{$*$};
\draw[red] (6,0) -- (7,0);
 \draw (10,-3) node[draw,shape=circle,scale=0.5,fill=red]{};
\end{tikzpicture}
\end{center}
\begin{center}
\begin{tikzpicture}[scale=.6, transform shape]
\draw[red]  (-4,0) node[black,above]{$X_1$} --  (-2,0) node[black,above]{$X_3$};
\draw[gray]  (2,0) node[black,above]{$X_4$} --  (4,0) node[black,above]{$X_2$};
\draw[red] (-2,0) -- (2,0);
 \draw (-4,0) node[draw,shape=circle,scale=0.5,fill=red]{};
 \draw (-2,0) node[draw,shape=circle,scale=0.5,fill=red]{};
 \draw (2,0) node[draw,shape=circle,scale=0.5,fill=red]{};
 \draw (4,0) node[draw,shape=circle,scale=0.5,fill=red]{};
\draw[red] (-2,0) -- (0,-3);
\draw[gray] (2,0) -- (0,-3) node[black,below]{$*$};
\draw[red] (4,0) -- (3,0);
 \draw (0,-3) node[draw,shape=circle,scale=0.5,fill=red]{};
\draw[gray]  (6,0) node[black,above]{$X_1$} --  (8,0) node[black,above]{$X_3$};
\draw[red]  (12,0) node[black,above]{$X_4$} --  (14,0) node[black,above]{$X_2$};
\draw[red] (8,0) -- (12,0);
 \draw (6,0) node[draw,shape=circle,scale=0.5,fill=red]{};
 \draw (8,0) node[draw,shape=circle,scale=0.5,fill=red]{};
 \draw (12,0) node[draw,shape=circle,scale=0.5,fill=red]{};
 \draw (14,0) node[draw,shape=circle,scale=0.5,fill=red]{};
\draw[red] (8,0) -- (10,-3);
\draw[gray] (12,0) -- (10,-3) node[black,below]{$*$};
\draw[red] (6,0) -- (7,0);
 \draw (10,-3) node[draw,shape=circle,scale=0.5,fill=red]{};
\end{tikzpicture}
\end{center}
\begin{center}
\begin{tikzpicture}[scale=.6, transform shape]
\draw[red]  (-4,0) node[black,above]{$X_1$} --  (-2,0) node[black,above]{$X_3$};
\draw[red]  (2,0) node[black,above]{$X_4$} --  (4,0) node[black,above]{$X_2$};
\draw[gray] (-2,0) -- (2,0);
 \draw (-4,0) node[draw,shape=circle,scale=0.5,fill=red]{};
 \draw (-2,0) node[draw,shape=circle,scale=0.5,fill=red]{};
 \draw (2,0) node[draw,shape=circle,scale=0.5,fill=red]{};
 \draw (4,0) node[draw,shape=circle,scale=0.5,fill=red]{};
\draw[red] (-2,0) -- (0,-3);
\draw[gray] (2,0) -- (0,-3) node[black,below]{$*$};
 \draw (0,-3) node[draw,shape=circle,scale=0.5,fill=red]{};
\draw[red]  (6,0) node[black,above]{$X_1$} --  (8,0) node[black,above]{$X_3$};
\draw[red]  (12,0) node[black,above]{$X_4$} --  (14,0) node[black,above]{$X_2$};
\draw[gray] (8,0) -- (12,0);
 \draw (6,0) node[draw,shape=circle,scale=0.5,fill=red]{};
 \draw (8,0) node[draw,shape=circle,scale=0.5,fill=red]{};
 \draw (12,0) node[draw,shape=circle,scale=0.5,fill=red]{};
 \draw (14,0) node[draw,shape=circle,scale=0.5,fill=red]{};
\draw[gray] (8,0) -- (10,-3);
\draw[red] (12,0) -- (10,-3) node[black,below]{$*$};
 \draw (10,-3) node[draw,shape=circle,scale=0.5,fill=red]{};
\end{tikzpicture}
\end{center}
There are also 6 spanning forests  $\textcolor{red}{F} \in \T_3$.
The graph polynomials are, according to \eqref{utree},\eqref{ftree}:
\bena
\U_G &=& \alpha_{13} \alpha_{3*} + \alpha_{13}\alpha_{34} + \alpha_{13} \alpha_{42} + \alpha_{42} \alpha_{2*}
+ \alpha_{23} \alpha_{2*} + \alpha_{23} \alpha_{3*} \\
\F_G &=& (1-Z_{12}) \alpha_{13}\alpha_{24} \alpha_{34} + \non\\
&& \ \ + \ \alpha_{13} \alpha_{3*} \alpha_{4*}
+ \alpha_{13} \alpha_{3*} \alpha_{34}
+ \alpha_{13} \alpha_{34} \alpha_{4*}
+ \alpha_{13} \alpha_{3*} \alpha_{24} \non\\
&& \ \ + \ \alpha_{3*} \alpha_{4*} \alpha_{42}
+ \alpha_{34} \alpha_{4*} \alpha_{42}
+ \alpha_{3*} \alpha_{34} \alpha_{42}
+ \alpha_{13} \alpha_{*4} \alpha_{42}      \eena
Thus, in summary, we have shown in this section that:
\begin{thm}\label{graphpoly}
\begin{itemize}
\item The graph polynomial $\U_G$ is homogeneous of degree $V$ in the Feynman parameters $\alpha_{ij}, i,j \in \{*,1,\dots,V+E\}$.
Each monomial has coefficient $+1$. The monomials are related to spanning forests of the graph $G^*$, from the set
\bena
\T_{E+1} &=& \{\text{spanning forests of $G^*$ with $E+1$ trees,}\non\\
&& \text{each containing precisely one vertex from $*,1,\dots,E$} \} \ ,
\eena
see eq.~\eqref{utree}.
\item The graph polynomial $\F_G$ is homogeneous of degree $V+1$ in the Feynman parameters $\alpha_{ij}, i,j \in \{*,1,\dots,V+E\}$. Each monomial has coefficient $+ 1$ or $(1-Z_{ij}) = H^2 (X_i - X_j)^2/2$,
which is $\ge 0$ for points $X_i \in S^D$, but which can become negative for points
$\in dS_D$. The monomials are related to spanning forests of the graph $G^*$ from
\ben
\T_{E} = \bigcup_{r,s \in \{*,1,\dots,E\}} \T_E(r,s) \ ,  
\een 
where $\T_E(r,s)$ is the set of all spanning forests with $E$ trees such that 
each tree contains precisely one vertex from $*,1,\dots,E$, except for one 
tree connecting the vertices $r,s$. The formula for $\F$ is eq.~\eqref{ftree}.
\end{itemize}
\end{thm}
For completeness, let us record the expression for $\F_G$ in terms of other coordinate systems, which are
obtained by simply substituting the corresponding expressions for the point-pair invariant, see table~\ref{table2}. Letting
the external points $X_i, i=1,\dots,E$ be at equal time, i.e. $=(t, {\bf x}_i)$, the expression for
$\F_G$ in the cosmological chart is:
\ben
\F_G =  H^2 \e^{2Ht} \sum_{1 \le i < j \le E} \ ({\bf x}_i - {\bf x}_j)^2 \sum_{F \in \T_E(i,j)}  \prod_{(kl) \in F} \alpha_{kl}
+\sum_{F \in \T_E}   \prod_{(kl) \in F} \alpha_{kl} \ .
\een
 Letting the points
$X_i, i=1,\dots,E$ be at equal time in the static chart, i.e. $=(\eta, r_i, \hat x_i)$,
and letting $Hr_i = \cos \theta_i, \theta_i \in (0,\pi)$, the expression in the static chart is
\ben
\F_G = \sum_{1 \le i \neq j \le E} \ (\sin \theta_i \sin \theta_j + \cos \theta_i \cos \theta_j \ \hat x_i \cdot \hat x_j) \sum_{F \in \T_E(i,j)}  \prod_{(kl) \in F} \alpha_{kl}
+ \sum_{1 \le i \le E} \ \sum_{F \in \T_E(i,*)}   \prod_{(kl) \in F} \alpha_{kl} \ .
\een
 In terms of the signed squared geodesic
distances $\sigma_{ij}$, the polynomial $\F_G$ is obtained by replacing
$1-Z_{ij}=2 \sin^2 ( \frac{H}{2}\sqrt{\sigma_{ij}} )$.

\subsection{Mellin-Barnes representation}\label{sec:mb}

Using the structure of the polynomials $\U,\F$ derived in the previous subsection, one can derive another ``Mellin-Barnes-type'' representation of the Feynman integral $I_G$, which now describe in the present section.
As in the previous sections, we
will assume in this subsection that all points $X_i \in S^D$ are in the Euclidean section of the complexified
deSitter space, meaning that $1-Z_{ij} \in (0,\infty)$ for all $i,j=1,\dots, E$.
The Mellin-Barnes representation will allow us to obtain the analytic continuation
to $dS_D$, as we explain below in subsec.~\ref{sec:anal}, and it will also
allow us to derive the exponential decay of the correlators below in sec.~\ref{sec:nohair}.

To state our result, we recall that the monomials in $\U$ resp. $\F$ were labeled by forests
$F$ from $\T_{E+1}$ resp. $\T_{E}$. Each such forest is a union of $E+1$ resp. $E$ trees
within a graph $G^*$ that is canonically associated with $G$, see the previous subsection for the
precise definition. There is one distinguished trivial forest in this collection, called
$F=\Phi$, given by
\ben\label{Phidef}
\Phi = \{ ((E+1)*), \dots, ((E+V)*) \} \in \T_{E+1} \  ,
\een
Our Mellin-Barnes formula will involve an integral over multiple paths in the complex
plane over variables $w_F$. There is one such variable for each forest $F$, except for
the forest $F = \Phi$. Furthermore, it involves the functions $H_n$ defined by the
multiple integral
\bena\label{Hndef}
&&H_n(z) = 2^{-2z} \ \left( \prod_{l=1}^{n-1} \int_{-i\infty}^{i\infty} \frac{ds_l}{2\pi i} \right) \\
&& \times \
\Gamma(\tfrac{D-1}{2}+i\rho+z-\sum_{l=1}^{n-1} s_l)\Gamma(\tfrac{D-1}{2}-i\rho+z-\sum_{l=1}^{n-1} s_l) \Gamma(-z+\sum_{l=1}^{n-1} s_l)
\Gamma(-z+\sum_{l=1}^{n-1} s_l-\tfrac{D-2}{2}) \non\\
&&
\times \ \prod_{l=1}^{n-1} \Gamma(\tfrac{D-1}{2}+i\rho+s_l)\Gamma(\tfrac{D-1}{2}-i\rho+s_l) \Gamma(-s_l)
\Gamma(-s_l-\tfrac{D-2}{2}) \ . \non
\eena
The $s_l$-integrations in $H_n(z)$ are along paths parallel to the imaginary axis in the strip~\eqref{zcontour}. $H_n(z)$ is analytic in the strip $-\frac{n(D-1)}{2} < \R(z) < -\frac{n(D-2)}{2}$.

\medskip
\noindent
The somewhat lengthy proof of the following result is given in appendix~\ref{app:d}:

\begin{thm}\label{mbthm}
The Feynman integral $I_G(X_1, \dots, X_E)$ for non-coinciding
Euclidean configurations of external points $X_1, \dots, X_E \in S^D$ is given by
\ben\label{ig4}
\boxed{
\\
 I_G =  \ K_G \ \int_{\vec w}
\Gamma_G ( \ \vec w \ ) \
\prod_{1 \le r \neq s \le E} ( 1-Z_{rs} )^{\sum_F w_F} \  ,
\\
}
\een
where the sum $\sum_F$ in the exponent is over all forests in $\T_E(r,s)$, i.e. those connecting the vertices $X_r$
and $X_s$, see the following picture for an example.
$K_G$ is the numerical constant in eq.~\eqref{kgdef}, and $Z_{rs}$ are the point-pair invariants $H^2 X_r \cdot X_s$. Furthermore, $\Gamma_G(\vec w)$ is the meromorphic kernel
\ben\label{gammag}
\boxed{
\\
\Gamma_G (\vec w ) := \frac{
\Gamma(\frac{D+1}{2} + \sum_F w_F) \ \prod_F \ \Gamma(-w_F) \ \prod_{(ij) \notin \Phi} \ H_{n_{ij}}(\sum_{F \owns (ij)} w_F)
}
{
\Gamma(\frac{D+1}{2} + \sum_{F \in \T_E} w_F) \ \prod_{(ij) \in \Phi}  \Gamma(\frac{D+1}{2} + \sum_{F /\!\!\!\!\!\owns (ij)} w_F) \ \prod_{(ij) \notin \Phi} \Gamma(-\sum_{F \owns (ij)} w_F)}
\\
}
\een
with $n_{ij} \ge 1$ the number of lines between a pair of vertices in $G$, and
with the forest $\Phi$  omitted in any sum/product over forests $F$.  The integration $\int_{\vec w} =
\prod_{F \neq \Phi} \int \frac{dw_F}{2\pi i}$ is over paths
which are asymptotically parallel
imaginary axis which:
\begin{itemize}
\item
leave the poles of $\Gamma(-w_F) < 0$ for $F \neq \Phi$ to the left,
\item
leave any poles of $\Gamma(\tfrac{D+1}{2} + \sum_{F \neq \Phi} w_F)$ to the right,
\item
leave the
$-\mn_0 \pm i\rho-\frac{n_{ij}(D-1)}{2}$ series of poles of $H_{n_{ij}}(\sum_{F \owns (ij)} w_F)$ to the left
 and the $+\mn_0-\frac{n_{ij}(D-2)}{2}$ series of poles to the right.
\end{itemize}
\end{thm}
Eq.~\eqref{ig4} is one of the key results of this paper. It expresses the Feynman integral $I_G$ as a multiple Mellin-integral.
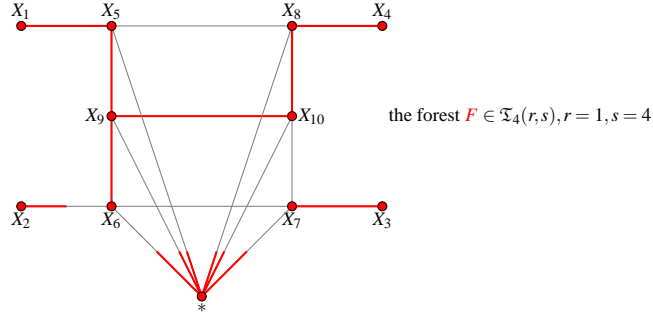
\begin{center}
\begin{tikzpicture}[scale=.60, transform shape]
\draw[red,thick]  (-4,4) node[black,above]{$X_1$} --  (-2,4) node[black,above]{$X_5$};
\draw[gray]  (-4,0) node[black,below]{$X_2$}--  (-2,0) node[black,below]{$X_6$};
\draw[red,thick]  (2,4) node[black,above]{$X_8$} --  (4,4) node[black,above]{$X_4$};
\draw[gray]  (2,0) node[black,below]{$X_7$} --  (4,0) node[black,below]{$X_3$};
\draw[red,thick] (-2,4) -- (-2,0);
\draw[red,thick] (2,4) -- (2,2);
\draw[gray] (2,2) -- (2,0);
\draw[gray] (-2,0) -- (2,0);
\draw[red,thick] (-2,2) -- (2,2);
\draw[gray] (-2,4) -- (2,4);
\draw[gray] (0,-2) -- (2,0);
\draw (-2,2) node[black,left]{$X_9$};
\draw (2,2) node[black,right]{$X_{10}$};
\draw (4,2) node[black,right]{the forest \textcolor{red}{$F$} $\in \T_4(r,s), r=1,s=4$};
\draw[gray] (-2,4) -- (0,-2);
\draw[gray] (-2,2) -- (0,-2);
\draw[gray] (-2,0) -- (0,-2);
\draw[gray] (2,4) -- (0,-2);
\draw[gray] (2,2) -- (0,-2);
\draw[red,thick] (1,-1) -- (0,-2);
\draw (0,-2) node[black,below]{$*$};
\draw[red,thick] (0,-2) -- (-1,-1);
\draw[red,thick] (0,-2) -- (-0.5,-1);
\draw[red,thick] (0,-2) -- (-0.33,-1);
\draw[red,thick] (0,-2) -- (0.33,-1);
\draw[red,thick] (0,-2) -- (0.5,-1);
\draw[red,thick] (-4,0) -- (-3,0);
\draw[red,thick] (4,0) -- (2,0);
\draw (-4,0) node[draw,shape=circle,scale=0.5,fill=red]{};
\draw (-2,0) node[draw,shape=circle,scale=0.5,fill=red]{};
\draw (-2,2) node[draw,shape=circle,scale=0.5,fill=red]{};
\draw (-2,4) node[draw,shape=circle,scale=0.5,fill=red]{};
\draw (-4,4) node[draw,shape=circle,scale=0.5,fill=red]{};
\draw (2,2) node[draw,shape=circle,scale=0.5,fill=red]{};
\draw (2,4) node[draw,shape=circle,scale=0.5,fill=red]{};
\draw (4,4) node[draw,shape=circle,scale=0.5,fill=red]{};
\draw (4,0) node[draw,shape=circle,scale=0.5,fill=red]{};
\draw (0,-2) node[draw,shape=circle,scale=0.5,fill=red]{};
\draw (2,0) node[draw,shape=circle,scale=0.5,fill=red]{};
\end{tikzpicture}
\end{center}
The Feynman integral $I_G$ is of the form of a so-called ``(generalized) $H$-function'' (see appendix~\ref{app:A}), and the conditions for {\em absolute convergence} of the contour integrals over $\vec w$ stated there are fulfilled in $I_G$, as we also show in appendix~\ref{app:d}. This implies in particular, as we will discuss in more detail in subsec.~\ref{sec:anal}, that $I_G(X_1, \dots, X_E)$ defines an analytic function (with cuts) in a subset of configurations in the complex deSitter spacetime.

\subsection{Example}

We will now illustrate the formula \eqref{ig4} for the deSitter space Feynman integral $I_G$
 in the case of the simple graph $G$ with $E$ external lines
and $V=1$ given by the following picture for $E=7$:
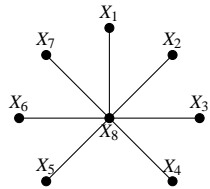
\begin{center}
\begin{tikzpicture}[scale=.6, transform shape]
\draw (0,0) node[draw,shape=circle,scale=0.5,fill=black]{};
 \draw (0,2) node[draw,shape=circle,scale=0.5,fill=black]{};
 \draw (1.4,1.4) node[draw,shape=circle,scale=0.5,fill=black]{};
 \draw (2,0) node[draw,shape=circle,scale=0.5,fill=black]{};
 \draw (1.4,-1.4) node[draw,shape=circle,scale=0.5,fill=black]{};
  \draw (-1.4,-1.4) node[draw,shape=circle,scale=0.5,fill=black]{};
  \draw (-2,0) node[draw,shape=circle,scale=0.5,fill=black]{};
   \draw (-1.4,1.4) node[draw,shape=circle,scale=0.5,fill=black]{};
 \draw[black]  (0,0) node[black,below]{$X_8$} --  (0,2) node[black,above]{$X_1$};
 \draw[black]  (0,0) --  (1.4,1.4) node[black,above]{$X_2$};
 \draw[black]  (0,0) --  (2,0) node[black,above]{$X_3$};
 \draw[black]  (0,0) --  (1.4,-1.4) node[black,above]{$X_4$};
 \draw[black]  (0,0) --  (-1.4,-1.4) node[black,above]{$X_5$};
 \draw[black]  (0,0) --  (-2,0) node[black,above]{$X_6$};
 \draw[black]  (0,0) --  (-1.4,1.4) node[black,above]{$X_7$};
\end{tikzpicture}
\end{center}
The integration variable is $X_{E+1}$, which is $X_8$ in the above example. For $E=4$, the corresponding Feynman integral $I_G$ gives the
first order perturbative correction to the 4-point function $\langle \phi(X_1) \dots \phi(X_4) \rangle_{0,\lambda}$.
As always, the corresponding graph $G^*$ has one more vertex $*$ and is shown in the following picture:
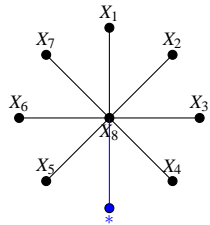
\begin{center}
\begin{tikzpicture}[scale=.6, transform shape]
\draw (0,0) node[draw,shape=circle,scale=0.5,fill=black]{};
 \draw (0,2) node[draw,shape=circle,scale=0.5,fill=black]{};
 \draw (1.4,1.4) node[draw,shape=circle,scale=0.5,fill=black]{};
 \draw (2,0) node[draw,shape=circle,scale=0.5,fill=black]{};
 \draw (1.4,-1.4) node[draw,shape=circle,scale=0.5,fill=black]{};
 \draw (0,-2) node[draw,shape=circle,scale=0.5,fill=blue]{};
  \draw (-1.4,-1.4) node[draw,shape=circle,scale=0.5,fill=black]{};
  \draw (-2,0) node[draw,shape=circle,scale=0.5,fill=black]{};
   \draw (-1.4,1.4) node[draw,shape=circle,scale=0.5,fill=black]{};
 \draw[black]  (0,0) node[black,below]{$X_8$} --  (0,2) node[black,above]{$X_1$};
 \draw[blue]  (0,0) --  (0,-2) node[blue,below]{$*$};
 \draw[black]  (0,0) --  (1.4,1.4) node[black,above]{$X_2$};
 \draw[black]  (0,0) --  (2,0) node[black,above]{$X_3$};
 \draw[black]  (0,0) --  (1.4,-1.4) node[black,above]{$X_4$};
 \draw[black]  (0,0) --  (-1.4,-1.4) node[black,above]{$X_5$};
 \draw[black]  (0,0) --  (-2,0) node[black,above]{$X_6$};
 \draw[black]  (0,0) --  (-1.4,1.4) node[black,above]{$X_7$};
\end{tikzpicture}
\end{center}
The possible forests in the sets
$\T_E$ and $\T_{E+1}$ except for $\Phi = \{(*(E+1))\}$ are depicted in the following pictures.
There is one integration variable $w_{ij}, i,j
\in \{*,1,\dots,E\}$ for each of the forests $\{(i (E+1)), ((E+1) j) \}$ connecting vertex $i$ with $j$. Furthermore,
there is one integration variable $w_i, i \in \{1, \dots, E\}$ for each of the forests $\{(i (E+1))\}$ connecting vertex $E+1$ with
with $i$. The collection of these integration variables is called $\vec w$ as in the previous subsection.
\begin{center}
\begin{tikzpicture}[scale=.6, transform shape]
 \draw (0,2) node[draw,shape=circle,scale=0.5,fill=red]{};
 \draw (1.4,1.4) node[draw,shape=circle,scale=0.5,fill=red]{};
 \draw (2,0) node[draw,shape=circle,scale=0.5,fill=red]{};
 \draw (1.4,-1.4) node[draw,shape=circle,scale=0.5,fill=red]{};
 \draw (0,-2) node[draw,shape=circle,scale=0.5,fill=red]{};
  \draw (-1.4,-1.4) node[draw,shape=circle,scale=0.5,fill=red]{};
  \draw (-2,0) node[draw,shape=circle,scale=0.5,fill=red]{};
   \draw (-1.4,1.4) node[draw,shape=circle,scale=0.5,fill=red]{};
 \draw[gray]  (0,0) --  (0,2) node[black,above]{$X_1$};
 \draw[gray]  (0,0) --  (0,-2) node[black,below]{$*$};
 \draw[red]  (0,0) --  (1.4,1.4) node[red,above]{$X_2$};
 \draw[gray]  (0,0) --  (2,0) node[black,above]{$X_3$};
 \draw[gray]  (0,0) --  (1.4,-1.4) node[black,above]{$X_4$};
 \draw[gray]  (0,0) --  (-1.4,-1.4) node[black,above]{$X_5$};
 \draw[red]  (0,0) --  (-2,0) node[red,above]{$X_6$};
 \draw[gray]  (0,0) --  (-1.4,1.4) node[black,above]{$X_7$};
 \draw[red]  (0,1) --  (0,2) node[black,above]{$X_1$};
 \draw[red]  (0,-1) --  (0,-2) node[black,below]{$*$};
 \draw[red]  (1,0) --  (2,0) node[black,above]{$X_3$};
 \draw[red]  (0.7,-0.7) --  (1.4,-1.4) node[black,above]{$X_4$};
 \draw[red]  (-0.7,-0.7) --  (-1.4,-1.4) node[black,above]{$X_5$};
 \draw[red]  (-0.7,0.7) --  (-1.4,1.4) node[black,above]{$X_7$};
 \draw (0,0) node[draw,shape=circle,scale=0.5,fill=red]{};
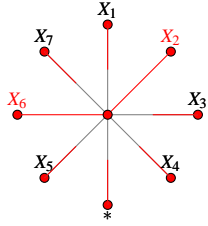
 \draw (-3,-3) node[black,right]{A forest in $\T_7(2,6)$, corresponding to $w_{26}$.};
\draw (10,2) node[draw,shape=circle,scale=0.5,fill=red]{};
 \draw (11.4,1.4) node[draw,shape=circle,scale=0.5,fill=red]{};
 \draw (12,0) node[draw,shape=circle,scale=0.5,fill=red]{};
 \draw (11.4,-1.4) node[draw,shape=circle,scale=0.5,fill=red]{};
 \draw (10,-2) node[draw,shape=circle,scale=0.5,fill=red]{};
  \draw (8.6,-1.4) node[draw,shape=circle,scale=0.5,fill=red]{};
  \draw (8,0) node[draw,shape=circle,scale=0.5,fill=red]{};
   \draw (8.6,1.4) node[draw,shape=circle,scale=0.5,fill=red]{};
 \draw[gray]  (10,0) --  (10,2) node[black,above]{$X_1$};
 \draw[gray]  (10,0) --  (10,-2) node[black,below]{$*$};
 \draw[red]  (10,0) --  (11.4,1.4) node[red,above]{$X_2$};
 \draw[gray]  (10,0) --  (12,0) node[black,above]{$X_3$};
 \draw[gray]  (10,0) --  (11.4,-1.4) node[black,above]{$X_4$};
 \draw[gray]  (10,0) --  (8.6,-1.4) node[black,above]{$X_5$};
 \draw[gray]  (10,0) --  (8,0) node[black,above]{$X_6$};
 \draw[gray]  (10,0) --  (8.6,1.4) node[black,above]{$X_7$};
 \draw[red]  (10,1) --  (10,2) node[black,above]{$X_1$};
 \draw[red]  (10,-1) --  (10,-2) node[black,below]{$*$};
 \draw[red]  (11,0) --  (12,0) node[black,above]{$X_3$};
 \draw[red]  (10.7,-0.7) --  (11.4,-1.4) node[black,above]{$X_4$};
 \draw[red]  (9.3,-0.7) --  (8.6,-1.4) node[black,above]{$X_5$};
 \draw[red]  (9,0) --  (8,0);
 \draw[red]  (9.3,0.7) --  (8.6,1.4) node[black,above]{$X_7$};
 \draw (10,0) node[draw,shape=circle,scale=0.5,fill=red]{};
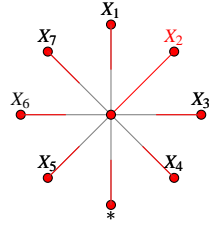
 \draw (7,-3) node[black,right]{A forest in $\T_8$, corresponding to $w_{2}$.};
\end{tikzpicture}
\end{center}
The kernel $\Gamma_G$ in eq.~\eqref{gammag} becomes
\bena
&&\Gamma_G(\vec w) = \frac{
\Gamma(\frac{D+1}{2}+ \sum_k z_k - \sum_{k\neq l} w_{kl}) \ \prod_{k<l} \Gamma(-w_{kl})
\ \prod_k \Gamma(-w_k) \Gamma(-z_{k}+w_k+\sum_{l\neq k} w_{kl})
}{
\Gamma(\frac{D+1}{2} + \sum_{k<l} w_{kl} + \sum_k w_k) \Gamma(\frac{D+1}{2} + \sum_k z_k -
\sum_k w_{k} - \sum_{l\neq k} w_{kl} )
}
\non\\
&&\hspace{3cm} \prod_k 2^{-2z_k} \Gamma(\tfrac{D-1}{2} + i\rho + z_k )
\Gamma(\tfrac{D-1}{2} - i\rho + z_k )
\Gamma(-\tfrac{D-2}{2} - z_k )
\eena
for this simple graph, where $z_k = w_k + w_{k*} + \sum_{l \neq k} w_{kl}$.
All summations/products are between $1,...,E$.
The resulting integral~\eqref{ig4} can be simplified somewhat in this simple example, by changing
variables to $z_k$, and making repeated use of the formula\footnote{This is a version
of the $_1 {}F_2$-summation formula [cf. eq.~\eqref{f12gauss}], written in Mellin-Barnes form.}
\ben
\int_{-i\infty}^{i\infty} \frac{d\nu}{2\pi i} \frac{\Gamma(b+\nu) \Gamma(d-\nu)}{\Gamma(c+\nu) \Gamma(e-\nu)}
= \frac{\Gamma(c-b-d+e-1) \Gamma(b+d)}{\Gamma(e-d) \Gamma(c+e-1) \Gamma(c-b)}
\een
valid for $\R(c+e-b-d-1) > 0$ to carry out the $w_k$ integrations, using also the
duplication formula $\Gamma(x) \Gamma(x+1/2) = 2^{-2x+1} \sqrt{\pi} \Gamma(2x)$. We do not
give this somewhat lengthy calculation here but only record the final result, which is:
\bena\label{example}
I_G &=&  \frac{2^{D} K_G}{\sqrt{\pi}} \prod_l \int_{z_l} \ \Gamma((D-1)/2+i\rho+z_l)\Gamma((D-1)/2-i\rho+z_l)
\Gamma(-z_l-(D-2)/2) \non\\
&& \vspace{1cm} \times \ \prod_{k<l} \int_{w_{kl}} \ \frac{\prod_{k<l} \Gamma(-w_{kl}) \ \prod_l \Gamma(-z_l + \sum_{k \neq l} w_{kl}) \ \Gamma(D/2+\sum_l z_l-\sum_{k \neq l} w_{kl})}{
\Gamma(D+\sum_l z_l)
} \non\\
&& \vspace{1cm} \times \  \prod_{k<l} \left( \frac{1-Z_{kl}}{2} \right)^{w_{kl}} \ .
\eena
 In this expressions, all sums/products are again between $1, ..., E$, and
the $z_l$-integration contours are parallel to the imaginary axis in the strip~\eqref{zcontour}, and the
$w_{kl}$-integration contours are parallel to the imaginary axis with $\R(w_{kl}) < 0$.
The constant $K_G$ is as defined in eq.~\eqref{kgdef}, with $V=1$ there. Our formula agrees with that found by~\cite{marolf2}
using another method.

\subsection{Analytic continuation}\label{sec:anal}

So far, we have focussed on the Feynman integrals $I_G$ for non-coinciding ($X_i \neq X_j$)
Euclidean configurations $(X_i \in S^D)$ of points, i.e.
ones such that all point-pair invariants $Z_{ij} \in [-1,1)$. To obtain the correlation functions $\langle
\phi(X_1) \dots \phi(X_E) \rangle_{\lambda,0}$ as distributions for deSitter configurations $(X_i \in dS_D)$,
we must, as already indicated above in eq.~\eqref{ac}, analytically continue through the complex
deSitter configurations $(X_i \in dS_D^\mc)$. We will outline this in the present subsection.

The free field 2-point function (i.e. $\lambda = 0$) was seen above in sec.~\ref{sec2} to be analytic on at least the
domain $T_2 = \{(X_1,X_2) \in (dS_D^\mc)^2 \mid \I(X_1-X_2) \in V^+ \}$. This domain by itself does not
contain real deSitter configurations $(X_1, X_2) \in (dS_D)^2$; those lie on the boundary of this domain. In
fact, the free 2-point function is defined as the distributional boundary value as we approach this boundary, see
item~2) of sec~\ref{sec2}, and footnote~\ref{foot1}. The full domain of analyticity of the free field 2-point
function is larger than $T_2$, and can be obtained e.g. by acting on $T_2$ with complex
deSitter isometries from $O(D,1)^\mc$. Alternatively, we may observe that the two-point function is, where it is analytic,
a function only of $Z$. As $(X_1, X_2)$ varies through $T_2$, $Z$ varies through the cut complex plane
$\mc \setminus [1,\infty)$. The two-point function is analytic for any configuration such that $Z$ is
in this cut complex plane, and this includes many real deSitter configurations, such as when
$(X_1,X_2) \in (dS_D)^2$ and spacelike. One sees more generally that, using Wick's theorem, the
free-field $E$-point functions are analytic at least in the domain
\ben
T_E := \{ (X_1, \dots, X_E) \in (dS_D^\mc)^E \mid  \I(X_j-X_{j+1}) \in V^+ \} \ .
\een
Bros and Moschella (see ref.~1 of~\cite{bros}) suggested that this should be contained in the domain of analyticity
also for interacting quantum field theories. Again, real deSitter configurations $(X_1, \dots,X_E) \in (dS_D)^E$
are only on the boundary of $T_E$, and the $E$-point correlation functions on real deSitter are defined as distributional
boundary values~\cite{bros}. They also showed how the domain $T_E$ can then be enlarged using
complex deSitter isometries, the ``edge-of-the-wedge-theorem'' etc., after which it also contains many real configurations.
We will now consider whether $T_E$ is indeed a domain of analyticity for the correlators in the interacting theory, in the
context of perturbation theory. Since the $E$-point functions $\langle \phi(X_1) \dots \phi(X_E) \rangle_{0,\lambda}$
are given in terms of the Feynman integrals $I_G$, we must analyze those.

For a configuration of $E$ points $(X_1, \dots, X_E) \in T_E$, the point pair invariants are all in the cut complex plane
\ben
Z_{rs} \in \mc \setminus [1,\infty) \quad \text{for $r \neq s$.}
\een
Indeed, letting e.g. $r<s$, we can write
\ben
\I( X_r - X_s ) = \sum_{j=r}^{s-1} \underbrace{\I(X_j - X_{j+1})}_{\in V^+} \in V^+ \ ,
\een
and by looking at the cases $\R(X_r-X_s) \in \bar V^+  \cup \bar V^-, \notin \bar V^+ \cup \bar V^-$ separately,
we get from this $H^2 (X_r-X_s)^2 \notin [-1, -\infty)$,
so $Z_{rs} = H^2 X_r \cdot X_s \notin [1, \infty)$.
Even though there are many dependencies among the $Z_{rs}$, the Feynman integrals $I_G$ can be viewed as functions of the $Z_{rs}, r \neq s$, considered as independent variables, which are defined when $Z_{rs} \in [-1,1)$. We must ask whether it can be analytically extended to $Z_{rs}$ in the cut complex plane, for then $I_G$, viewed as depending on $X_1, \dots, X_E$, is also defined as an analytic function on $T_E$. For this, we can look at the explicit form of $I_G$ as functions of the $Z_{rs}$, viewed as {\em independent} variables. The most useful representation of $I_G$ for this purpose is~\eqref{ig4}.

That formula gives $I_G$ as a $H$-function, whose
analyticity domain is described in appendix~\ref{app:A}. According to eq.~\eqref{Deltadef} there, we must calculate
the real numbers $\Delta_F$ associated with each integration variable $w_F$ in eq.~\eqref{ig4},
in terms of which the domain of analyticity is then (at least)
\ben
|{\rm arg}(1-Z_{rs})| < \frac{1}{2} \pi \inf\{\Delta_F \mid
F \in \T_E(r,s) \} \ .
\een
Looking at the explicit form of the kernel $\Gamma_G$ in eq.~\eqref{gammag}, we derive in appendix~\ref{app:d}, step 7), that $\Delta_F \ge 2$ for any forest $F$, from which it then follows
that $I_G$ is indeed analytic in the $Z_{rs}$ in the cut complex plane. When viewed as a function
$I_G(X_1, \dots, X_E)$ of the external complex deSitter configuration, $I_G$ is hence analytic at least on $T_E \subset
(dS_D^\mc)^E$, but even for all complex deSitter configurations such that the $Z_{rs}$ are all
in the cut complex plane. This set contains many real deSitter configurations in
$(dS_D^\mr)^E$, e.g. if all points are mutually spacelike. To define $I_G$ on the entire space
of real deSitter configurations, it must be understood as a distributional boundary value.

In order to define this distributional boundary value, one can again make use of the representation~\eqref{ig4} together with a general result in distribution
theory, see e.g. ch.~IX of~\cite{hor}. This general result states that if $u(z)$ is an analytic function that is defined on an open subset of $\mc^n$ of the form $U + iC$, where $U \subset \mr^n$ is open and $C$ is an open convex cone in $\mr^n$,
and if $u(x + iy)$ does not increase too fast as $C \owns y \to 0$ in the sense that
\ben\label{condition*}
|u(x + iy)| \le {\rm cst.} \ |y|^{-N} \quad \text{for all $y \in C$, some $N$,}
\een
then $u$ possess a distributional boundary value for $C \owns y \to 0$. The same applies if $u$ is only defined and analytic for $y$ in a neighborhood of the tip in $C$. When $u$ is defined on a complex manifold, then the boundary
value is defined in each chart and then pieced together by a partition of unity. Now let $(X_1 +iY_1, \dots, X_E+iY_E) \in (dS_D^\mc)^E \cap T_E$ be a complex deSitter configuration, with $X_i, Y_i$ real. Then, $Y_r - Y_s \in V^+$
for $r<s$, and because
\ben
|\xi^2| \ge |(\I \xi)^2| \quad \text{if $\I \xi \in V^+$} \ ,
\een
we get $|1-Z_{rs}| \ge \frac{1}{2} |(Y_r - Y_s)^2|$. Because $I_G$ is a generalized $H$-function that is
defined by an absolutely convergent multiple contour integral, we can use this to estimate for $(X_1 + iY_1,
\dots, X_E + iY_E)$ in a bounded subset of $T_E$:
\bena\label{ineq}
 |I_G(X_1+iY_1, \dots, X_E+iY_E)|&\le& K_G \prod_{1 \le r \neq s \le E} |1-Z_{rs}|^{\inf \sum_F \R(w_F)} \int_{\vec w} |\Gamma_G(\vec w)| \\
&\le& {\rm cst.} \ \prod_{1 \le r < s \le E} |(Y_r-Y_s)^2|^{-N} \ , \non
\eena
where the $\Gamma_G$ is the integral kernel~\eqref{gammag}, and the sum is over all forests $F$ in $\T_E(r,s)$
connecting vertex $r$ with vertex $s$. In the second line we  have used the fact that the multiple contour
integral is absolutely convergent, and we have set $-N = \inf \sum_F \R(w_F)$.
Let $C_{E-1} \subset (\mr^{D+1})^{E-1}$ be a convex cone with small opening
angle around the vector $(e,\dots,e)$ with $e=(1,0,\dots,0) \in \mr^{D+1}$. Then, for
$(Y_1-Y_2, \dots, Y_{E-1}-Y_E) \in C_{E-1}$, we have
\ben
\prod_{1 \le r < s \le E} |(Y_r-Y_s)^2|^{-N} \le {\rm cst.} \ [|Y_1-Y_2|^2+\dots+|Y_{E-1}-Y_E|^2]^{-M}
\een
for some $M$, where $| \ . \ |$ denotes the Euclidean norm of a vector.
Combining this inequality with eq.~\eqref{ineq} then evidently gives
\ben
|I_G(X_1+iY_1, \dots, X_E+iY_E)| \le {\rm cst.} \ [|Y_1-Y_2|^2+\dots+|Y_{E-1}-Y_E|^2]^{-M}
\een
for a bounded set of $(X_1+iY_1, \dots, X_E+iY_E) \in (dS^\mc_D)^E$ such that
$(Y_1, Y_1-Y_2, \dots, Y_{E-1}-Y_E) \in \mr^{D+1} \times C_{E-1}$. Expressing this result in any coordinate system
$x+iy$ of $(dS_D^\mc)^{E}$ centered around $(X_1, \dots, X_E)$ then leads to the desired inequality~\eqref{condition*} for a suitable cone $C$ related to $\mr^{D+1} \times C_{E-1}$.

Thus, the upshot of our discussion is that, for real deSitter configurations, $I_G$ is the distributional boundary
value of an analytic function. The boundary value prescription given above can be summarized as saying that
[compare eq.~\eqref{ig4}]
\ben
 I_G(X_1, \dots, X_E) =  \ 2K_G \ \int_{\vec w}
\Gamma_G ( \vec w ) \
\prod_{1 \le r < s \le E} \left( \frac{(X_r - X_s +i(s-r)\epsilon \ e)^2}{2} \right)^{\sum_{F}  w_F}
\een
where the $X_i$ are now in the {\em real} deSitter space, and where $e = (1,0,\dots,0)$. As usual we mean here that $I_G$ has to be smeared with a testfunction $f(X_1, \dots, X_E)$ first for $\epsilon>0$, and then we send $\epsilon \to 0$. The sum over $F$ in the exponent runs as usual over all $E$-forests in the graph $G^*$ such that the external
leg associated with $X_r$ is connect to that with $X_s$ by a tree in the forest.

\section{Quantum no-hair/exponential clustering}\label{sec:nohair}

We will now argue that the connected correlation functions have exponential decay
in timelike directions, and a corresponding decay in spacelike directions. As above, we consider a principal
series scalar field $(c = -(D-1)/2 + i\rho, \rho \in \mr)$, and we restrict attention to
configurations $(X_1, \dots, X_E)$ in the real deSitter spacetime such that no point is on each other's
lightcone. For general configurations, we would have to take into the distributional nature of the correlation functions. This case could also be dealt with using our methods, but for simplicity
we will leave it aside.

Thus, we take one point, $X_r$, and move it away from the other points to infinity
either in timelike directions so that $Z_{rs} \to + \infty$ for any $s \neq r$,
or in spacelike directions so that $Z_{rs} \to -\infty$ for any $s \neq r$.\footnote{Note that,
since deSitter space has horizons, we could also let $X_r$ go to ${\mathscr I}^+$ in such a way
that $X_r$ does not become eventually time-like to any of the $X_s$ for $s \neq r$, and so in
general it need not be true that $Z_{rs} \to \infty$ for all $s$.} We will argue that
the correlation functions satisfy, to all orders in the coupling constant $\lambda$:
\ben\label{decay1}
\langle \phi(X_1) \cdots \phi(X_E) \rangle_{\lambda, 0}^C = O(|Z_{rs}|^{-\xi}) \quad
\text{as any $|Z_{rs}| \to \infty$,}
\een
for any $\xi < (D-1)/2$ and any $s = 1, \dots, \hat r, \dots, E$. Here, $Z_{rs}$ is the point pair invariant, which for timelike related points is given by proper time separation $\tau_{rs}$ between the two points $X_r, X_s$ as $Z_{rs} = \cosh H \tau_{rs} \sim \e^{H|\tau_{rs}|}$, and similarly for spacelike related points, see eq.~\eqref{ztau}.
For large spatial separation, e.g. in the
cosmological chart, see~\eqref{cosmc}, this implies:
\ben\label{decay2}
\langle \phi(t,{\bf x}_1) \cdots \phi(t,{\bf x}_E) \rangle_{\lambda, 0}^C =
O(|{\bf x}_r-{\bf x}_s|^{-2\xi}) \quad
\text{as any $|{\bf x}_r-{\bf x}_s| \to  \infty$}.
\een
For a scalar field in the complementary series, our decay result~\eqref{decay1}
would hold for $\xi$ with $\xi < (D-1)/2 - \sqrt{(D-1)^2/4-m^2/H^2}$,
and we would get a correspondingly weaker decay. Again, to keep things as simple as possible,
we will not consider this case here.

We demonstrate~\eqref{decay1} in the remainder of this section. Evidently, since the connected correlation function is
a sum of contributions from connected Feynman graphs $G$, it is sufficient to demonstrate the exponential decay for each such
contribution, i.e.
\ben\label{igdecay}
I_G = O(|Z_{rs}|^{-\xi}) \ ,
\een
for $|Z_{rs}| \to \infty$.
Our main tool is the Mellin-Barnes formula~\eqref{ig4} for $I_G$. The desired exponential
decay~\eqref{igdecay} will be obtained by looking at the location of the contours in this formula.
Indeed the Mellin-Barnes formula immediately gives for $X_r$ going to infinity in time- or spacelike
directions:
\ben
|I_G(X_1, \dots, X_G)| \le {\rm cst.} \prod_{s:s \neq r} |Z_{rs}|^{\sup \sum_F \R(w_F)} \int_{\vec w}
|\Gamma_G(\vec w)|
\le {\rm cst.} \prod_{s:s \neq r} |Z_{rs}|^{\sup \sum_F \R(w_F)}  \ ,
\een
where the sum is over all forests $F \in \T_E(r,s)$ in the graph $G^*$ consisting of $E$ trees, one of which
connects the vertex $X_s$ with $X_r$, see subsec.~\ref{app:c} for the precise definitions. The
contour integral is absolutely convergent, as discussed in sec.~\eqref{sec:anal}. The supremum
is taken along the integration paths followed by the variables $\vec w$, see subsec.~\ref{sec:mb}. These
contours can be somewhat complicated, so to get an estimate of the supremum, it is best deform the contours
followed by the $w_F$ for such forests to straight lines parallel to the imaginary axis. There are many
ways of doing this.

Let us first suppose, for ease of notation, that $E=2$ and $r=1$, so that $s=2$, and let us
also suppose, that a pair of vertices in $G$ is connected by at most one line [so that $n_{ij} = 1$
for all $i,j$ in eq.~\eqref{gammag}]. Take any forest $P \in \T_2(1,2), P \cap \Phi = \emptyset$
[cf.~eq.~\eqref{Phidef}] and move its path of integration to be at $\R(w_{P}) = cst.$ with
\ben
-\frac{D-1}{2} < \R(w_{P}) < -\frac{D-1}{2} + \epsilon \  \ .
\een
For each $(kl) \notin P, (kl) \notin \Phi$, choose a forest $F(kl) \in \T_{3}$ containing $(kl)$, but $F(kl) \cap P = \emptyset$. We deform the integration path for the corresponding variable $w_{F(kl)}$ so that it is the same as that for $w_P$. We deform all other $w_F$'s so as to run along paths with $-\epsilon<\R(w_F)<0$. The paths have been chosen such that
any pole at $\mn_0$ of $\Gamma(-w_F)$ remains to the right, such that the
$-\mn_0 \pm i\rho-(D-1)/2$ series of poles of each $H_{1}(\sum_{F \owns (kl)} w_F)$ remain to the left
 and the $+\mn_0-(D-2)/2$ series of poles remains to the right of the integration paths. Furthermore, 
 $\sup\{ \R(w_F) \mid F \in \T_2(1,2)\} \le -\xi$ for sufficiently small $\xi$, thus showing exponential 
 decay.
 
However, when deforming the contours, we will cross some of the poles of
$\Gamma(\frac{D+1}{2} + \sum_F w_F)$ to the left. To avoid this from happening, we choose another forest $Q \in \T_{3}$
 and we translate the contour followed by $w_Q$ to the right by the appropriate amount.
 The price that we pay is that now
 we cross some poles of $\Gamma(-w_Q)$, and we pick up corresponding residue.
 Each such pole term looks exactly like~\eqref{ig4}, but it
 has no integration over $w_Q$, which is set to some $n \in \mn_0$.
 There is one such term for each pole crossed. But since
all the variables $w_F$ appearing in the exponent of the factors $(1-Z_{rs})$ are associated with
$F$'s from $\T_2$, they are not affected by this, as $Q \in \T_3$. As a consequence,
the pole terms can be estimated in the
same way as before. The general case is dealt with in exactly the same way; without loss of generality
we may again assume that $r=1, s=2$. The forests $F(kl), Q$ are now in $\T_{E+1}$.  When
some $n_{ij} > 1$, a similar argument can be made.

\section{Conclusions and outlook}

In this paper, we have:

\begin{enumerate}
\item
Given parametric formulae for Feynman integrals for massive scalar fields
in deSitter spacetime for an arbitrary graph $G$, see eq.~\eqref{ig4}.
These integrals are the building blocks of the perturbative expansion of deSitter correlation
functions (in the Bunch Davis state $=$ Hartle Hawking state $=$ Euclidean vacuum state). 
The parametric integrals involve a multiple
 contour integral over parameters which are in correspondence with
 tree graphs, and more generally forests, within the graph $G$. An alternative parametric representation
 in terms of certain graph polynomials associated with $G$ was also given in eq.~\eqref{ig1}.
 Our parametric representations have certain features in common with similar expression known in
 Minkowski quantum field theory. However, they are different at least in that (a) we work in position space and (b)
 our forests and effective graphs associated with $G$ involve a ``virtual vertex'', $*$, that is, at some level, a reflection
 of the fact that the loop integrals are restricted to the deSitter hyperboloid. We were not able
 to find a ``momentum space version'' of our formulae, mainly due to the fact that no useful and simple momentum space expression for the
Feynman propagator appears to be known in deSitter.

\item We have found that the Feynman integral belong to a class of special functions called
``generalized $H$-functions''~\cite{panda}, which are a generalization of the hypergeometric function.

\item
We have discussed the renormalization and convergence of our parametric form of the Feynman integrals.

\item
We have discussed the analytic properties of the Feynman integrals/correlators, in particular how
to go from Euclidean to Lorentizan configurations via analytic continuation.

\item
We have used the parametric representation to show the exponential decay in timelike directions of
the deSitter correlation functions, to arbitrary orders in perturbation theory. As we have explained, these
results show that the correlation functions of essentially {\em any} state will approach that of the vacuum
at late times, at an exponential rate. This statement may be viewed as a quantum analog of the cosmic no
hair theorem, for a test scalar field with a positive mass.
\end{enumerate}

It would be very interesting to try to generalize our results in the following directions:

\begin{enumerate}
\item
Look at exponential decay for other types of fields, in particular to massless fields such as
(i) massless scalar/spin-$\frac{1}{2}$-fields (ii) Yang-Mills-fields (iii) the graviton. For
the free spin-$\frac{1}{2}$-propagators, see~\cite{camporesi}, for a general
discussion of the renormalization theory of the Yang-Mills-field on general curved spacetimes
see~\cite{hollands2}, for free Yang-Mills- and ghost- in deSitter see~\cite{higuchi2}, and for
free graviton propagators see e.g.~\cite{woodard,higuchi,higuchi2}.

\item
Look at the decay of correlation functions in the Schwarzschild-deSitter
spacetime, depicted in the following conformal diagram:
    
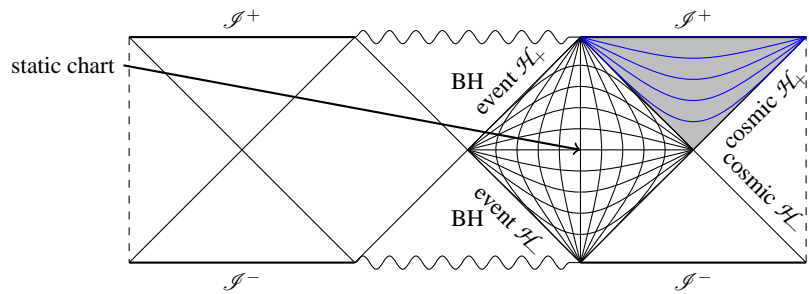
\begin{center}
\begin{tikzpicture}[scale=.75, transform shape]
\filldraw[fill=gray!50] (2,2) -- (6,2) -- (4,0);
\draw[black,thick] (-6,-2) -- node[black,below]{${\mathscr I}^-$}  (-2,-2);
\draw[black,thick] (-6,2) -- node[black,above]{${\mathscr I}^+$}  (-2,2);
\draw[black,thick] (2,-2) -- node[black,below]{${\mathscr I}^-$}  (6,-2);
\draw[black,thick] (2,2) -- node[black,above]{${\mathscr I}^+$}  (6,2);
\draw (-6,2) -- (-4,0) -- (-6,-2);
\draw[dashed] (-6,2) -- (-6,-2);
\draw (0,0) -- (-2,2) -- (-4,0) -- (-2,-2) --   (0,0);
\draw (0,0) -- node[black,above, sloped]{event ${\mathcal H}_+$}(2,2) -- (4,0) -- (2,-2) -- node[black,below, sloped]{event ${\mathcal H}_-$}(0,0);
\draw (-2,2) decorate[decoration=snake] {--(2,2)};
\draw (0,1.5) node[black,below]{BH};
\draw (0,-1.5) node[black,above]{BH};
\draw (-2,-2)  decorate[decoration=snake] {-- (2,-2)};
\draw (4,0) --node[black,below, sloped]{cosmic ${\mathcal H}_+$}(6,2);
\draw (4,0) --node[black,above, sloped]{cosmic ${\mathcal H}_-$}(6,-2);
\draw[dashed] (6,-2)  -- (6,2);
\draw (0,0) .. controls (2,2) and (2,2) .. (4,0);
\draw (0,0) .. controls (2,1.5) and (2,1.5) .. (4,0);
\draw (0,0) .. controls (2,1) and (2,1) .. (4,0);
\draw (0,0) .. controls (2,0.5) and (2,0.5) .. (4,0);
\draw (0,0) .. controls (2,0) and (2,0) .. (4,0);
\draw (0,0) .. controls (2,-0.5) and (2,-0.5) .. (4,0);
\draw (0,0) .. controls (2,-1) and (2,-1) .. (4,0);
\draw (0,0) .. controls (2,-1.5) and (2,-1.5) .. (4,0);
\draw (0,0) .. controls (2,-2) and (2,-2) .. (4,0);
\draw[->, thick] (-6.1,1.5) node[black,left] {static chart} -- (2,0);
\draw (2,-2) .. controls (1.5,0) and (1.5,0) .. (2,2);
\draw (2,-2) .. controls (1,0) and (1,0) .. (2,2);
\draw (2,-2) .. controls (.5,0) and (.5,0) .. (2,2);
\draw (2,-2) .. controls (0,0) and (0,0) .. (2,2);
\draw (2,-2) .. controls (2.5,0) and (2.5,0) .. (2,2);
\draw (2,-2) .. controls (3,0) and (3,0) .. (2,2);
\draw (2,-2) .. controls (3.5,0) and (3.5,0) .. (2,2);
\draw (2,-2) .. controls (4,0) and (4,0) .. (2,2);
\draw (2,-2) -- (2,2);
\draw[blue] (2,2) .. controls (4,2) and (4,2) .. (6,2);
\draw[blue] (2,2) .. controls (4,1.5) and (4,1.5) .. (6,2);
\draw[blue] (2,2) .. controls (4,1) and (4,1) .. (6,2);
\draw[blue] (2,2) .. controls (4,0.5) and (4,0.5) .. (6,2);
\draw[blue] (2,2) .. controls (4,0) and (4,0) .. (6,2);
\end{tikzpicture}
\end{center}
\ben
ds^2 = -f \ d\eta^2 +  f^{-1} \ dr^2 + r^2 d\omega_{D-2}^2 \ , \quad f = 1-\frac{r_0^{D-3}}{r^{D-3}}-H^2 r^2 \ .
\een
In the gray region, one would expect the decay of the fields/correlators to be essentially the same as in
pure deSitter spacetime. But in the static region, the influence of the both the black hole and cosmological
horizon will be felt, and
the likely outcome is less clear. A decay of the correlation functions in time-like directions ($|\eta_i-\eta_j| \to \infty$) in this region
would indicate a kind of quantum stability of the black hole, at least in the test-field approximation. Unfortunately, a systematic study along the lines of methods of this paper seems not possible,
because even the propagators of the free field theory in this background are not known in sufficiently explicit form. However, it might be possible to extract enough information about their asymptotic
 behavior  e.g. from known properties of quasi-normal modes in Schwarzschild-deSitter~\cite{cardoso}.
\end{enumerate}

\vspace{2cm}

{\bf Acknowledgements:} It is a pleasure to thank J.~Bros, H.~Epstein, A.~Ishibashi, Ch.~Jaekel, H. Kodama, D.~Marolf, I.~Morrison, and U.~Moschella
for stimulating discussions. I am very grateful to D.~Marolf and I.~Morrison for showing me their forthcoming paper~\cite{marolf2} (which is released simultaneously) before publication. I would also like to thank the Erwin-Schrodinger Institute, Vienna, for its hospitality and financial support in March-June 2010, and the cosmo-physics group at KEK (Japan) for its hospitality and financial support in January 2010. A considerable part of this work was completed during these visits.

\appendix

\section{Hypergeometric functions and their generalizations}\label{app:A}

In this section we give formulas for the Gauss hypergeometric function and
the free propagator referred to in the main text.
The hypergeometric function is defined as
\ben
\hF(a,b;c;z) = \sum_{n \ge 0} \frac{(a)_n(b)_n}{n! (c)_n} z^n
\een
where $(a)_n = \Gamma(a+n)/\Gamma(a)$. The series converges and
hence defines an analytic function in the open unit disk in the complex
$z$-plane. A value used in this paper is
\ben\label{f12gauss}
\hF(a,b;c,-1) = \frac{\Gamma(c-a-b)\Gamma(a)\Gamma(b)}{\Gamma(c-a) \Gamma(c-b)} \ .
\een
A transformation formula used in this paper is, for $m=1,2,\dots$
\bena
&&\hF(a,b;a+b-m;z) = \frac{\Gamma(m) \Gamma(a+b-m)}{\Gamma(a)\Gamma(b)} (1-z)^{-m}
\sum_{n=0}^{m-1} \frac{(a-m)_n (b-m)_n}{n! (1-m)_n} (1-z)^n \non\\
&& \ \ -(-1)^m \frac{\Gamma(a+b-m)}{\Gamma(a-m) \Gamma(b-m)} \sum_{n\ge 0}
\frac{(a)_n (b)_n}{n! (n+m)!} \non\\
&& \hspace{1cm} \times [\log (1-z) - \psi(n+1) - \psi(n+m+1) + \psi(a+n) + \psi(b+n)] (1-z)^n
\eena
where $\psi(z) = \Gamma'(z)/\Gamma(z)$. Application of this formula to the
vacuum 2-point function of the free Klein-Gordon field~\eqref{vac} gives for
even dimensions $D$:
\bena\label{hadarep}
&&\langle \phi(X_1) \phi(X_2) \rangle_0 =
\frac{H^{D-2}}{(4\pi)^{D/2})} \ \bigg(
\Gamma(D/2-1) \bigg[ \frac{H^2 (X_1-X_2)^2}{4} \bigg]^{-D/2+1} \\
&& \ \ \times \sum_{n=0}^{D/2-2} \frac{(-c-D/2+1)_n
(c+D/2)_n}{n!(2-D/2)_n} \bigg[ \frac{H^2 (X_1-X_2)^2}{4} \bigg]^{n} + \non\\
&& \ \ + \ \frac{(-1)^{D/2}}{\Gamma(-c-D/2+1)\Gamma(c+D/2)} \sum_{n \ge 0}
\frac{\Gamma(-c+n)\Gamma(c+D-1+n)}{n!\Gamma(D/2+n)} \ \bigg[ \frac{H^2 (X_1-X_2)^2}{4} \bigg]^{n} \non\\
&& \ \
\times [\log \frac{H^2 (X_1-X_2)^2}{4} - \psi(n+1) - \psi(n+D/2) + \psi(-c+n) + \psi(c+D-1+n)] \
\bigg) \ . \non
\eena
The following asymptotic formula is known for large values $|L|$ when $|{\rm arg}(Z-1)| < \pi$,
$|{\rm arg}(L)| < \pi$, see~\cite{jones}:
\bena\label{largeL}
&&\hF(a+L, b-L; c; (1-Z)/2) \sim 2^{(a+b-1)/2} \Gamma(c) \tau^{1/2} \frac{(\cosh \tau+1)^{(c-a-b)/2}}{(\cosh \tau-1)^{(c-1)/2}} \non\\
&& \ \ \times \{ \ \
I_{c-1}((L+(a-b)/2) \tau) + \non\\
&& \hspace{1cm}
+ I_{c-2}((L+(a-b)/2) \tau)[(c-1/2)(c-3/2)(\tau^{-1} - \cot \tau)] /(2L+a-b) + \non\\
&& \hspace{1cm} + (2c-a-b-1)(a+b-1) \tanh (\tau/2)
\ \ \}
\eena
where $\cosh \tau = Z$, and $I_\nu$ are the modified Bessel functions.
This asymptotic expansion is valid uniformly in $Z$ in the indicated domain.

There are many generalizations of the hypergeometric function. A very general class is
provided by the so-called (generalized) $H$-function of several variables, which is defined
by the Mellin-Barnes-type formula
\bena
&&H\left(
\begin{array}{cccc}
(A) & (B) & (a) & (b) \\
(C) & (D) & (c) & (d)
\end{array}
; z_1, \dots, z_n\right) = \left( \prod_{k=1}^n \int_{K_k} \frac{dw_k}{2\pi i} \right) \non\\
&& \frac{\prod_{j=1}^P \Gamma(a_j + \sum_i A_{ij} w_i) \ \prod_{j=1}^Q \Gamma(1-b_j - \sum_i B_{ij} w_i)}{
\prod_{j=1}^R \Gamma(c_j + \sum_i C_{ij} w_i) \ \prod_{j=1}^S \Gamma(1-d_j - \sum_i D_{ij} w_j)}
 z_1^{w_1} \cdots z_n^{w_n} \ .
\eena
The paths of integration $K_i$ are intended, if necessary, in such a way that all the poles of
$\Gamma(a_j + \sum_i A_{ij} w_i)$ are separated from the poles of $\Gamma(1-b_j - \sum_i B_{ij} w_i)$.
The parameters $A_{ij}, B_{ij}, C_{ij}, D_{ij} \in \mr$ are all assumed to be $\ge 0$, and $a_i, b_i, c_i, d_i \in \mc$.
A discussion of such kinds of functions including functional identities, asymptotics,
and their relation to various other
classes of special functions can be found e.g. in \cite{panda}.
The above integrals converge absolutely and define an analytic function
(at least) when the $z_i$ are not equal to zero and when
\ben
| {\rm arg}(z_i) | < \frac{\pi \Delta_i}{2}
\een
where
\ben\label{Deltadef}
\Delta_i := \sum_j A_{ij} - \sum_j C_{ij} + \sum_j B_{ij} - \sum_j D_{ij} \ .
\een
The formula reduces to the Gauss hypergeometric function and standard simple generalizations thereof
in special cases.

\section{Spherical harmonics, Gegenbauer polynomials, Kallen-Lehmann representation}
\label{app:B}

Here we give a ``spectral representation'' of the 2-point correlation function of an interacting field
analogous to the Kallen-Lehmann representation in Minkowski spacetime. The derivation of the formula
involves spherical harmonics in $D$-dimensions, so we briefly recall their basic properties, see e.g.~\cite{vilenken,axler}
for more details.

Spherical harmonics on the unit $S^D$ can be introduced via harmonic polynomials in the embedding space $\mr^{D+1}$.
A polynomial $P(X)$ on $\mr^{D+1}$ is called homogeneous of degree $h$ if $P(\lambda X) = \lambda^h P(X)$,
and it is called harmonic if it is a solution to the Laplace equation on $\mr^{D+1}$. The harmonic polynomials of
degree $h=L$ form a vector space of dimension $N(L,D) =
\frac{(2L+D-1)(L+D-2)!}{(D-1)!L!}$; spherical harmonics on $S^D$ of order $L$ are by definition just
the restriction of the harmonic polynomials to $S^D$. The spherical harmonics $Y_{Lj}(X), j = 1, ..., N(D,L)$
may be normalized so that
\ben
\sum_{Lm} Y_{Lm}(X_1)^* Y_{Lm}(X_2) = \delta (X_1, X_2) \ , \quad \int_{S^D} d\mu(X) \
Y_{Lm}(X)^* Y_{L'm'}(X) = \delta_{L,L'}\delta_{m,m'}
\een
where the $\delta$ function is that on $S^D$, defined with respect to the measure $d\mu$. Expressing the Laplacian
on $\mr^{D+1}$ in polar coordinates, on sees that the spherical
harmonics are eigenvalues of the Laplacian $\nabla^2$ on the $D$-sphere with eigenvalue $-L(L+D-1)$, so that $L$ may be viewed as the analog of the total angular momentum, and $m$ may be viewed as the analog of the magnetic quantum numbers. One
has
\ben\label{proj}
\sum_{m=1}^{N(D,L)} Y_{Lm}(X_1)^* Y_{Lm}(X_2) = \frac{2L+D-1}{{\rm vol}(S^{D-1})} \ C_L^{(D-1)/2}(Z) \, ,
\een
where $C^{\mu}_L$ are the Gegenbauer polynomials, where $Z$ is the point pair invariant.
The Gegenbauer polynomials are expressible in terms of a hypergeometric function,
\ben
C^{(D-1)/2}_L(Z) = \frac{\Gamma(L+D-1)}{\Gamma(D)\Gamma(L+1)} \ \hF \left( -L, L+D-1; D/2; \frac{1-Z}{2} \right) \ .
\een
Eq.~\eqref{proj} may be viewed as saying that the Gegenbauer polynomials are, up to normalization, the integral kernels
of the projector onto the eigenspace for the eigenvalue $-L(L+D-1)$ of the Laplacian on $S^D$. Since the dimension of this
eigenspace is equal to $N(D,L)$, one gets  the orthogonality relation
\ben
\int_{-1}^1 dZ \ (1-Z^2)^{D/2-1} \ C^{(D-1)/2}_L(Z) C^{(D-1)/2}_{L'}(Z) = N_{D,L} \ \delta_{L,L'}
\een
for $L,L' \in \mn_0$, with normalization factor
\ben
N_{D,L} = \frac{{\rm vol}(S^{D-1})}{{\rm vol}(S^D)} \ N(D,L) \ (2L+D-1)^{-2} \ .
\een
By the same argument, one gets the formula
\ben\label{euclhad}
\langle \phi(X_1) \phi(X_2) \rangle_0 = \frac{H^{D-2}}{{\rm vol}(S^{D-1})} \sum_{L=0}^\infty C_L^{(D-1)/2}(Z) \frac{2L+D-1}{-c(c+D-1) + L(L-D+1)}
\een
for the Euclidean Green's function of $(-\nabla^2+m^2)$ on the sphere $S^D$ of radius $H^{-1}$, where $c$ is as in the main text.
The above sum can be converted to a contour integral over $L$ with the help of a Watson-Sommerfeld transformation, as
observed in \cite{marolf1}:
\ben\label{euclhad1}
\langle \phi(X_1) \phi(X_2) \rangle_0 =   \int_C \frac{dL}{2\pi i} \ (2L+D-1) \ P_L \ \Delta_L(Z)
\een
where the contour $C$ is running parallel to the imaginary axis, leaving the poles in the denominator of
\ben
P_L := \frac{\pi^{1/2} \Gamma(D/2)}{\Gamma(D/2+1/2)} \  \frac{1}{-c(c+D-1) + L(L-D+1)}
\een
to the left, and the poles at $\mn_0$ of $\Delta_L$ to the right. The kernel
\ben\label{spec1}
\Delta_L(Z) = \frac{H^{D-2}}{(4\pi)^{D/2}} \frac{\Gamma(L+D-1)\Gamma(-L)}{\Gamma(D/2)} \ \hF \left( -L, L+D-1; D/2; \frac{1+Z}{2} \right)
\een
is equal to the free Euclidean Green's function for the mass parameter $M^2= -H^2 L(L+D-1)$.
$P_L$ is interpreted as the ``power spectrum'', or ``spectral density''. The original
sum is recovered if the contour integral is evaluated by means of the residue theorem; convergence
of the contour integral follows from eq.~\eqref{largeL}. Such representations
are closely related to the so-called Kallen-Lehmann representation in Minkowski-space, and
have been pioneered by \cite{bros} in deSitter spacetime.

A similar representation is also possible for the two-point function of the interacting theory.
In the context of perturbation theory, the analog of the above spectral representation \eqref{spec1}
for the 2-point function $\langle \phi(X_1) \phi(X_2) \rangle_{0,\lambda}$ for an interacting field has
spectral density given by a sum of contributions from individual Feynman diagrams $G$,
\ben\label{euclhad2}
P_{L,\lambda} =  \sum_{G} \frac{\lambda^V}{V!} \ {\rm sym}(G) \ P_{L,G} \ ,
\een
i.e. $P_{L,G}$ is the contribution from an individual Feynman diagram $G$. A formula for this can be
obtained straightforwardly using the orthogonality of the Gegenbauer polynomials, and our parametric formula~\eqref{ig4}
for $I_G$. To state the formula, let us define the graph $\tilde G$ to be the graph obtained from $G$
by closing off the two external legs, so that $\tilde G$ will have no external lines, and so that it has
one more internal line, called ``$\tilde l$''. Let $\tilde I_{G}(L)$ be the expression~\eqref{ig4},
with one $c \to L$ in the $H_{n_{12}}$-factor corresponding to the line $\tilde l$ which we closed off.
Then we have
\ben
P_{L,G} = P_{L,0} \ \frac{\pi^{1/2} \Gamma(D/2)}{\Gamma(D/2+1/2)}
\frac{\sin \pi L}{\pi}
\ \frac{(2L + D-1) \ \tilde I_G(L)}{-c(c+D-1) + L(L-D+1)}
\een
The contour $C$ in the spectral representation must now run parallel to the imaginary $L$-axis asymptotically,
and it must separate the poles of $P_{L,\lambda}$ from those of $\Delta_L$ (at $\mn_0$).
The spectral representation provides an alternative way of performing the analytic continuation of the
two-point function from the sphere to deSitter spacetime.

\section{Proof of thm.~\ref{mbthm}}\label{app:d}

We here give the lengthy proof of thm.~\ref{mbthm}. Recall from subsec.~\ref{app:c}
that $\U,\F$ are polynomials of parameters $\alpha_{ij}$, where there is one
parameter for each edge $(ij) \in G^*$, with $G^*$ the graph associated with $G$.
In this appendix, we find it convenient to set
\ben
\alpha_{ij} =
\begin{cases}
\e^{u_i} & \text{if $j=*, i=E+1, \dots, E+V$}\\
\e^{v_{ij}} & \text{otherwise if $i,j \neq *$,}
\end{cases}
\een
and we also set $H=1$.
The parameters $u_i, v_{ij} \in \mr$ are real. The main tool in the proof will
be the well-known Mellin-Barnes (MB) identity
\bena\label{mbformula}
&& (A_1 + \dots + A_{N+1})^{\omega}   \\
&=& \ \frac{1}{\Gamma(-\omega)}  \left( \prod_{i=1}^N \int_{K_i} \frac{dw_i}{2\pi i} \right) \Gamma(-\omega+\sum_{i=1}^N w_i)
A_{N+1}^{\omega - \sum w_i } \prod_{i=1}^N \Gamma(-w_i)  \prod_{i=1}^N A_i^{w_i}  \, , \non
\eena
for $\R(\omega) < 0, A_i \ge 0$ and for $N$ contours $K_i$ going parallel to the imaginary axis such that
\ben
\R(w_i) < 0, \quad \R \left[\omega-\sum w_i \right] < 0 \ .
\een
After these preliminaries, we come to the proof, which is divided into several steps.

\medskip
\noindent
{\bf Step 1:}
For $\R(\omega) < -V(D+1)/2$, $0>\R(z_j)>-(D+1)/2$ the following integrals are absolutely convergent:
\bena
&&\int_X \left[ \sum_{j,k} \e^{v_{ij}}(X_j-X_k)^2 + \sum_j \e^{u_j} X_j^2   \right]^\omega \non\\
&=&\pi^{V(D+1)} \frac{\Gamma(-\omega-\frac{V(D+1)}{2})}{\Gamma(-\omega)} \frac{\F(\vec u, \vec v)^{\omega+V(D+1)/2}}{\U(\vec u, \vec v)^{\omega + (V+1)(D+1)/2}}\\
&=& \int_X \int_{\vec z} \frac{\Gamma(-\omega+\sum_j z_j) \ \prod_j \Gamma(-z_j)}{\Gamma(-\omega)}
\left[\sum_{j,k} \e^{v_{ij}}(X_j-X_k)^2 \right]^{\omega-\sum_j z_j}  \ \prod_j \e^{u_j z_j} \ (X_j^2)^{z_j} \non
\ .
\eena
Here, there is one complex variable $z_j$ for each variable $u_j$, and the integral $\int_X$ denotes
an integral over $(\mr^{D+1})^V$ against the variables $X_{E+1}, \dots, X_{E+V}$. The second equality sign,
follows from the MB-formula, whereas the first one is obtained e.g. by manipulations similar to those
in sec.~\ref{sec:parametric}. Because the integrals are absolutely convergent, we may change the order
of the integrals in the last expression. Multiplying also both sides by $\Gamma(-\omega)$, we trivially get
\bena
&&\pi^{V(D+1)} \Gamma(-\omega-\frac{V(D+1)}{2}) \frac{\F(\vec u, \vec v)^{\omega+V(D+1)/2}}{\U(\vec u, \vec v)^{\omega + (V+1)(D+1)/2}}\\
&=& \int_{\vec z} \Gamma(-\omega+\sum_j z_j) \prod_j \Gamma(-z_j) \ \e^{\vec u \cdot \vec z} \ \int_X
\prod_j \ (X_j^2)^{z_j} \ \left[\sum_{j,k} \e^{v_{ij}}(X_j-X_k)^2 \right]^{\omega-\sum_j z_j} \   \non \ .
\eena

\medskip
\noindent
{\bf Step 2:}
We now apply an inverse Fourier transform $\int d\vec u \e^{-i\vec u \vec x}$ to both sides. The integrals
are absolutely convergent and given an analytic function in $\vec x$ for
\ben
\frac{D+1}{2} > \I(x_j) > 0 \ ,
\een
and we get, redefining also $\omega - \sum_j ix_j \to \omega$:
\bena
&&\pi^{V(D+1)} \frac{\Gamma(-\omega-\sum_j ix_j - \frac{V(D+1)}{2})}{\Gamma(-\omega) \prod_j \Gamma(-ix_j)}
\int_{\vec u}
\e^{-i\vec u \cdot \vec x} \ \frac{\F(\vec u, \vec v)^{\omega+\sum_j ix_j + V(D+1)/2}}{\U(\vec u, \vec v)^{\omega  + \sum_j ix_j + (V+1)(D+1)/2}}\\
&=&  \ \int_X\prod_j (X_j^2)^{ix_j} \
\left[\sum_{j,k} \e^{v_{ij}}(X_j-X_k)^2 \right]^{\omega} \   \non \ .
\eena
This equation is valid now for
\ben
\R(\omega)<0 \ ,
\een
and for $\I(x_j)$ in the above range.

\medskip
\noindent
{\bf Step 3:} We integrate both sides of the last equation against $(2\pi)^{-V} \int d\vec x$ over
a contour of constant $\I(x_j)$ in the above range. We also use that for such a contour $\int dx_j \ (X_j)^{ix_j}
= 2\pi \ \delta(X_j^2-1)$ in the sense of distributions. We are allowed to use this distributional identity because
the integrand on the right side is in $L^1 \cup C^\infty$ w.r.t. the variables $X_{E+1}, \dots, X_{E+V}$. Using
also that the measure on the sphere $S^D$ is $d\mu(X_j) = 2^{-1} \ \delta(X_j^2 - 1) \ d^{D+1} X_j$, we therefore get
\bena
&&\left( \prod_j \int_{S^D} d\mu(X_j) \right) \ \left[\sum_{j,k} \e^{v_{ij}}(X_j-X_k)^2 \right]^{\omega} = \\
&&2^{-V} \pi^{V(D+1)} \int_{\vec x} \frac{\Gamma(-\omega-\sum_j ix_j - \frac{V(D+1)}{2})}{\Gamma(-\omega) \prod_j \Gamma(-ix_j)}
\int_{\vec u}
\e^{-i\vec u \cdot \vec x} \ \frac{\F(\vec u, \vec v)^{\omega+\sum_j ix_j + V(D+1)/2}}{\U(\vec u, \vec v)^{\omega  + \sum_j ix_j + (V+1)(D+1)/2}}\non \ ,
\eena
which is valid at least for $\R(\omega)<0$ and $0<\I(x_j)<(D+1)/2$.

\medskip
\noindent
{\bf Step 4:} For $\R(\omega) < 0$ it is legal to use the MB formula on the integrand on the left side, which
gives
\ben
\left[\sum_{j,k} \e^{v_{ij}}(X_j-X_k)^2 \right]^{\omega} = \int_{\vec z} \frac{\prod_{j,k} \Gamma(-z_{jk})}{
\Gamma(-\omega)}
\ \e^{\vec v \cdot \vec z} \ t_{\vec z}(X_1, \dots, X_{V+E}) \ .
\een
Here, there is one integration variable $z_{jk}$ for each $v_{jk}$, except for one distinguished
but arbitrarily chose edge $(jk) = e \in \E G^*$. We must have $\R(z_{jk}) < 0$ along the integration
contours for {\em all} $z_{kl}$, and $\sum_{k,l} z_{kl} = \omega$. $t_{\vec z}$ is as in eq.~\eqref{tdef}.
We now take $\prod_j \int d\mu(X_j)$ of both sides of the equation. For sufficiently small $\epsilon > 0$ and $-\epsilon < \R(z_{jk}) < 0$,
the left and right side are in $L^1$ with respect to the integration over $X_{E+1}, \dots, X_{V+E} \in S^D$.
The integrals over $\vec z$ on the left side is also absolutely convergent due to standard estimates on the
$\Gamma$-function (Stirling's formula). Thus, we can exchange the integrals over $\vec X$ and $\vec z$, and
we obtain
\ben
\left( \prod_j \int_{S^D} d\mu(X_j) \right) \
\left[\sum_{j,k} \e^{v_{ij}}(X_j-X_k)^2 \right]^{\omega} =
\int_{\vec z} \frac{\prod_{j,k} \Gamma(-z_{jk})}{
\Gamma(-\omega)}
\ \e^{\vec v \cdot \vec z} \ \left( \prod_j \int_{S^D} d\mu(X_j) \right) \ t_{\vec z}(X_1, \dots, X_{V+E}) \ . \non
\een
We take an inverse Fourier transform $\int d\vec v \ \e^{-i\vec v \cdot \vec y}$ of this, where there is
one variable $y_{jk}$ for each of the variables $v_{jk}$. Then using the result of the previous step 3),
we get
\bena\label{step5}
&&\left( \prod_j \int_{S^D} d\mu(X_j) \right) \ t_{i\vec y}(X_1, \dots, X_{V+E}) = \\
&&2^{-V} \pi^{V(D+1)} \int_{\vec v}\int_{\vec x} \frac{\Gamma(-\sum_j ix_j -\sum_{j,k} iy_{jk}- \frac{V(D+1)}{2})}{\prod_j \Gamma(-ix_j) \prod_{j,k} \Gamma(-iy_{jk})}
\int_{\vec u}
\e^{-i\vec u \cdot \vec x-i\vec v \cdot \vec y} \ \frac{\F(\vec u, \vec v)^{\sum_j ix_j + \sum_{j,k} iy_{jk} + V(D+1)/2}}{\U(\vec u, \vec v)^{\sum_j ix_j +  \sum_{j,k} iy_{jk} + (V+1)(D+1)/2}}\non \ ,
\eena
which is valid at least for $\epsilon > \I(y_{jk}) > 0, \frac{D+1}{2} > \I(x_j) > 0$, and for
\ben\label{previouscond}
\sum_{j,k} \I(y_{jk}) + \sum_j \I(x_j) > \frac{V(D+1)}{2} \  .
\een

\medskip
\noindent
{\bf Step 5:} We now change the order of integration $\int_{\vec v}\int_{\vec x} \to \int_{\vec x} \int_{\vec v}$. This will be justified a posteriori in the end, because all integrals in the final expression will turn out to be
absolutely convergent\footnote{Indeed, we may replace the numerator in
eq.~\eqref{step5} by $\Gamma(-\sum_j i(1+\epsilon)x_j -\sum_{j,k} iy_{jk}- \frac{V(D+1)}{2})$.
This will make all integrals absolutely convergent by eq.~\eqref{Deltadef}, so we are free to exchange the order of integration. In the end, we get absolutely convergent integrals depending on $\epsilon$. At this stage,
we can take $\epsilon \to 0$ under those integrals, the limit being uniform. We omit the details.}. We then get the Fourier transform $\int_{\vec u, \vec v}
\e^{-i\vec u \cdot \vec x-i\vec v \cdot \vec y} \F^{\omega+V(D+1)/2}/\U^{\omega+(V+1)(D+1)/2}$,
where now  $\omega = \sum iy_{jk} + \sum ix_j$. We next rewrite this
by bringing the integrand of this into the form of a Mellin-integral, using the MB-identity, so that the Fourier transform can be read off.
The MB-identity can be applied provided that exponents of both $\U,\F$ have negative real part, i.e. that
\ben\label{zcontour}
-\frac{(V+1)(D+1)}{2} < \R(\omega)  < -\frac{V(D+1)}{2}  \ ,
\een
which is compatible with our previous condition~\eqref{previouscond}.
There is basically one Mellin-Barnes integration parameter for each term in $\U, \F$. As explained above
in subsec.~\ref{app:c},
the monomials from $\F$ are labeled by elements $F$ of the set
\ben
\T_E = \bigcup_{i,j \in \{*,1,\dots,E\}} \T_E(i,j) \ .
\een
Thus, $F \in \T_E$ if it is a spanning forest with $E$ trees. One of these trees connects a pair $i,j \in \{*,1, \dots, E\}$,
whereas the remaining $E-1$ trees each only have precisely one element from this set. There is one independent integration parameter $w_F$ for each such forest $F$, except for one dependent parameter corresponding to one particular forest.
Similarly, the monomials in $\U$ are labeled by forests $F \in \T_{E+1}$, see eq.~\eqref{utree}. There is again one
Mellin-Barnes integration parameter for each such monomial, except for one forest, which is dependent.
The two dependent parameters are related to the independent ones by the formula
\ben\label{wffree}
\sum_{F \in \T_E} w_F = \frac{V(D+1)}{2} + \omega \ ,
\een
for $\F$ and
\ben\label{wffree1}
\sum_{F \in \T_{E+1}} w_F = - \frac{(V+1)(D+1)}{2} - \omega \ ,
\een
for $\U$.
Furthermore, for any forest from either $\T_E$ or $\T_{E+1}$, we must have  $\R(w_F) < 0$.
Let us define the quantities
\ben\label{qij}
q_{rs}(\vec w) = \sum_{F \in \T_E(r,s)} w_F \ ,
\een
and
\bena \label{zkldef}
ix_{j}(\vec w) &=&  \sum_{F: \ F \owns (*j)} w_F  \ , \\
iy_{jk}(\vec w) &=& \sum_{F: \ F \owns (jk)} w_F \ .
\eena
Then using the MB-identity, we get
\bena
&& \Gamma(-\omega-\tfrac{V(D+1)}{2}) \frac{\F(\vec u, \vec v)^{\omega + V(D+1)/2}}{\U(\vec u, \vec v)^{\omega + (V+1)(D+1)/2}} \non\\
&=&  \int_{\vec w}
\frac{\prod_F \Gamma(-w_F)}{\Gamma(\omega+\frac{(V+1)(D+1)}{2})}  \
\prod_{r \neq s}^E (1-Z_{rs})^{q_{rs}(\vec w)} \exp[i\vec x \cdot \vec u + i\vec y \cdot \vec v] \, .
\eena
To read off the Fourier transform, we need to convert the integral over $\vec w$ into an integral
over the quantities $\vec x, \vec y$ as defined in
eq.~\eqref{qij}, and a set of complementary variables which we call $\vec h$. Thus, we are looking for
a map $(\vec x, \vec y, \vec h) \mapsto \vec w(\vec x, \vec y, \vec h)$ which must satisfy the following properties:
\begin{enumerate}
\item From eq.~\eqref{zkldef} together with eq.~\eqref{wffree}, we must have
\ben
\sum_{F \in \T_E} w_F(\vec x, \vec y,\vec h) = V(D+1)/2 + \sum_j ix_j + \sum_{j,k} iy_{jk} \ .
\een
\item
From eq.~\eqref{zkldef} together with eq.~\eqref{wffree1}, we must have
\ben
\sum_{F \in \T_{E+1}} w_F(\vec x, \vec y, \vec h) = -(V+1)(D+1)/2 - \sum_j ix_j - \sum_{j,k} iy_{jk} \ .
\een
\item
From eq.~\eqref{zkldef}, we must have $x_{j} = \sum_{F \owns (*j)} w_F(\vec x, \vec y, \vec h)$
and $y_{jk} = \sum_{F \owns (jk)} w_F(\vec x, \vec y, \vec h)$ and $\omega = \sum ix_j + \sum iy_{jk}$ .
\item The map $(\vec x,\vec y, \vec h) \mapsto \vec w(\vec x, \vec y,\vec h)$ should be linear in $\vec h$,
and have maximum possible rank compatible with the previous constraints, i.e. $|\T_E \cup \T_{E+1}| - |\E G^*| -1$.
\end{enumerate}
There are many ways choosing such a map.
One way is to pick, for any $(ij) \in \E G^*$, a suitable
corresponding forest $\Phi_{ij} \in \T_{E+1}$ such that $\Phi_{ij} \owns (ij)$, and e.g. one forest $F_0 \in \T_{E}$.
Then the variables $\vec h$ are defined as those variables $w_F$ such that $F \notin \{ F_0, \Phi_{ij} \mid (ij) \in \E G^*\}$,
and the variables $\vec x, \vec y$ are as defined in eq.~\eqref{qij}. The choice of the forests must be made in
such a way that the transformation $(\vec x, \vec y,\vec h) \to \vec w$ is invertible (it is automatically linear). It is not difficult to see that there are indeed (many) ways of achieving this. Furthermore, for a suitable
choice of the $\vec h$ contours, the $\vec x$ and $\vec y$ contours may be chosen so as
to be at $\epsilon > \I(y_{jk}) > 0, \frac{D+1}{2} > \I(x_j) > 0$, and this is compatible
with the range for which eq.~\eqref{step5} is valid. Hence, we can read off the
Fourier transform as:
\bena
&&\Gamma(\omega+\tfrac{V(D+1)}{2}) \int_{\vec u, \vec v}
\e^{-i\vec u \cdot \vec x-i\vec v \cdot \vec y} \ \frac{\F(\vec u, \vec v)^{\omega + V(D+1)/2}}{\U(\vec u, \vec v)^{\omega+(V+1)(D+1)/2}} \non \\
&=& \int_{\vec h}
\frac{\prod_F \Gamma(-w_F(\vec x, \vec y, \vec h))}{\Gamma(\omega+\frac{(V+1)(D+1)}{2})}  \
\prod_{r \neq s}^E (1-Z_{rs})^{q_{rs}(\vec x, \vec y, \vec h)} \ .
\eena
Then it follows from eq.~\eqref{step5} that:
\bena\label{step51}
&&\left( \prod_j \int_{S^D} d\mu(X_j) \right) \ t_{i\vec y}(X_1, \dots, X_{V+E}) =
2^{-V} \pi^{V(D+1)} \int_{\vec x} \frac{1}{\prod_j \Gamma(-ix_j) \prod_{j,k} \Gamma(-iy_{jk})} \non\\
&&\hspace{2cm} \times \ \int_{\vec h}
\frac{\prod_F \Gamma(-w_F(\vec x, \vec y, \vec h))}{\Gamma(\sum_j ix_j +\sum_{j,k} iy_{jk} +\frac{(V+1)(D+1)}{2})}  \
\prod_{r \neq s}^E (1-Z_{rs})^{q_{rs}(\vec x, \vec y, \vec h)} \ .
\eena
This formula is valid for $\epsilon > \I(y_{jk}) > 0, \frac{D+1}{2} > \I(x_j) > 0$.

\medskip
\noindent
{\bf Step 6:} We next analytically continue this result to a larger set of $\vec y$. From sec.~\ref{sec:ren},
we know that the left side is analytic in $i\vec y$ for which $\Delta_G(i\vec y) \cap -\mn_0 = \emptyset$.
When we analytically continue the right side, we must move the $\vec x,\vec h$ contours to new contours in such a way
that (i) the poles of $\Gamma(-w_F(\vec x, \vec y, \vec h)$, (ii) the poles of $\Gamma(-ix_j)$, and (iii) the
poles of $\Gamma(iy_{jk})$ all stay to the right. After that, we can assume that we have continued to
$\vec y$ in the range
\ben
\frac{n_{jk}(D-1)}{2} > \I(y_{jk}) > \frac{n_{jk}(D-2)}{2} \ .
\een
We multiply the result of step~5) by $\prod_{j,k} H_{n_{jk}}(iy_{jk})$ [cf.~\eqref{Hndef}] and integrate over
all $y_{jk}$, along contours in the above range. The result is
\bena\label{step6}
&&\int_{\vec y} \prod_{j,k} H_{n_{jk}}(iy_{jk})
\left( \prod_j \int_{S^D} d\mu(X_j) \right) \ t_{i\vec y}(X_1, \dots, X_{V+E}) =
2^{-V} \pi^{V(D+1)} \int_{\vec x,\vec y} \frac{\prod_{j,k} H_{n_{jk}}(iy_{jk})}{\prod_j \Gamma(-ix_j) \prod_{j,k} \Gamma(-iy_{jk})} \non\\
&&\hspace{2cm} \times \ \int_{\vec h}
\frac{\prod_F \Gamma(-w_F(\vec x, \vec y, \vec h))}{\Gamma(\sum_j ix_j +\sum_{j,k} iy_{jk} +\frac{(V+1)(D+1)}{2})}  \
\prod_{r \neq s}^E (1-Z_{rs})^{q_{rs}(\vec x, \vec y, \vec h)} \ .
\eena
The left side, is up to a constant, the Feynman integral $I_G$, by eq.~\eqref{ig}.

\medskip
\noindent
{\bf Step 7:}
In order to get the desired formula for $I_G$, we need to change the variables of integration from
$\vec x, \vec y, \vec h$ back to $\vec w$,
which by definition consists of all $w_F, F \in \T_E \cup \T_{E+1}$, except for the forest $F=\Phi$
given by
\ben
\Phi = \{ ((E+1)*), \dots, ((E+V)*) \} \in \T_{E+1} \
\een
After a straightforward calculation, one then obtains the formula~\eqref{ig4} stated in the theorem. The
contour integrals in that formula define a generalized $H$-function as described in appendix~\ref{app:A}.
The criteria for absolute convergence stated in eq.~\eqref{Deltadef} are verified as follows. Let
$\Delta_F$ be the parameter in eq.~\eqref{Deltadef} corresponding to the integration variable $w_F$. We note that
$|F \setminus \Phi|$ is the number of times that $w_F$ appears in the sum $\sum_{(ij) \notin \Phi} \sum_{F \owns (ij)} w_F$, whereas $|\Phi \setminus F|$ is the number of times that it appears in the sum $\sum_{(ij) \in \Phi} \sum_{F / \!\!\!\! \owns (ij)} w_F$. Consequently, e.g. when $F \in \T_{E+1}$
\ben
\Delta_F = 2 + 3|F \setminus \Phi| - |\Phi \setminus F| = 2(1 + |F \setminus \Phi|) \ge 2 \ ,
\een
where in the second step we have used that $|F|=|\Phi|=V$ so that $|F \setminus \Phi| = |\Phi \setminus F|$.
In the other case when $F \in \T_E$, we similarly have $|F|=V+1$, $|\Phi |= V$ and hence
$|F \setminus \Phi| = |\Phi \setminus F| + 1$, implying again
\ben
\Delta_F = 1 + 3|F \setminus \Phi| - |\Phi \setminus F| = 2(1 + |F \setminus \Phi|) \ge 2 .
\een
Therefore, by the criteria stated in appendix~\ref{app:A},
the integrations over the $\vec w$ in eq.~\eqref{ig4} are absolutely convergent
for any $|{\rm arg}(1-Z_{rs})| < \pi \le \pi \inf \{ \Delta_F \mid F \in \T_E \}/2$.
This also justifies a posteriori performing the integrations
over $\vec y$ in step~5),6) and the changes of integration variables/order of integration.


\begin{thebibliography}{99}
\bibitem{allen}
B. Allen: ``Vacuum states in deSitter space,'' Phys. Rev. {\bf D}32, 3136 (1985)

\bibitem{axler}
S.~Axler, P.~Bourdon, and W.~Ramey:
{\em Harmonic Function Theory},
Springer, New York, 2001.


\bibitem{anderson}
  M.~T.~Anderson,
  ``Existence and stability of even dimensional asymptotically de Sitter
  spaces,''
  Annales Henri Poincare {\bf 6}, 801 (2005)



\bibitem{binoth1}
  T.~Binoth, J.~P.~Guillet and G.~Heinrich,
  ``Reduction formalism for dimensionally regulated one-loop N-point
  integrals,''
  Nucl.\ Phys.\  B {\bf 572}, 361 (2000)
  [arXiv:hep-ph/9911342].



\bibitem{birke}
  L.~Birke and J.~Fr\" ohlich,
  ``KMS, etc,''
  Rev.\ Math.\ Phys.\  {\bf 14}, 829 (2002)
  [arXiv:math-ph/0204023].

\bibitem{weinzierl2}
 C.~Bogner and S.~Weinzierl,
  ``Feynman graph polynomials,''
  Int.\ J.\ Mod.\ Phys.\  A {\bf 25}, 2585 (2010)
  [arXiv:1002.3458 [hep-ph]].

\bibitem{weinzierl1}
  C.~Bogner and S.~Weinzierl,
  ``Resolution of singularities for multi-loop integrals,''
  Comput.\ Phys.\ Commun.\  {\bf 178}, 596 (2008)

\bibitem{bros1}
  J.~Bros, H.~Epstein and U.~Moschella,
  ``Particle decays and stability on the de Sitter universe,''
  Annales Henri Poincare {\bf 11}, 611 (2010);   J.~Bros, H.~Epstein and U.~Moschella,
  ``Lifetime of a massive particle in a de Sitter universe,''
  JCAP {\bf 0802}, 003 (2008)

\bibitem{bros}
  J.~Bros, H.~Epstein and U.~Moschella,
  ``Analyticity properties and thermal effects for general quantum field
  theory on de Sitter space-time,''
  Commun.\ Math.\ Phys.\  {\bf 196}, 535 (1998);   J.~Bros and U.~Moschella,
  ``Two-point Functions and Quantum Fields in de Sitter Universe,''
  Rev.\ Math.\ Phys.\  {\bf 8}, 327 (1996)

\bibitem{brunetti}
  R.~Brunetti and K.~Fredenhagen,
  ``Microlocal analysis and interacting quantum field theories:
  Renormalization on physical backgrounds,''
  Commun.\ Math.\ Phys.\  {\bf 208}, 623 (2000)

  \bibitem{holfred}
 R.~Brunetti, K.~Fredenhagen and S.~Hollands,
  ``A remark on alpha vacua for quantum field theories on de Sitter space,''
  JHEP {\bf 0505}, 063 (2005)

\bibitem{bfk}
  R.~Brunetti, K.~Fredenhagen and M.~Kohler,
  ``The microlocal spectrum condition and Wick polynomials of free fields on
  curved spacetimes,''
  Commun.\ Math.\ Phys.\  {\bf 180}, 633 (1996)
  [arXiv:gr-qc/9510056].

\bibitem{freddue}
  R.~Brunetti, M.~Duetsch and K.~Fredenhagen,
  ``Perturbative Algebraic Quantum Field Theory and the Renormalization
  Groups,''
  arXiv:0901.2038 [math-ph].

\bibitem{bunch}
  T.~S.~Bunch,
  ``BPHZ Renormalization Of Lambda Phi**4 Field Theory In Curved Space-Time,''
  Annals Phys.\  {\bf 131}, 118 (1981).

\bibitem{buchholz}
  D.~Buchholz and J.~Schlemmer,
  ``Local Temperature in Curved Spacetime,''
  Class.\ Quant.\ Grav.\  {\bf 24}, F25 (2007)

\bibitem{camporesi}
 R.~Camporesi and A.~Higuchi,
  ``On The Eigen Functions Of The Dirac Operator On Spheres And Real Hyperbolic
  Spaces,''
  J.\ Geom.\ Phys.\  {\bf 20}, 1 (1996)

\bibitem{cardoso}
 V.~Cardoso, J.~Natario and R.~Schiappa,
  ``Asymptotic quasinormal frequencies for black holes in  non-asymptotically
  flat spacetimes,''
  J.\ Math.\ Phys.\  {\bf 45}, 4698 (2004)
  [arXiv:hep-th/0403132].

\bibitem{eg}
H. Epstein, V. Glaser: ``The role of locality in quantum field theory,''
Ann. Poincare Theor. Phys. {\bf A19} 211-295 (1973)

\bibitem{friedlaender}
F. G. Friedlander, {\it The wave equation on a curved space-time}, Cambridge University Press (1975)

\bibitem{friedrich}
H. Friedrich: ``Existence and structure of $n$-geodesically complete of future complete solutions
of Einstein's equations with smooth asymptotic structure,'' Commun. Math. Phys. {\bf 107}, 587 (1986);
``Existence and structure of past asymptotically simple solutions of Einstein's field equations
with positive cosmological constant,'' J. Geom. Phys. {\bf 3} 101 (1986)

\bibitem{garcia}
  J.~M.~Gracia-Bondia and S.~Lazzarini,
  ``Connes-Kreimer-Epstein-Glaser renormalization,''
  arXiv:hep-th/0006106.

\bibitem{gibbons}
  G.~W.~Gibbons and S.~W.~Hawking,
  ``Cosmological Event Horizons, Thermodynamics, And Particle Creation,''
  Phys.\ Rev.\  D {\bf 15}, 2738 (1977).

\bibitem{goldstein}
K. Goldstein: ``A note on $\alpha$-vacua and interacting field theory on deSitter
spacetime,'' Nucl. Phys. {\bf B}699, 325 (2003)

\bibitem{binoth2}
  G.~Heinrich and T.~Binoth,
  ``A general reduction method for one-loop N-point integrals,''
  Nucl.\ Phys.\ Proc.\ Suppl.\  {\bf 89}, 246 (2000)
  [arXiv:hep-ph/0005324].

\bibitem{hepp}
  K.~Hepp,
  ``Proof of the Bogolyubov-Parasiuk theorem on renormalization,''
  Commun.\ Math.\ Phys.\  {\bf 2}, 301 (1966).

\bibitem{higuchi}
A. Higuchi and S. S. Kouris: ``The covariant graviton propagator in deSitter spacetime,''
Class. Quant. Grav.~{\bf 18} 4317 (2001); ``On the scalar sector of the covariant graviton
two-point function in deSitter spacetime,'' Class. Quant. Grav.~{\bf 18} 2933 (2001);
A.~Higuchi and R.~H.~Weeks, ``The physical graviton two-point function in deSitter
spacetime with $S^3$ spatial sections,'' Class. Quant. Grav.~{\bf 20} 3005 (2003);

\bibitem{higuchi1}
A. Higuchi: ``Tree level vacuum instability in an interacting field theory in deSitter spacetime,''
arXiv:0809.1255; ``Decay of the free-theory vacuum of scalar field theory in deSitter spacetime
in the interaction picture,'' Class. Quant. Grav.~{\bf 26} 072001 (2009)

\bibitem{higuchi2}
  A.~Higuchi, Y.~C.~Lee and J.~R.~Nicholas,
  ``More on the covariant retarded Green's function for the electromagnetic
  field in de Sitter spacetime,''
  Phys.\ Rev.\  D {\bf 80}, 107502 (2009);   M.~Faizal and A.~Higuchi,
  ``On the FP-ghost propagators for Yang-Mills theories and perturbative
  quantum gravity in the covariant gauge in de Sitter spacetime,''
  Phys.\ Rev.\  D {\bf 78}, 067502 (2008)
  [arXiv:0806.3735 [gr-qc]].


\bibitem{hollandswald1}
  S.~Hollands and R.~M.~Wald,
  ``Local Wick polynomials and time ordered products of quantum fields in
  curved spacetime,''
  Commun.\ Math.\ Phys.\  {\bf 223}, 289 (2001)

\bibitem{hollandswald2}
  S.~Hollands and R.~M.~Wald,
  ``Existence of local covariant time ordered products of quantum fields in
  curved spacetime,''
  Commun.\ Math.\ Phys.\  {\bf 231}, 309 (2002)

\bibitem{hollandswald3}
  S.~Hollands and R.~M.~Wald,
  ``Axiomatic quantum field theory in curved spacetime,''
  Commun.\ Math.\ Phys.\  {\bf 293}, 85 (2010)

\bibitem{hollands1}
  S.~Hollands,
  ``The operator product expansion for perturbative quantum field theory in
  curved spacetime,''
  Commun.\ Math.\ Phys.\  {\bf 273}, 1 (2007)

\bibitem{hollands2}
  S.~Hollands,
  ``Renormalized Quantum Yang-Mills Fields in Curved Spacetime,''
  Rev.\ Math.\ Phys.\  {\bf 20}, 1033 (2008)
  [arXiv:0705.3340 [gr-qc]].

\bibitem{hor}
L. Hormander: {\em The analysis of linear partial differential operators I''} 2nd edition, Springer Verlag
(1990)

\bibitem{ishibashi}
  H.~Kodama and A.~Ishibashi,
  ``A master equation for gravitational perturbations of maximally  symmetric
  black holes in higher dimensions,''
  Prog.\ Theor.\ Phys.\  {\bf 110}, 701 (2003)
  [arXiv:hep-th/0305147].

\bibitem{iz}
C. Itzykson et J. B. Zuber, {\it Quantum Field Theory}, Dover Publ. (2005)

\bibitem{jaekel}
C. Jaekel, in preparation

\bibitem{jones}
D. S. Jones,
``Asymptotics of the hypergeometric function,''
Mathematical Methods in the Applied Sciences {\bf 24} 369 (2001)

\bibitem{kaywald}
 B.~S.~Kay and R.~M.~Wald,
  ``Theorems on the Uniqueness and Thermal Properties of Stationary,
  Nonsingular, Quasifree States on Space-Times with a Bifurcate Killing
  Horizon,''
  Phys.\ Rept.\  {\bf 207}, 49 (1991).

\bibitem{keller1}
 K.~J.~Keller,
  ``Euclidean Epstein-Glaser Renormalization,''
  J.\ Math.\ Phys.\  {\bf 50}, 103503 (2009)
  [arXiv:0902.4789 [math-ph]].


\bibitem{keller2}
 K.~J.~Keller,
  ``Dimensional Regularization in Position Space and a Forest Formula for
  Regularized Epstein-Glaser Renormalization,''
  arXiv:1006.2148 [math-ph].


\bibitem{rivasseau}
 T.~Krajewski, V.~Rivasseau, A.~Tanasa and Z.~Wang,
  ``Topological Graph Polynomials and Quantum Field Theory, Part I: Heat Kernel
  Theories,''
  arXiv:0811.0186 [math-ph].

 \bibitem{rivasseau2}
J. Magnen, V. Rivasseau: ``Constructive $\phi^4$ field theory without tears,''
Annales Henri Poincare {\bf 9}, 403-424 (2008).
  [arXiv:0706.2457 [math-ph]].


\bibitem{marolf1}
 D.~Marolf and I.~A.~Morrison,
  ``The IR stability of de Sitter: Loop corrections to scalar propagators,''
  arXiv:1006.0035 [gr-qc].

\bibitem{marolf2}
 D.~Marolf and I.~A.~Morrison, ``The IR-stability of deSitter QFT: results at
 all orders.'' arXiv:1010.5327 [gr-qc]

\bibitem{moretti}
 V.~Moretti,
  ``Proof of the symmetry of the off-diagonal Hadamard/Seeley-deWitt's
  coefficients in C(infinity) Lorentzian manifolds by a 'local Wick
  rotation',''
  Commun.\ Math.\ Phys.\  {\bf 212}, 165 (2000)

\bibitem{mottola}
  E.~Mottola,
  ``Particle Creation In De Sitter Space,''
  Phys.\ Rev.\  D {\bf 31}, 754 (1985);  P.~Mazur and E.~Mottola,
  ``Spontaneous Breaking Of De Sitter Symmetry By Radiative Effects,''
  Nucl.\ Phys.\  B {\bf 278}, 694 (1986);   I.~Antoniadis and E.~Mottola,
  ``Graviton Fluctuations In De Sitter Space,''
  J.\ Math.\ Phys.\  {\bf 32}, 1037 (1991).

\bibitem{osterwalder}
  K.~Osterwalder and R.~Schrader,
  ``Axioms For Euclidean Green's Functions. 1,''
  Commun.\ Math.\ Phys.\  {\bf 31}, 83 (1973).
  ``Axioms For Euclidean Green's Functions. 2,''
  Commun.\ Math.\ Phys.\  {\bf 42}, 281 (1975).

\bibitem{panda}
H. M. Srivastava, and R. Panda: ``Some bilateral generating functions for a class of
generalized hypergeometric polynomials,'' J. Reine angew. Math. {\bf 283}, 265 (1976);
Expansion theorems for the $H$-function of several complex variables,'' J. Reine angew. Math. {\bf 288}, 129 (1976);
B.~L.~Mathur, ``Some results concerning a special function of several complex variables,''
Indian J. pure appl. Math. {\bf 12}(8), 1001 (1980); J.~Prathima and T.M.~Vasudevan~Nambisan: ``Multiple Mellin-transforms of
generalized $H$-function of $r$ variables,'' Adv. Theor. Appl. Math. {\bf 5}, 91 (2010)

\bibitem{pinter}
  G.~Pinter,
  ``The Hopf algebra structure of Connes and Kreimer in Epstein-Glaser
  renormalization,''
  Lett.\ Math.\ Phys.\  {\bf 54}, 227 (2000)

\bibitem{polyakov}
A. M. Polyakov: ``DeSitter space and eternity,'' Nucl. Phys. {\bf B797}
199 (2008); ``Decay of vacuum energy,'' arXiv:0912.5503



\bibitem{rad}
  M.~J.~Radzikowski,
  ``Micro-Local Approach To The Hadamard Condition In Quantum Field Theory On
  Curved Space-Time,''
  Commun.\ Math.\ Phys.\  {\bf 179}, 529 (1996).



\bibitem{sanders}
  K.~Sanders,
  ``Equivalence of the (generalised) Hadamard and microlocal spectrum condition
  for (generalised) free fields in curved spacetime,''
  Commun.\ Math.\ Phys.\  {\bf 295}, 485 (2010)
  [arXiv:0903.1021 [math-ph]].

\bibitem{stroh}
  A.~Strohmaier, R.~Verch and M.~Wollenberg,
  ``Microlocal analysis of quantum fields on curved spacetimes: Analytic
  wavefront sets and Reeh-Schlieder theorems,''
  J.\ Math.\ Phys.\  {\bf 43}, 5514 (2002)
  [arXiv:math-ph/0202003].

\bibitem{woodard}
 N.~C.~Tsamis and R.~P.~Woodard,
  ``Quantum Gravity Slows Inflation,''
  Nucl.\ Phys.\  B {\bf 474}, 235 (1996);
  N.~C.~Tsamis and R.~P.~Woodard,
  ``The quantum gravitational back-reaction on inflation,''
  Annals Phys.\  {\bf 253}, 1 (1997);
  ``Strong infrared effects in quantum gravity,''
  Annals Phys.\  {\bf 238}, 1 (1995);
  ``The Structure of perturbative quantum gravity on a De Sitter background,''
  Commun.\ Math.\ Phys.\  {\bf 162}, 217 (1994);
  N.~C.~Tsamis and R.~P.~Woodard,
  ``Physical Green's functions in quantum gravity,''
  Annals Phys.\  {\bf 215}, 96 (1992).




\bibitem{tutte}
W. T. Tutte, {\it Graph Theory} Cambridge University Press (2001)



\bibitem{tanaka}
 Y.~Urakawa and T.~Tanaka,
  ``Natural selection of inflationary vacuum required by infra-red regularity
  and gauge-invariance,''
  arXiv:1009.2947 [hep-th]; ``IR divergence does not affect the gauge-invariant curvature perturbation,''
  arXiv:1007.0468 [hep-th];
  ``Influence on observation from IR divergence during inflation -- Multi field
  inflation --,''
  Prog.\ Theor.\ Phys.\  {\bf 122}, 1207 (2010); ``Influence on Observation from IR Divergence during Inflation. I,''
  Prog.\ Theor.\ Phys.\  {\bf 122}, 779 (2009)


\bibitem{vilenken}
N. Y. Vilenken and A. U. Klimyk: {\it ``Representations of Lie-Groups and Special functions,''}
vol. 1--3, Dodrecht, Kluwer Acad. Publ., 1991





\bibitem{zimmermann}
  W.~Zimmermann,
  ``Convergence of Bogolyubov's method of renormalization in momentum space,''
  Commun.\ Math.\ Phys.\  {\bf 15}, 208 (1969)
  [Lect.\ Notes Phys.\  {\bf 558}, 217 (2000)].

\end{thebibliography}
\end{document}